\documentclass[aps,prb,twocolumn,longbibliography]{revtex4-2}
\usepackage{graphicx}
\usepackage{float}
\usepackage{bm}
\usepackage{amsmath,amssymb}
\usepackage[cal=boondoxo]{mathalfa}
\allowdisplaybreaks
\usepackage[bookmarksnumbered,bookmarksopen]{hyperref}

\newcommand* {\braket}[1]{\langle {#1} \rangle}
\newcommand* {\bra}[1]{\langle {#1} |}
\newcommand* {\ket}[1]{| {#1} \rangle}
\newcommand* {\vek}[1]{{\bm{\mathrm{#1}}}}
\newcommand* {\vekc}[1]{{\bm{\mathcal{#1}}}}
\newcommand* {\vekt}[1]{\bm{\mathrm{#1}}'}
\newcommand* {\frack}[2]{{\textstyle\frac{#1}{#2}}}
\DeclareMathOperator{\trace}{tr}

\usepackage{multirow}
\usepackage{array}
\newcolumntype {L}{>{$}l<{$}}             
\newcolumntype {C}{>{$}c<{$}}             
\newcolumntype {R}{>{$}r<{$}}             
\newcolumntype {s}[1]{@{\hspace*{#1}}}    
\newcolumntype {S}[1]{@{\extracolsep{#1}}} 
\newcommand* {\tvek}[2][c]{\left( \begin{array}{s{0.15em}#1s{0.15em}}
     #2\end{array} \right)}

\newcommand* {\Hspace}[1][]{H (\vek{r} #1)}
\newcommand {\T}{\mskip 0.4\thinmuskip}
\newcommand* {\seitz}[2]{\lbrace #1 \T | \T #2 \rbrace}
\newcommand* {\group}{G} 
\newcommand* {\spacegroup}{\mathcal{G}} 
\newcommand* {\pointgroup}{\mathcal{G}_0} 
\newcommand* {\kgroup}[1][\vek{k}]{\mathcal{G}_{#1}} 
\newcommand* {\sitegroup}[1][W]{\mathcal{G}_{#1}} 
\newcommand* {\tshift}{\vek{\tau}} 
\newcommand* {\Tshift}{T} 
\newcommand* {\spacesym}{\mathcal{T}} 
\newcommand* {\pwbase}{\mathcal{q}} 
\newcommand* {\pwBloch}{{Q}} 
\newcommand* {\pwsBloch}{\mathcal{Q}} 

\begin{document}
\title{Effective Dynamics of 2D Bloch Electrons in External Fields
Derived From Symmetry}

\author{E. A. Fajardo}
\altaffiliation[On leave from ]{Department of Physics, Mindanao State University--Main Campus, Marawi City, Lanao del Sur, Philippines 9700}
\affiliation{Department of Physics, Northern Illinois University,
DeKalb, Ilinois 60115, USA}

\author{R. Winkler}
\affiliation{Department of Physics, Northern Illinois University,
DeKalb, Illinois 60115, USA}
\affiliation{Materials Science Division, Argonne National
Laboratory, Argonne, Illinois 60439, USA}

\date{4 September 2019}

\begin{abstract}
We develop a comprehensive theory for the effective dynamics of Bloch electrons based on symmetry.  We begin with a scheme to systematically derive the irreducible representations (IRs) characterizing the Bloch eigenstates in a crystal.  Starting from a tight-binding (TB) approach, we decompose the TB basis functions into localized symmetry-adapted atomic orbitals and crystal-periodic symmetry-adapted plane waves.  Each of these two subproblems is independent of the details of a particular crystal structure and it is largely independent of the relevant aspects of the other subproblem, hence permitting for each subproblem an independent universal solution.  Taking monolayer MoS$_2$ and few-layer graphene as examples, we tabulate the symmetrized $p$ and $d$ orbitals as well as the symmetrized plane-wave spinors relevant for these crystal structures.  The symmetry-adapted basis functions block-diagonalize the TB Hamiltonian such that each block yields eigenstates transforming according to one of the IRs of the group of the wave vector $\kgroup$.

For many crystal structures, it is possible to define multiple distinct coordinate systems such that for wave vectors $\vek{k}$ at the border of the Brillouin zone the IRs characterizing the Bloch states depend on the coordinate system, i.e., these IRs of $\kgroup$ are not uniquely determined by the symmetry of a crystal structure.  The different coordinate systems are related by a coordinate shift that results in a rearrangement of the IRs of $\kgroup$ characterizing the Bloch states.  We illustrate this rearrangement with three coordinate systems for MoS$_2$ and trilayer graphene.

The freedom to choose different distinct coordinate systems can simplify the symmetry analysis of the Bloch states.  Given the IRs of the Bloch states in one coordinate system, a rearrangement lemma yields immediately the IRs of the Bloch states in the other coordinate systems.  The rearrangement of the IRs in different coordinate systems does not affect observable physics such as selection rules or the effective Hamiltonians for the dynamics of Bloch states in external fields.

Using monolayer MoS$_2$ as an example, we combine the symmetry analysis of its bulk Bloch states with the theory of invariants to construct a generic multiband Hamiltonian for electrons near the $\vek{K}$ point of the Brillouin zone.  The Hamiltonian includes the effect of spin-orbit coupling, strain and external electric and magnetic fields. Invariance of the Hamiltonian under time reversal yields additional constraints for the allowed terms in the Hamiltonian and it determines the phase (real or imaginary) of the prefactors.
\end{abstract}

\maketitle


\section{Introduction}

Near a band extremum, the electron dynamics in a crystalline solid resembles the dynamics of free electrons in the absence of the periodic crystal potential. In the multiband envelope-function approximation (EFA) the electrons are characterized by an $N \times N$ Hamiltonian $\mathcal{H}$ for $N$-component spinors conceptually similar to relativistic electrons described by the Dirac equation \cite{lut55, kit63, bir74, bas88, win03}.  The simplest approach within the EFA is the effective-mass approximation (EMA) that represents the electron dynamics by a Schr\"odinger equation with effective mass $m^\ast$ reflecting the curvature of the band dispersion $E(\vek{k})$.  External electric and magnetic fields $\vekc{E}$ and $\vekc{B}$ break the lattice periodicity of the crystal structure.  It is an important advantage of EFA and EMA that they allow one to incorporate the field $\vekc{E}$ by adding the corresponding scalar potential $\Phi$ to the diagonal of the Hamiltonian, and the operator of crystal momentum $\hbar\vek{k} = -i\hbar\nabla$ is replaced by $-i\hbar\nabla + e\vek{A}$, where $\vek{A}$ is the vector potential for the magnetic field $\vekc{B}$.  Other perturbations such as spin-orbit coupling, strain and electron-phonon coupling can likewise be included in the Hamiltonian \cite{bir74}.  This is a major reason why EFA and EMA are very popular for theoretical studies of both bulk semiconductors (e.g., Refs.~\cite{dre55, lip70, bir74, suz74, hen74, tre79, ran79}) and semiconductor quantum structures (e.g., Refs.~\cite{bas88, ivc97a, win03, hau04, net09, ihn10, kor15}).

The form of the Hamiltonian $\mathcal{H}$ depends on the symmetry of the crystal structure and more specifically on the symmetry of the bulk electronic states that are included in $\mathcal{H}$ \cite{lut56, kan66, bir74}.  The relevant symmetry group for states with wave vector $\vek{k}$ is the point group $\kgroup$ which includes those symmetry elements of the crystallographic point group $\pointgroup$ (crystal class) which either leave $\vek{k}$ unchanged or map $\vek{k}$ onto an equivalent wave vector.  The symmetry of individual states at $\vek{k}$ is characterized by the respective irreducible representations (IRs) of $\kgroup$ according to which these states transform.  The general form of the Hamiltonian $\mathcal{H}$ including its dependence on, e.g., spin-orbit coupling, strain and  external fields can then be derived from its invariance under $\kgroup$ \cite{lut56, bir74}. Here we develop a general theory to determine the IRs of Bloch functions for a given wave vector $\vek{k}$, focusing for clarity on symmorphic space groups.  Using a tight-binding (TB) approach along with the fact that the atomic orbitals are localized in the vicinity of the atomic sites we demonstrate that the TB basis functions can be factorized into localized symmetry-adapted atomic orbitals and crystal-periodic symmetry-adapted plane waves. Each of these two subproblems permits a universal classification, independent of the details of a particular crystal structure and also largely independent of the other subproblem. The symmetrized atomic orbitals depend only on the angular momentum of the atomic orbitals and the point group $\kgroup$ of the wave vector~$\vek{k}$; but these orbitals are independent of the specific type of atom and the details of the crystal structure.  The symmetrized plane waves form discrete Bloch functions that depend on the wave vector $\vek{k}$ and the Wyckoff positions of the atoms in a crystal structure; but they are independent of the type of atoms occupying these positions.  The symmetry-adapted basis functions block-diagonalize the TB Hamiltonian such that each block yields eigenstates transforming according to one of the IRs of the group of the wave vector $\kgroup$.

Given the symmetry group $\group$ of a quantum system, the IRs of $\group$ are generally assumed to provide a distinct label for the eigenstates of the Hamiltonian, as noted by Wigner: ``\emph{The representation of the group of the Schr\"odinger equation which belongs to a particular eigenvalue is uniquely determined up to a similarity transformation}.'' (Ref.\ \cite{wig59}, p.~110, highlighting adopted from Ref.\ \cite{wig59}).  This uniqueness of the IRs is immediately relevant for many physical properties of a physical system that depend on the symmetry of its electronic states.  For example, the Wigner-Eckart theorem allows one to express the selection rules for optical transitions in terms of the IRs of the initial and final states between which a transition occurs \cite{wig59}.  Similarly, the EFA Hamiltonians $\mathcal{H}$ depend on the IRs of the bands described by $\mathcal{H}$, as noted above.
We demonstrate that the IRs characterizing the Bloch eigenstates in certain crystals including transition metal dichalcogenides (TMDCs) are \emph{not unique}, but they depend on the coordinate system used to describe the space group symmetries of these materials \cite{bir66, mor68, cor71, cor72}.  We show that distinct valid coordinate systems are related by a coordinate shift that defines a rearrangement representation.  The IRs of the electronic states in the different coordinate systems are then related via a rearrangement lemma that facilitates the symmetry analysis of Bloch states.  Also, we show how important physics including optical selection rules and EFA Hamiltonians $\mathcal{H}$, despite the rearrangement of band IRs, does not depend on the coordinate system being used.

Our general theory applies to any crystalline material.  For a detailed example, we focus on a monolayer of the TMDC MoS$_2$.  TMDCs are of the general form $TX_2$, where $T$ is a transition-metal such as Mo or W and $X$ is a chalcogen which can be S, Se, or Te. Three-dimensional (3D) bulk $TX_2$ consists of covalently bonded 2D monolayers coupled vertically by weak van der Waals forces \cite{mat73oct}, making it possible to obtain monolayers via, e.g., mechanical exfoliation \cite{mak10}. Electronic band structure calculations have shown that bulk 2H-MoS$_2$ is a semiconductor \cite{mat73oct}. More recently, optical spectroscopy \cite{mak10} and theoretical studies \cite{ell11, kad12, yun12} found that decreasing the number of layers changes the fundamental gap from indirect to direct in the limit of a single monolayer. The spin-dependent dispersion of monolayer TMDCs has been studied using TB \cite{xia12, cap13} and $\vek{k} \cdot \vek{p}$ methods \cite{kor13, kor15}.  See Refs.~\cite{jar14, aja16, man17, wan18} for general reviews of 2D TMDCs.  In this paper, we combine our symmetry analysis for the bulk Bloch states in monolayer  MoS$_2$ with the theory of invariants \cite{bir74} to derive a generic multiband EFA Hamiltonian for electrons near the $\vek{K}$ point of the Brillouin zone (BZ).  The Hamiltonian includes the effect of strain, external electric and magnetic fields, spin and valley degrees of freedom.  For comparison, we also perform a symmetry analysis for few-layer graphene \cite{net09} which confirms earlier work \cite{win10, win15}.
We note that our work expands on the theory of IRs for point and space groups \cite{streitwolf71, bir74, dre08}.  It is conceptually rather different from recent work on band representations \cite{bra17, bra18, can18}.

In Sec.~\ref{sec:bloch-sym}, we develop the general theory of the symmetry of TB Bloch functions.  The decomposition of TB wave functions is discussed in Sec.~\ref{sec:tb-decompose} followed by detailed discussions of the symmetrized atomic orbitals (Sec.~\ref{sec:trafo:atom:orbitals}) and symmetrized plane waves (Sec.~\ref{sec:trafo-plane-wave}). The rearrangement of the IRs of Bloch states under a change of the coordinate system is discussed in Sec.~\ref{sec:band-amb}.  We use the general formalism of Sec.~\ref{sec:bloch-sym} to derive the symmetry of bulk Bloch states in monolayer TMDCs (Sec.~\ref{sec:bloch-sym-mos2}) such as MoS$_2$ and to few-layer graphene (Sec.~\ref{sec:bloch-sym-graphene}).
We show in Sec.~\ref{sec:sel-rules} how optical selection rules are not affected by the rearrangement of IRs under a change of coordinate system.  In Sec.~\ref{sec:theory-inv} we derive the generic invariant expansion of the EFA Hamiltonian $\mathcal{H}$ for MoS$_2$.  Section~\ref{sec:conclusion} contains our conclusions.

\section{Symmetry of Bloch Functions}
\label{sec:bloch-sym}

Very generally, the eigenstates of a Hamiltonian transform according to an IR of the symmetry group of the Hamiltonian.  In band theory it is thus an important goal to determine the IRs of the energy bands $E_n (\vek{k})$ and corresponding Bloch functions $\Psi_{n \vek{k}} (\vek{r})$, where the symmetry group $\kgroup$ at a given wave vector $\vek{k}$ is called the point group of the wave vector $\vek{k}$.  In this section we discuss a general method for determining the transformation properties of Bloch functions with a certain wave vector $\vek{k}$, which allows us to determine the corresponding IRs of $\kgroup$. We also discuss a rearrangement lemma for the IRs characterizing the Bloch functions in a crystal. Applications to specific materials such as monolayer MoS$_2$ will be discussed in subsequent sections.

\subsection{The group of the wave vector}
\label{sec:kgroup}

In the following, we will repeatedly need to evaluate the action of a point symmetry operation $g$ on a plane wave $\exp( i \vek{k} \cdot \vek{r})$.  Here, $g$ can be represented via an orthogonal $3 \times 3$ matrix $\vek{g}$. (In the context of quasi-2D materials discussed below $\vek{g}$ becomes a $2 \times 2$ matrix.)  Thus we have
\begin{subequations}
  \label{eq:Rj-trans}
  \begin{equation}\label{eq:plane-matrix}
    g \, \exp( i \vek{k} \cdot \vek{r})
    = \exp( i \vek{k} \cdot \vek{g} \cdot \vek{r})
    \equiv \exp( i \vek{k} \cdot \vekt{r})
  \end{equation}
  with
  \begin{equation}
    \vekt{r} = \vek{g} \cdot \vek{r} .
  \end{equation}
\end{subequations}
Note that when transforming the position vector $\vek{r}$, the wave vector $\vek{k}$ is a fixed parameter characterizing the plane wave $\exp( i \vek{k} \cdot \vek{r})$ that does not change under $g$.  Nonetheless, since $g$ is an orthogonal transformation, we can also write Eq.~(\ref{eq:plane-matrix}) as
\begin{subequations}
\label{eq:k-trans}
\begin{align}
    g \, \exp( i \vek{k} \cdot \vek{r})
    & = \exp [ i (\vek{g}^{-1} \cdot \vek{k}) \cdot \vek{r} ] \\
    & = \exp( i \vekt{k} \cdot \vek{r})
\end{align}
with
\begin{equation}
  \label{eq:k-trans-prime}
   \vekt{k} = \vek{g}^{-1} \cdot \vek{k} .
\end{equation}
\end{subequations}
Thus we can evaluate $g$ either by transforming the position vector $\vek{r}$ or by inversely transforming the wave vector~$\vek{k}$.

In the group theory of crystallographic space groups, the point-group symmetries $g$ of Bloch functions $\Psi_\vek{k} (\vek{r})$ with wave vector $\vek{k}$ form the point group $\kgroup$ of the wave vector $\vek{k}$ \cite{streitwolf71, bir74, dre08}.  Given the point group $\pointgroup$ of a crystal structure, the group $\kgroup$ is defined by the condition that it contains the symmetry elements of $\pointgroup$ that map $\vek{k}$ onto a vector $\vekt{k}$ such that
\begin{equation}
  \label{eq:vec-bg}
  \vekt{k} = \vek{g}^{-1} \cdot \vek{k} = \vek{k} + \vek{b}_g ,
\end{equation}
where $\vek{b}_g$ is a reciprocal lattice vector with the possibility $\vek{b}_g = 0$.  Indeed, since $g$ represents point group operations, we can have $\vek{b}_g \ne 0$ only if $\vek{k}$ is from the border of the BZ.  For positions $\vek{r} = \vek{a}$ that are lattice vectors we have
\begin{equation}
  \label{eq:lattice-trans}
    g \, \exp( i \vek{k} \cdot \vek{a})
    = \exp( i \vekt{k} \cdot \vek{a})
    = \exp( i \vek{k} \cdot \vek{a})
\end{equation}
by definition of $\kgroup$.

\subsection{Tight-binding Hamiltonian}
\label{sec:tb}

We denote the TB basis functions (that are Bloch functions) as
\begin{equation}\label{eq:tb-basis}
  \Phi_{\nu\vek{k}}^{W \mu} (\vek{r}) = \frac{e^{i \vek{k} \cdot \vek{r}}}{\sqrt{N}}\sum_j e^{-i \vek{k} \cdot (\vek{r}-\vek{R}_j^{W \mu})} \, \phi_\nu^W (\vek{r} -\vek{R}_j^{W \mu}) ,
\end{equation}
where $\phi_\nu^W (\vek{r}-\vek{R}_j^{W \mu})$ are the atomic orbitals of type $\nu$ centered about the positions $\vek{R}_j^{W \mu}$ of the atoms. The label $W$ denotes the Wyckoff letter of the atomic positions of the crystal structure \cite{burns2013}. The label $\mu$ has values $\mu = 1, \dots, m$, where $m$ is the multiplicity of $W$. The index $j$ labels the unit cells of the crystal structure; it runs through the positions in a Bravais lattice. The matrix elements of the TB Hamiltonian can then be written as
\begin{subequations}\label{eq:tb-ham}
  \begin{align}
    H (\vek{k})_{\nu \nu'}^{W \mu W' \mu'} & = \int \Phi_{\nu\vek{k}}^{W \mu \ast} (\vek{r}) \, H \, \Phi_{\nu' \vek{k}}^{W' \mu'} (\vek{r}) \, d^3r\\
    & = \epsilon_\nu^W \delta_{\nu \nu'} \, \delta_{W W'} \, \delta_{\mu \mu'} + \sideset{}{'}\sum_{jj'} t_{\nu \nu' jj'}^{W W' \mu \mu'} ,
    \label{eq:tb-ham-b}
  \end{align}
\end{subequations}
where
\begin{subequations}
\begin{align}
  \epsilon_\nu^W & \equiv \int \phi_\nu^{W \ast} (\vek{r} -\vek{R}_j^{W \mu}) \, H \, \phi_\nu^W (\vek{r} -\vek{R}_j^{W \mu}) \, d^3r \\
  & = \int \phi_\nu^{W \ast} (\vek{r}) \, H \, \phi_\nu^W (\vek{r}) \, d^3r
\end{align}
\end{subequations}
denotes the on-site energies for the atomic orbitals (that do not depend on the indices $\mu$ and $j$) and
\begin{align}
  t_{\nu \nu' jj'}^{W W' \mu \mu'}
  & \equiv \; e^{-i \vek{k} \cdot (\vek{R}_j^{W \mu} - \vek{R}_{j'}^{W' \mu'})} \,
  \nonumber\\ & \times \int \phi_\nu^{W \ast} (\vek{r} -\vek{R}_j^{W \mu}) \, H \, \phi_{\nu'}^{W'} (\vek{r} -\vek{R}_{j'}^{W' \mu'}) \, d^3r
 \label{eq:hop-int}
\end{align}
are the hopping integrals.  The prime on the summation sign in Eq.\ (\ref{eq:tb-ham-b}) indicates that the sum excludes the on-site term $\epsilon_\nu^W$.  The TB approximation implies that this sum is restricted to $n$th-nearest neighbors with a small value of $n$.  The hopping integrals can be written in terms of the Slater-Koster parameters \cite{sla54} for hopping integrals between atomic orbitals at positions $\vek{R}_j^{W \mu}$ and $\vek{R}_{j'}^{W' \mu'}$.

\subsection{Decomposition of TB wave functions}
\label{sec:tb-decompose}

Generally, the basis functions (\ref{eq:tb-basis}) for a given wave vector $\vek{k}$ transform according to a representation $\Gamma^\Phi_{\vek{k} W}$ of the group of the wave vector $\kgroup$ that need not be irreducible.  (Here the generic superscript $\Phi$ accounts for the fact that multiple atomic orbitals with different indices $\nu$ may transform jointly according to the same representation.  By definition of Wyckoff letters $W$, orbitals at different positions $\mu$ of a given Wyckoff letter $W$ transform jointly according to one representation.)  As a consequence of the Wigner-Eckart theorem \cite{wig59,eck30} and the fact that $H$ transforms according to the identity representation $\Gamma_1$ of $\kgroup$,  the hopping integrals (\ref{eq:hop-int}) vanish when the product of the representations $\Gamma^\Phi_{\vek{k} W}$ and $\Gamma^{\Phi'}_{\vek{k} W'}$ does not contain the identity representation. In the following, we present a general scheme for transforming the set of basis functions (\ref{eq:tb-basis}) into a symmetry-adapted set of basis functions, where each function transforms irreducibly under $\kgroup$, so that two such basis functions only couple when they both transform according to the same IR of $\kgroup$.  This scheme is based on a decomposition of the basis functions into symmetry-adapted plane waves and symmetry-adapted atomic orbitals.

We denote
\begin{equation}\label{eq:orbital-tilde}
  \tilde{\phi}_{\nu \vek{k}}^W (\vek{r} -\vek{R}_j^{W \mu})
  \equiv e^{-i \vek{k} \cdot (\vek{r} -\vek{R}_j^{W \mu})}
  \, \phi_\nu^W (\vek{r} -\vek{R}_j^{W \mu}),
\end{equation}
so that Eq.\ (\ref{eq:tb-basis}) becomes
\begin{equation}
  \Phi_{\nu \vek{k}}^{W \mu} (\vek{r}) =  \frac{e^{i \vek{k} \cdot \vek{r}}}{\sqrt{N}}  \sum_j \tilde{\phi}_{\nu \vek{k}}^W (\vek{r} - \vek{R}_j^{W \mu}).
\end{equation}
Assuming for conceptual simplicity that the atomic orbitals $\phi_\nu^W$ are localized over a region much smaller than the nearest-neighbor distance
\footnote{This assumption facilitates a discussion of the symmetry of the basis functions (\ref{eq:tb-basis}).  Functions (\ref{eq:tb-basis}) with a finite overlap between nearest neighbors that are needed in a TB calculation must have the same symmetry as the simplified functions discussed here.  The symmetry of these functions cannot change discontinuously when the overlap between the atomic orbitals is switched off.},
the functions $\tilde{\phi}_{\nu \vek{k}}^W (\vek{r} - \vek{R}_j^{W \mu})$ are only nonzero for $\vek{r}$ close to $\vek{R}_j^{W \mu}$.  Hence, in the vicinity of any atomic position $\vek{R}_{j}^{W \mu}$, i.e., for $\vek{r} \equiv \vek{R}_{j}^{W \mu} + \delta \vek{r}$ with small $\delta \vek{r}$, we have
\begin{equation}
  \label{eq:tilde-phi-approx}
  \tilde{\phi}_{\nu \vek{k}}^W (\vek{r} - \vek{R}_j^{W \mu})
  \approx \phi_\nu^W (\vek{r} - \vek{R}_j^{W \mu}).
\end{equation}
Therefore, the TB basis function $\Phi_{\nu \vek{k}}^{W \mu} (\vek{r})$ can be approximated as
\begin{subequations}
  \label{eq:tb-basis-approx}
  \begin{align}
    \Phi_{\nu \vek{k}}^{W \mu} (\vek{r})
    & \approx \frac{e^{i \vek{k} \cdot \vek{r}}}{\sqrt{N}}
      \sum_j  \phi_\nu^W (\vek{r} - \vek{R}_j^{W \mu}) \\
    & = e^{i \vek{k} \cdot \vek{r}} \, \mathcal{A}_\nu^{W \mu} (\vek{r})
  \end{align}
\end{subequations}
with atomic functions
\begin{equation}
  \mathcal{A}_\nu^{W \mu} (\vek{r})
  = \frac{1}{\sqrt{N}} \sum_j  \phi_\nu^W (\vek{r} - \vek{R}_j^{W \mu})
\end{equation}
independent of the wave vector $\vek{k}$.  For strongly localized atomic orbitals and positions $\vek{r} = \vek{R}_j^{W \mu} + \delta\vek{r}$ we have
\begin{equation}\label{eq:func-a}
  \mathcal{A}_\nu^{W \mu} (\vek{r} = \vek{R}_j^{W \mu} + \delta\vek{r})
  = \frac{1}{\sqrt{N}} \sum_j \phi_\nu^W (\delta\vek{r})
  \propto \phi_\nu^W (\delta\vek{r}),
\end{equation}
that is, near an atomic site $\vek{R}_j^{W \mu}$, the atomic functions $\mathcal{A}_\nu^{W \mu} (\vek{r})$ are simply proportional to $\phi_\nu^W (\delta\vek{r})$, independent of the index $\mu$.  Therefore, the atomic functions $\mathcal{A}_\nu^{W \mu} (\vek{r} = \vek{R}_j^{W \mu} + \delta\vek{r})$ have the same symmetry properties as the atomic orbitals $\phi_\nu^W (\delta\vek{r})$.

The plane wave $e^{i \vek{k} \cdot \vek{r}}$ for positions $\vek{r} = \vek{R}_j^{W \mu} + \delta \vek{r}$ near an atomic site $\vek{R}_j^{W \mu}$ is approximately given by
\begin{equation}
  \label{eq:plane_wave_approx}
  \exp [i \vek{k} \cdot (\vek{R}_j^{W \mu} + \delta \vek{r})]
  \approx \exp (i \vek{k} \cdot \vek{R}_j^{W \mu})
  \equiv q_\vek{k} (\vek{R}_j^{W \mu}) ,
\end{equation}
where $q_\vek{k} (\vek{R}_j^{W \mu})$ denotes the plane wave with wave vector $\vek{k}$ associated with the Wyckoff position $\vek{R}_j^{W \mu}$ for fixed $W$ and $\mu$, but $j$ runs over all positions in a Bravais lattice.  These discrete quantities $q_\vek{k} (\vek{R}_j^{W \mu})$ will be discussed in more detail in Sec.~\ref{sec:trafo-plane-wave}. The TB basis function thus can be factorized (ignoring normalization)
\begin{equation}
  \label{eq:tb-basis-factorized}
  \Phi_{\nu \vek{k}}^{W \mu} (\vek{r}
  = \vek{R}_j^{W \mu} + \delta \vek{r} )
  \approx q_\vek{k} (\vek{R}_j^{W \mu}) \, \phi_\nu^W (\delta \vek{r}).
\end{equation}
This expression will be analyzed further in the following sections.

As a side remark, we note that the eigenfunctions of the TB Hamiltonian (\ref{eq:tb-ham}) for the band $n$ and wave vector $\vek{k}$ expressed in terms of the basis functions (\ref{eq:tb-basis}) take the form
\begin{equation}\label{eq:tb-eigenfun}
  \Psi_{n \vek{k}} (\vek{r})
  = \sum_{W \mu \nu} \psi_{\nu n\vek{k}}^{W \mu} \,
    \Phi_{\nu \vek{k}}^{W \mu} (\vek{r})
\end{equation}
with expansion coefficients $\psi_{\nu n \vek{k}}^{W \mu}$.  These eigenfunctions permit a factorization similar to Eq.\ (\ref{eq:tb-basis-factorized})
\begin{equation}
  \Psi_{n \vek{k}} (\vek{r}
  = \vek{R}_j^{W \mu} + \delta \vek{r})
  \approx \sum_{W \mu} q_\vek{k} (\vek{R}_j^{W \mu})
  \sum_\nu \psi_{\nu n \vek{k}}^{W \mu} \; \phi_\nu^W (\delta \vek{r}) .
\end{equation}
However, the discussion of the TB wave functions $\Psi_{n \vek{k}} (\vek{r})$ is greatly simplified if instead of the basis functions (\ref{eq:tb-basis}) we use symmetry-adapted basis functions to be discussed in the following.

\subsection{Symmetry-adapted basis functions}
\label{sec:sym-adapted-basis}

The main advantage of the approximate expression (\ref{eq:tb-basis-factorized}) lies in the fact that the function (\ref{eq:tb-basis-factorized}) has the same symmetry properties as the TB basis function (\ref{eq:tb-basis}).  Yet the factorization in Eq.\ (\ref{eq:tb-basis-factorized}) allows one to discuss the symmetry of the plane waves $q_\vek{k} (\vek{R}_j^{W \mu})$ (characterized by a representation $\Gamma^q_{\vek{k} W}$ of $\kgroup$, see Sec.~\ref{sec:trafo-plane-wave}) separate from the symmetry of the atomic orbitals $\phi_\nu^W (\delta \vek{r})$ (characterized by a representation $\Gamma^\phi_{\vek{k} W}$ of $\kgroup$, see Sec.~\ref{sec:trafo:atom:orbitals}).  Often, these representations are irreducible, though generally they can be reducible.  The TB basis functions (\ref{eq:tb-basis}) at one Wyckoff letter $W$ then transform according to the product representation
\begin{equation}
  \label{eq:ir-basis-product}
  \Gamma^\Phi_{\vek{k} W} \equiv \Gamma^q_{\vek{k} W} \times \Gamma^\phi_{\vek{k} W}.
\end{equation}
The representation $\Gamma^\Phi_{\vek{k} W}$ may be reducible (even if $\Gamma^q_{\vek{k} W}$ and $\Gamma^\phi_{\vek{k} W}$ are irreducible), giving the decomposition
\begin{equation}
  \Gamma^\Phi_{\vek{k} W} = \sum_I a_I \Gamma_I ,
\end{equation}
where $a_I \ge 0$ are the multiplicities with which the IRs $\Gamma_I$ of $\kgroup$ appear in $\Gamma^\Phi_{\vek{k} W}$.  Generally, the IRs $\Gamma_I$ contained in $\Gamma^\Phi_{\vek{k} W}$ give symmetry-adapted basis functions
\begin{equation}
  \label{eq:tb-basis-sym}
  \Phi_\vek{k}^{W I \alpha \beta} (\vek{r})
  = \sum_{\mu\nu} \mathcal{C}_{I \alpha \beta}^{W \mu \nu \vek{k}} \;
  \Phi_{\nu\vek{k}}^{W \mu} (\vek{r}) ,
\end{equation}
where  $\alpha = 1, 2, \ldots, a_I$ and $\beta$ labels the different functions transforming jointly according to $\Gamma_I$. The coefficients $\mathcal{C}_{I \alpha \beta}^{W \mu \nu \vek{k}}$ represent the weights with which the functions $\Phi_{\nu\vek{k}}^{W \mu} (\vek{r})$ contribute to the symmetry-adapted basis functions $\Phi_\vek{k}^{W I \alpha \beta} (\vek{r})$.  Given symmetry-adapted plane waves transforming according to the IR $\Gamma_J$ and atomic orbitals transforming according to the IR $\Gamma_{J'}$ (see discussion below), the expansion coefficients $\mathcal{C}_{I \alpha \beta}^{W \mu \nu \vek{k}}$ are given by the Clebsch-Gordan coefficients for coupling $\Gamma_J$ and $\Gamma_{J'}$ to obtain $\Gamma_I$.

For the symmetry-adapted basis functions (\ref{eq:tb-basis-sym}), the matrix elements (\ref{eq:tb-ham}) of the TB Hamiltonian become block-diagonal with respect to different IRs $\Gamma_I$
\begin{equation}
  \label{eq:tb-ham-sym}
  H (\vek{k})_{\alpha \alpha' \beta \beta'}^{W W' I I'}
  = \delta_{I I'} \Bigl[
  \epsilon_{I \alpha}^W \, \delta_{W W'} \,
  \delta_{\alpha \alpha'} \, \delta_{\beta \beta'}
  + \sideset{}{'}\sum_{jj'} t_{\beta \beta' \, j j'}^{W W' I \, \alpha \alpha'}
  \Bigr] .
\end{equation}
Here
\begin{equation}
  \epsilon_{I \alpha}^W \equiv \int \Phi_\vek{k}^{W I \alpha \beta \ast} (\vek{r}) \, H \, \Phi_\vek{k}^{W I \alpha \beta} (\vek{r}) \, d^3r
\end{equation}
denotes the on-site energies.  These can always be made diagonal in the index $\alpha$ by a suitable definition of the $a_I$ sets of basis functions (\ref{eq:tb-basis-sym}) transforming according to $\Gamma_I$.  The hopping matrix elements become
\begin{equation}
 \label{eq:hop-int-sym}
  t_{\beta \beta' \, j j'}^{W W' I \, \alpha \alpha'}
  \equiv \int \Phi_\vek{k}^{W I \alpha \beta \ast} (\vek{r}) \, H \, \Phi_\vek{k}^{W' I \alpha' \beta'} (\vek{r}') \, d^3r
\end{equation}
with $\vek{r} = \vek{R}_j^{W \mu} + \delta \vek{r}$ and $\vek{r}' = \vek{R}_{j'}^{W' \mu'} + \delta \vek{r}$.  We remark that in actual TB models it may happen that a block (\ref{eq:tb-ham-sym}) can be further decomposed into subblocks if symmetry-allowed couplings between distant neighbors are ignored within the TB approximation.

Each block (\ref{eq:tb-ham-sym}) of the TB Hamiltonian yields eigenfunctions
\begin{equation}
  \label{eq:tb-eigen-sym}
  \Psi_{n \vek{k}}^{I \beta} (\vek{r})
  = \sum_{W, \alpha} C_{n \vek{k}}^{W I \alpha}
  \Phi_\vek{k}^{W I \alpha \beta} (\vek{r})
\end{equation}
with expansion coefficients $C_{n \vek{k}}^{W I \alpha \beta}$ that transform irreducibly according to the $\beta$th component of the IR $\Gamma_I$ of $\kgroup$.  To proceed, we discuss first the symmetry of the atomic orbitals $\phi_\nu^W (\delta\vek{r})$, followed by a discussion of the symmetry of the plane waves $q_\vek{k} (\vek{R}_j^{W \mu})$.

\subsection{Transformation of atomic orbitals $\phi_\nu^W (\delta\mathbf{r})$}
\label{sec:trafo:atom:orbitals}

We can study the symmetry of the atomic function $\mathcal{A}_\nu^{W \mu} (\vek{r})$ in the vicinity of the atomic positions $\vek{R}_j^{W \mu}$ by only looking at the orbitals $\phi_\nu^W (\delta \vek{r})$.  Obviously, this problem is independent of the index $\mu$, the band index $n$, and the wave vector $\vek{k}$ though, of course, we are generally interested in the transformational behavior of these orbitals with respect to the group $\kgroup$ of the wave vector~$\vek{k}$.  For symmetry operations $g \in \kgroup$ we have
\begin{equation}
  \label{eq:ao-trafo}
  g \, \phi_\nu^W (\delta\vek{r})
  = \phi_\nu^W (g \, \delta \vek{r})
  = \sum_{\nu'} \mathcal{D}^\phi_{\vek{k} W} (g)_{\nu\nu'}
    \phi_{\nu'}^W (\delta \vek{r}) ,
\end{equation}
where $\Gamma^\phi_{\vek{k} W} = \{\mathcal{D}^\phi_{\vek{k} W} (g): g \in \kgroup\}$ is the representation describing the transformation of the atomic orbitals $\phi_\nu^W$.  As usual, we assume that the atomic orbitals are characterized by some orbital angular momentum $l=0, 1, 2, \ldots$, so that the matrices $\mathcal{D}^\phi_{\vek{k} W} (g)$ acquire a block structure corresponding to different values of $l$, indicating that there exists no mixing between atomic orbitals with different angular momenta under symmetry transformations $g \in \kgroup$.

In general, the representation $\Gamma^\phi_{\vek{k} W}$ is reducible
\footnote{The $2l+1$ atomic orbitals for a given magnitude $l$ of orbital angular momentum transform according to an IR of the full rotation group.  The compatibility tables for decomposing the IRs of the full rotation group into IRs of the crystallographic point groups are listed in Ref.~\cite{kos63}.  Here we require a more complete solution of this problem where we also construct the linear combinations of angular-momentum eigenstates that transform irreducibly according to the different IRs of $\kgroup$.  This problem does not depend on, e.g., the principal quantum numbers of these orbitals.  We note that for sufficiently large $l$ (for any $\kgroup$ if $l \ge 6$) these $2l+1$ atomic orbitals contribute to all (even or odd if $\kgroup$ includes inversion) IRs of $\kgroup$, which limits possibilities to decompose a TB Hamiltonian into blocks describing parts of the full band structure.}.
The projection operators (\ref{eq:projection-op}) yield symmetry-adapted atomic orbitals $\phi^W_{I \beta} (\delta\vek{r})$ transforming like component $\beta$ of the IR $\Gamma_I$ of $\kgroup$,
\begin{subequations}
  \label{eq:proj-ao}
  \begin{align}
    \phi^W_{I \beta} (\delta\vek{r})
    & \equiv \Pi_{I \beta} \, \phi_\nu^W (\delta\vek{r}) \\
    & = \frac{n_I}{h}\sum_g \mathcal{D}_I (g)^\ast_{\beta \beta}
        \: g \, \phi_\nu^W (\delta\vek{r}) \\
    & = \sum_{\nu'} c^{W \nu'}_{I \beta} \phi_{\nu'}^W (\delta \vek{r})
  \end{align}
  with expansion coefficients
  \begin{equation}
    c^{W \nu'}_{I \beta}
    = \frac{n_I}{h} \sum_g \mathcal{D}_I (g)^\ast_{\beta \beta}
    \, \mathcal{D}^\phi_{\vek{k} W} (g)_{\nu\nu'} .
  \end{equation}
\end{subequations}
We note that this analysis applies to the spinless case when the angular part of the atomic orbitals is given by the usual spherical harmonics $Y_l^m$.  It can likewise be used in the spin-dependent case when the angular part of the atomic orbitals is given by spin-angular functions and the projection operators $\Pi_{I \beta}$ project on the double-group representations of $\kgroup$.

\subsection{Transformation of plane waves $q_\mathbf{k} (\mathbf{R}_j^{W \mu})$} \label{sec:trafo-plane-wave}

While the IRs of TB eigenfunctions (\ref{eq:tb-eigenfun}) depend on the band index $n$, the symmetry of the plane waves $q_\vek{k} (\vek{R}_j^{W \mu})$ can be discussed independent of the index $n$.
Very generally, for a given Wyckoff letter $W$, the positions $\vek{R}_j^{W \mu}$ transform among themselves under the operations of the space group. Hence, using the matrix $\vek{g}$ defined in Eq.~(\ref{eq:Rj-trans}), we have
\begin{equation}\label{eq:g-R-trans}
  g \, \vek{R}_j^{W \mu}
  = \vek{g} \cdot \vek{R}_j^{W \mu}
  \equiv \vek{R}_{j'}^{W \mu'} ,
\end{equation}
i.e., a symmetry transformation $g \in \kgroup$ generally changes both $\mu$ and $j$.  We rewrite this as
\begin{equation}\label{eq:g-R}
  \vek{g} \cdot \vek{R}_j^{W \mu}
  = \vek{R}_j^{W \mu'}
    + \left[ \vek{R}_{j'}^{W \mu'} - \vek{R}_j^{W \mu'} \right] ,
\end{equation}
where the term in square brackets is a lattice vector. Applying the operation $g$ to the plane wave $q_\vek{k} (\vek{R}_j^{W \mu})$ yields
\begin{subequations}\label{eq:pw-phase}
\begin{align}
  g \, q_\vek{k} (\vek{R}_j^{W \mu})
  & = \exp( i \vek{k} \cdot \vek{R}_{j'}^{W \mu'})\\
  & = q_\vek{k} (\vek{R}_j^{W \mu'})
    \exp [ i \vek{k} \cdot (\vek{R}_{j'}^{W \mu'} - \vek{R}_j^{W \mu'})] .
\end{align}
\end{subequations}
Hence, the symmetry operation $g$ generally maps the plane wave $q_\vek{k} (\vek{R}_j^{W \mu})$ onto $q_\vek{k} (\vek{R}_j^{W \mu'})$ multiplied by a phase factor.

To analyze the mappings (\ref{eq:pw-phase}) further, we interpret the discrete plane waves $\{ q_\vek{k} (\vek{R}_j^{W \mu}) : \mu = 1, \ldots, m \}$ as basis vectors in an $m$-dimensional vector space.  Relative to this basis, we can express arbitrary points as $m$-component spinors.  It facilitates this analysis to introduce $m$-component base spinors $\{\pwbase_\vek{k} (\vek{R}_j^{W \mu}) : \mu = 1, \ldots, m \}$ with components $\nu$ equal to $\delta_{\mu\nu}$.  Using these base spinors, a plane wave at positions $\vekc{R}_j^W \equiv \{\vek{R}_j^{W \mu}: \mu = 1, \ldots, m\}$ becomes
\begin{equation}
  \label{eq:pw-spinor}
  \pwBloch_\vek{k} (\vekc{R}_j^W) = \sum_\mu \pwbase_\vek{k} (\vek{R}_j^{W \mu})
  = \tvek{ 1 \\[-1ex] \vdots \\ 1} .
\end{equation}
Applying a symmetry transformation $g$ to $\pwBloch_\vek{k} (\vekc{R}_j^W)$ gives
\begin{equation}
  \label{eq:pw-spinor-map}
  g \, \pwBloch_\vek{k} (\vekc{R}_j^W)
  = \mathcal{D}^q_{\vek{k} W} (g) \, \pwBloch_\vek{k} (\vekc{R}_j^W) ,
\end{equation}
where $\mathcal{D}^q_{\vek{k} W} (g)$ is an $m \times m$ matrix corresponding to the group element $g$.  Each row and each column of  $\mathcal{D}^q_{\vek{k} W} (g)$ has only one nonzero matrix element which, according to Eq.\ (\ref{eq:pw-phase}), is given by
\begin{equation}\label{eq:pw-rep}
  \mathcal{D}^q_{\vek{k} W} (g)_{\mu' \mu}
  = \exp [ i \vek{k} \cdot (\vek{R}_{j'}^{W \mu'} - \vek{R}_j^{W \mu'})] ,
\end{equation}
where the $g$ dependence on the right-hand side is given by Eq.\ (\ref{eq:g-R}).

We show in the following that $\Gamma^q_{\vek{k} W} \equiv \{ \mathcal{D}^q_{\vek{k} W} (g) : g \in \kgroup\}$ defines an $m$-dimensional representation of $\kgroup$, i.e., for any two group elements $g_1, g_2 \in \kgroup$ we have
\begin{equation}\label{eq:pw-rep-elements}
  \mathcal{D}^q_{\vek{k} W} (g_2) \mathcal{D}^q_{\vek{k} W} (g_1)
  = \mathcal{D}^q_{\vek{k} W} (g_2 g_1).
\end{equation}
Given a position  $\vek{R}_{j_0}^{W \mu_0}$ and using Eq.\ (\ref{eq:g-R}), we have
\begin{subequations}
  \begin{align}
    g_1 \vek{R}_{j_0}^{W \mu_0}
    & = \vek{g}_1 \cdot \vek{R}_{j_0}^{W \mu_0} \nonumber \\
    & \equiv \vek{R}_{j_1}^{W \mu_1}
    = \vek{R}_{j_0}^{W \mu_1}
    + \left[ \vek{R}_{j_1}^{W \mu_1} - \vek{R}_{j_0}^{W \mu_1} \right] , \\
    g_2 \vek{R}_{j_0}^{W \mu_1}
    & = \vek{g}_2 \cdot \vek{R}_{j_0}^{W \mu_1} \nonumber \\
    & \equiv \vek{R}_{j_2}^{W \mu_2}
    = \vek{R}_{j_0}^{W \mu_2}
    + \left[ \vek{R}_{j_2}^{W \mu_2} - \vek{R}_{j_0}^{W \mu_2} \right] ,
  \end{align}
\end{subequations}
so that the phase (\ref{eq:pw-rep}) for the group elements $g_1$ and $g_2$ becomes
\begin{subequations}\label{pw-phase-g1-g2}
  \begin{align}
    \mathcal{D}^q_{\vek{k} W} (g_1)_{\mu_1 \mu_0}
    & = \exp [ i \vek{k} \cdot
      (\vek{R}_{j_1}^{W \mu_1} - \vek{R}_{j_0}^{W \mu_1})] , \\
    \mathcal{D}^q_{\vek{k} W} (g_2)_{\mu_2 \mu_1}
    & = \exp [ i \vek{k} \cdot
    (\vek{R}_{j_2}^{W \mu_2} - \vek{R}_{j_0}^{W \mu_2})].
  \end{align}
\end{subequations}
The product $g_2 \, g_1$, i.e., the transformation $g_1$ followed by $g_2$, is also an element of $\kgroup$, giving
\begin{widetext}
\begin{subequations}\label{eq:g-trans-prod}
  \begin{align}
    g_2 \, g_1 \, q_\vek{k} (\vek{R}_{j_0}^{W \mu_0})
    & = \exp(i \vek{k} \cdot g_2 \vek{R}_{j_1}^{W \mu_1})\\
    & = \exp(i \vek{k} \cdot g_2 \vek{R}_{j_0}^{W \mu_1})
       \exp[i \vek{k} \cdot g_2 (\vek{R}_{j_1}^{W \mu_1}
            - \vek{R}_{j_0}^{W \mu_1})]\\
    & = \exp(i \vek{k} \cdot \vek{R}_{j_0}^{W \mu_2})
       \exp[i \vek{k} \cdot ( \vek{R}_{j_2}^{W \mu_2}
            - \vek{R}_{j_0}^{W \mu_2})]
       \, \exp[i \vek{k} \cdot g_2 (\vek{R}_{j_1}^{W \mu_1}
       - \vek{R}_{j_0}^{W \mu_1})] .
  \end{align}
\end{subequations}
\end{widetext}
As $\vek{R}_{j_1}^{W \mu_1} - \vek{R}_{j_0}^{W \mu_1}$ is a lattice vector, it follows from Eq.\ (\ref{eq:lattice-trans})
\begin{equation}
  g_2 \, g_1 \, q_\vek{k} (\vek{R}_{j_0}^{W \mu_0})
  =  q_\vek{k} (\vek{R}_{j_0}^{W \mu_2})
  \, \mathcal{D}^q_{\vek{k} W} (g_2 \, g_1)_{\mu_2 \mu_0}
\end{equation}
with
\begin{equation}\label{eq:prod-phase-D}
  \mathcal{D}^q_{\vek{k} W} (g_2 \, g_1)_{\mu_2 \mu_0} = \exp[i \vek{k} \cdot ( \vek{R}_{j_2}^{W \mu_2} - \vek{R}_{j_0}^{W \mu_2} + \vek{R}_{j_1}^{W \mu_1} - \vek{R}_{j_0}^{W \mu_1})] .
\end{equation}
This confirms Eq.\ (\ref{eq:pw-rep-elements}).

We remark that for $\vek{k} = 0$ the representation $\Gamma^q_{\vek{k} W}$ is known as permutation representation \cite{mcw63} or equivalence representation \cite{dre08}, where it characterizes the permutations of $m$ objects under the symmetry operations $g \in \pointgroup$.

Using the fact that $g$ is an orthogonal transformation, we can also write Eq.\ (\ref{eq:pw-rep}) as
\begin{subequations}
  \label{eq:g-orth}
\begin{align}
  \mathcal{D}^q_{\vek{k} W} (g)_{\mu' \mu}
  & = \exp [ i \vek{k} \cdot (\vek{R}_{j'}^{W \mu'} - \vek{R}_j^{W \mu'})]\\
  & = \exp [ i \vek{k} \cdot (\vek{g} \cdot \vek{R}_j^{W \mu} - \vek{R}_j^{W \mu'})]\\
  & = \exp \{ i [(\vek{g}^{-1} \cdot \vek{k}) \cdot \vek{R}_j^{W \mu} - \vek{k} \cdot \vek{R}_j^{W \mu'}] \}.
\end{align}
\end{subequations}
For $\vek{k}$ inside the Brillouin zone, where $\vek{g}^{-1} \cdot \vek{k} = \vek{k}$ by definition of $\kgroup$, the nonzero matrix elements of $\mathcal{D}^q_{\vek{k} W} (g)$ therefore become
\begin{equation}
  \label{eq:pw-phase-inside}
  \mathcal{D}^q_{\vek{k} W} (g)_{\mu' \mu}
  \equiv \exp [ i  \vek{k} \cdot ( \vek{R}_j^{W \mu} - \vek{R}_j^{W \mu'})].
\end{equation}
We see that the representation is generally nontrivial for $m>1$. However, for $m=1$, we have $\vek{R}_j^{W \mu} = \vek{R}_j^{W \mu'}$, so that $q_\vek{k} (\vek{R}_j^{W \mu})$ transforms according to the identity representation $\Gamma_1$.

\subsubsection{Wyckoff positions with multiplicity $m = 1$}

We consider first the special case of Wyckoff positions with multiplicity $m=1$.  Here we drop the index $\mu$, denoting atomic positions as $\vek{R}_j^W$.  This case is equivalent to atomic positions $\vek{R}_j^W$ forming a Bravais lattice.  Note also that multiplicities $m=1$ occur only for symmorphic space groups \cite{burns2013}.  Here, plane waves $\pwBloch_\vek{k} (\vek{R}_j^W) = \pwbase_\vek{k} (\vek{R}_j^W)$ transform according to the one-dimensional IR
\begin{equation}
  \label{eq:phase-D-1d-r}
  \mathcal{D}^q_{\vek{k} W} (g)
  = \exp [ i \, \vek{k} \cdot (\vek{R}_{j'}^W - \vek{R}_j^W)] .
\end{equation}
We saw in Eq.\ (\ref{eq:pw-phase-inside}) that for wave vectors $\vek{k}$ inside the BZ this becomes the identity representation $\Gamma_1$.  It is illuminating to rederive this result by writing Eq.\ (\ref{eq:phase-D-1d-r}) as
\begin{subequations}
  \begin{align}
    \mathcal{D}^q_{\vek{k} W} (g)
    & = \exp [ i \, (\vek{k} \cdot \vek{g} - \vek{k}) \cdot \vek{R}_j^W] \\
    & = \exp [ i \, (\vek{g}^{-1} \cdot \vek{k} - \vek{k}) \cdot \vek{R}_j^W] .
\end{align}
\end{subequations}
It follows from Eq.\ (\ref{eq:vec-bg}) that $\vekt{k} = \vek{g}^{-1} \cdot \vek{k} = \vek{k} + \vek{b}_g $ with a reciprocal lattice vector $\vek{b}_g$.  Thus Eq.\ (\ref{eq:phase-D-1d-r}) describing the effect of $g$ in real space is equivalent to
\begin{equation}
  \label{eq:phase-D-1d-k}
  \mathcal{D}^q_{\vek{k} W} (g)
  = \exp [ i \, \vek{b}_g \cdot \vek{R}_j^W]
\end{equation}
describing the effect of $g$ in reciprocal space.  Hence, for $m=1$ plane waves $\pwBloch_\vek{k} (\vek{R}_j^W)$ transform under $\kgroup$ in a nontrivial way only if the vector $\vek{k}$ is from the border of the BZ when $\vek{k}' = \vek{g}^{-1} \cdot \vek{k}$ and $\vek{k}$ can differ by a reciprocal lattice vector $\vek{b}_g \ne 0$.  Otherwise, $\vek{k}' = \vek{k}$ implies that $\pwBloch_\vek{k}^W (\vek{R}_j^W)$ transforms according to $\Gamma_1$ of $\kgroup$.

If a space group has Wyckoff positions $W \ne W'$ each with multiplicity $m=1$, we can compare the IRs of the plane waves at the positions $\vek{R}_j^W$ and $\vek{R}_j^{W'}$. We have
\begin{subequations}
  \begin{align}
    \mathcal{D}^q_{\vek{k} W} (g) & = \exp ( i \vek{b}_g \cdot \vek{R}_j^W ) ,\\
    \mathcal{D}^q_{\vek{k} W'} (g) & = \exp ( i \vek{b}_g \cdot \vek{R}_j^{W'} ).
  \end{align}
\end{subequations}
Hence
\begin{equation}
  \mathcal{D}^q_{\vek{k} W} (g) = \mathcal{D}^q_{\vek{k} W'} (g)
  \exp [ i \vek{b}_g \cdot (\vek{R}_j^W - \vek{R}_j^{W'}) ] .
\end{equation}
For $W \ne W'$, the vector $\vek{R}_j^W - \vek{R}_j^{W'}$ is not equal to a lattice vector, so that for nonzero $\vek{b}_g$ (i.e., for $\vek{k}$ on the boundary of the Brillouin zone) we generally have
\begin{equation}
\label{eq:ir-inequal-wyckoff}
\Gamma^q_{\vek{k} W} \ne \Gamma^q_{\vek{k} W'} .
\end{equation}
This implies that a nontrivial change of the coordinate system which requires a relabeling of the Wyckoff letters associated with atomic positions changes the IRs of the plane waves at these positions. This relabeling of IR assignments will be discussed in Sec.~\ref{sec:plane-multiple}.

As a simple example for Eq.\ (\ref{eq:ir-inequal-wyckoff}), consider the case where the positions $\vek{R}_j^W$ are equal to lattice vectors, i.e., one of the  positions $\vek{R}_j^W$ is at the origin of the coordinate system. Hence, Eq.\ (\ref{eq:phase-D-1d-k}) gives $\Gamma^q_{\vek{k} W} = \{ \mathcal{D}^q_{\vek{k} W} (g) =1 : g \in \kgroup \}$, i.e., the plane wave $\pwBloch_\vek{k} (\vek{R}_j^W)$ transforms according to the identity representation. On the other hand, the IR $\Gamma^q_{\vek{k} W'}$ at a different Wyckoff position $W'$ can never be the identity representation.

Examples of Wyckoff positions with multiplicity $m=1$ are the positions occupied by the Mo atoms and the center of the hexagon in monolayer MoS$_2$. The plane wave at these positions transforms according to different IRs. In a certain coordinate system where one of these two inequivalent positions are located at the origin, the plane waves at that Wyckoff position transform as the identity representation, while the plane waves at the other Wyckoff position transform according to a different IR. Another example is given by the inequivalent IRs of the plane waves at the positions of C atoms in the central layer of graphene with odd number of layers such as the trilayer graphene discussed in Sec.~\ref{sec:crystal-tlg}.

\subsubsection{Wyckoff positions with multiplicity $m > 1$}
\label{sec:plane-multiple}

For Wyckoff positions with multiplicity $m > 1$ the simple analysis based on Eq.\ (\ref{eq:vec-bg}) is not valid as it does not keep track of how positions $\vek{R}_j^{W \mu}$ are mapped onto each other by symmetry operations $g \in \kgroup$.  Instead we need to use the plane-wave spinors (\ref{eq:pw-spinor}).  For multiplicity $m > 1$, the representation $\Gamma^q_{\vek{k} W}$ characterizing the plane-wave spinors $\pwBloch_\vek{k} (\vekc{R}_j^W)$ is generally reducible.  Using the projection operator (\ref{eq:projection-op}), we can construct linear combinations $\pwsBloch_\vek{k}^{I \beta} (\vekc{R}_j^W)$ of the base spinors $\pwbase_\vek{k} (\vek{R}_j^{W \mu})$ transforming like component $\beta$ of the IR $\Gamma_I$ of $\kgroup$ contained in $\Gamma^q_{\vek{k} W}$
\begin{subequations}
  \label{eq:proj-plane-wave}
  \begin{align}
    \pwsBloch_\vek{k}^{I\beta} (\vekc{R}_j^W)
    & \equiv \Pi_\vek{k}^{I \beta} \, \pwBloch_\vek{k} (\vekc{R}_j^W) \\[1ex]
    & = \frac{n_I}{h}\sum_g \mathcal{D}_I (g)^\ast_{\beta \beta} \: g \, \pwBloch_\vek{k} (\vekc{R}_j^W)\\
    & = \frac{n_I}{h} \sum_g \mathcal{D}_I (g)^\ast_{\beta \beta} \, \mathcal{D}^q_{\vek{k} W} (g) \, \pwBloch_\vek{k} (\vekc{R}_j^W) ,
  \end{align}
\end{subequations}
where we used Eq.~(\ref{eq:pw-spinor-map}).  This yields
\begin{equation}\label{eq:plane-spinor}
  \pwsBloch_\vek{k}^{I\beta} (\vekc{R}_j^W)
  = \sum_m u_{\vek{k} W \mu}^{I\beta} \, \pwbase_\vek{k} (\vek{R}_j^{W \mu})
  = \tvek{ u_{\vek{k} W 1}^{I\beta} \\[-0.5ex] \vdots \\ u_{\vek{k} W m}^{I\beta} }
\end{equation}
with expansion coefficients
\begin{equation}
   u_{\vek{k} W \mu}^{I\beta}
   = \frac{n_I}{h} \sum_g \mathcal{D}_I (g)^\ast_{\beta \beta} \,
   \sum_{\tilde{\mu}} \mathcal{D}^q_{\vek{k} W} (g)_{\mu \tilde{\mu}}
\end{equation}
that completely characterize each symmetrized spinor $\pwsBloch_\vek{k}^{I \beta} (\vekc{R}_j^W)$.

Upon translation by a lattice vector $\vek{a}$, the plane waves $q_\vek{k} (\vek{R}_j^{W \mu})$ acquire a phase $\exp (i \vek{k} \cdot \vek{a})$
\begin{equation}
  q_\vek{k} (\vek{R}_j^{W \mu} + \vek{a})
  = \exp (i \vek{k} \cdot \vek{a}) \, q_\vek{k} (\vek{R}_j^{W \mu}) .
\end{equation}
This implies that $\pwbase_\vek{k} (\vek{R}_j^{W \mu})$ represents (for each $\mu$) a discrete Bloch function for wave vector $\vek{k}$.  Similarly, linear combinations of these base spinors including the spinors $\pwBloch_\vek{k} (\vekc{R}_j^W)$ and $\pwsBloch_\vek{k}^{I\beta} (\vekc{R}_j^W)$ are thus discrete Bloch functions for wave vector $\vek{k}$.  The expansion coefficients $u_{\vek{k} W \mu}^{I\beta}$ take the role of lattice-periodic functions for these discrete Bloch functions.

The projection (\ref{eq:proj-plane-wave}) is valid for all wave vectors $\vek{k}$ in the Brillouin zone (though trivial for $m=1$ when $\vek{k}$ is inside the Brillouin zone, as noted above).  In general, the projectors $\Pi_\vek{k}^{I\beta}$ decompose a plane-wave spinor $\pwBloch_\vek{k} (\vekc{R}_j^W)$ into multiple Bloch functions $\pwsBloch_\vek{k}^{I\beta} (\vekc{R}_j^W)$ corresponding to different IRs $\Gamma_I$ of $\kgroup$. Yet we often have sets of special positions $\vekc{R}_{W j}^{I \vek{k}} = \{\vek{R}_{W \mu j}^{I \vek{k}} : \mu = 1, \ldots, m\}$ within the unit cell where only the Bloch functions $\pwsBloch_\vek{k}^{I\beta} (\vekc{R}_j^W)$ for one IR $\Gamma_I$ are nonzero, but all other projections vanish.  This greatly simplifies further discussion of TB Bloch functions at positions $\vekc{R}_{W j}^{I \vek{k}}$.  The positions $\vekc{R}_{W j}^{I \vek{k}}$ are characterized by two different groups, the group describing the site symmetry \cite{burns2013} denoted as $\sitegroup$ and the group of the wave vector $\kgroup$.  Often the positions $\vekc{R}_{W j}^{I \vek{k}}$ with nontrivial $\kgroup$ coincide with Wyckoff positions with a nontrivial $\sitegroup$. For Wyckoff positions with multiplicity $m=1$, we always have $\sitegroup = \pointgroup$, so that $\kgroup \subseteq \sitegroup$. However, for $m>1$ we will find below that in general there is no simple relation between the group $\sitegroup$ characterizing such special positions $\vekc{R}_{W j}^{I \vek{k}}$ and the group of the wave vector $\kgroup$ for which this happens
\footnote{In SLG, the carbon atoms with multiplicity $m=2$ are characterized by the site symmetry group $\sitegroup = D_{3h}$, see Table~\ref{tab:examples-site-sym}.  Here, likewise, the group of the wave vector at the $\vek{K}$ point is $\kgroup[\vek{K}] = D_{3h}$ and symmetrized plane waves at $\vek{K}$ transform according to the two-dimensional IR $\Gamma_6$ of $D_{3h}$, see Table~\ref{tab:plane-wave-examples}.  On the other hand, in BLG, the carbon atoms at positions $\vek{R}_j^{d1},\vek{R}_j^{d2}$ have site symmetry $\sitegroup = C_{3v}$, whereas the symmetrized plane waves at $\vek{K}$ for these positions transform according to the two-dimensional IR $\Gamma_3$ of $D_3$.  Neither of the groups $C_{3v}$ and $D_3$ can be viewed as a subgroup of the other one.}.

The \emph{symmetrized plane waves} $\pwsBloch_\vek{k}^{I\beta} (\vekc{R}_j^W)$ including the positions $\vekc{R}_{W j}^{I \vek{k}}$ are universal features of each space group, independent of the ``atomistic realization'' of a space group in different crystal structures (e.g., the number and positions of atoms in a unit cell).  They apply both to spinless models and models that include the spin degree of freedom.  We note that the symmetrized plane waves  $\pwsBloch_\vek{k}^{I\beta} (\vekc{R}_j^W)$ introduced here in the context of the TB approximation for Wyckoff positions $\vekc{R}_j^W$ are conceptually different from the symmetrized plane waves discussed previously in the context of the nearly-free electron approximation, see, e.g., Refs.~\cite{streitwolf71, jon75, dre08}.

\subsection{Rearrangement of IRs of Bloch states under a change of the coordinate system}
\label{sec:band-amb}

The Bloch states in certain crystal structures are characterized by IRs of the group $\kgroup$ that depend in a nontrivial way on the location of the origin or the orientation of the coordinate system relative to the position of the atoms \cite{bir66, mor68, cor71, cor72}.  Cornwell \cite{cor71} has given a general discussion of the origin dependence of the symmetry labeling of electron states in such systems.  Here we review and extend these findings, focusing on symmorphic space groups and adopting a notation matching other parts of this study. We show that for different choices of the origin we get a rearrangement of the IRs of Bloch states.  We exploit this rearrangement lemma when discussing band symmetries for specific materials further below.

We consider a crystal structure with space group $\spacegroup$.  For the coordinate system $\vek{r} = (x,y,z)$, the lattice-periodic (single-electron) Hamiltonian is $\Hspace$.  The eigenfunctions of $\Hspace$ are Bloch functions $\Psi_{n \vek{k}}^{I \beta} (\vek{r})$, obeying the eigenvalue equation
\begin{equation}\label{eq:rar-eigen}
  \Hspace \, \Psi_{n \vek{k}}^{I \beta} (\vek{r})
  = E_n (\vek{k}) \, \Psi_{n \vek{k}}^{I \beta} (\vek{r})
\end{equation}
with energy $E_n (\vek{k})$. For a given wave vector $\vek{k}$, the index $I$ denotes the IR $\Gamma_I$ of the point group $\kgroup$ of the wave vector, to be discussed in more detail below. The eigenfunction $\Psi_{n \vek{k}}^{I \beta} (\vek{r})$ transforms according to the $\beta$th component of the IR $\Gamma_I$.  For brevity, we drop in this section the band index $n$.

We denote coordinate transformations using the Seitz notation as $\seitz{g}{\tshift}$, where $g$ is a (proper or improper) rotation that is followed by a translation $\tshift$.  We seek to identify a pure translation $\Tshift \equiv \seitz{\openone}{\tshift}$ of the coordinate system $\vek{r}$, where $\tshift$ equals a fraction of a lattice vector such that for the shifted, primed coordinated system $\vek{r}' = (x',y',z')$ the crystal structure has the same space group symmetry $\spacegroup$ as for the unprimed coordinate system $\vek{r} = (x,y,z)$.  The translation $\Tshift$ transforms the Bloch function $\Psi_{\vek{k}}^{I \beta} (\vek{r})$ into
\begin{equation}
  \Psi_{\vek{k}}^{I' \beta} (\vek{r}')
  \equiv \Tshift \, \Psi_{\vek{k}}^{I \beta} (\vek{r}).
\end{equation}
The index $I' \ne I$ will be justified below.
As $\Tshift = \seitz{\openone}{\tshift}$ commutes with primitive translations $\seitz{\openone}{\vek{a}}$ we get
\begin{equation}
  \seitz{\openone}{\vek{a}} \, \Psi_{\vek{k}}^{I' \beta} (\vek{r}')
  = \exp (-i \vek{k} \cdot \vek{a}) \, \Psi_{\vek{k}}^{I' \beta} (\vek{r}') ,
\end{equation}
so that the transformed Bloch function $\Psi_{\vek{k}}^{I' \beta} (\vek{r}')$ has, indeed, the same wave vector $\vek{k}$ as  $\Psi_{\vek{k}}^{I \beta} (\vek{r})$.  Furthermore, the Hamiltonian in the primed coordinate system becomes
\begin{equation}\label{eq:hamiltonian-translate}
  \Hspace['] = \Tshift \, \Hspace \, \Tshift^{-1} .
\end{equation}
Thus
\begin{subequations}
  \begin{align}
    \Hspace['] \, \Psi_{\vek{k}}^{I' \beta} (\vek{r}')
    & = \Tshift \, \Hspace \, \Psi_{\vek{k}}^{I \beta} (\vek{r}) \\
    & = E(\vek{k}) \, \Psi_{\vek{k}}^{I' \beta} (\vek{r}') ,
  \end{align}
\end{subequations}
so that the transformed function
$\Psi_{\vek{k}}^{I' \beta} (\vek{r}')$ is an eigenfunction of
$\Hspace[']$ with the same eigenvalue $E(\vek{k})$ as
$\Psi_{\vek{k}}^{I \beta} (\vek{r})$.

Given the space group $\spacegroup$ of the crystal, the invariance of
$\Hspace$ under the symmetry operations
$\spacesym \equiv \seitz{g}{\vek{a}} \in \spacegroup$ reads
\begin{equation}
  \label{eq:invar:Hspace}
  \spacesym \, \Hspace \, \spacesym^{-1} = \Hspace \quad \forall \spacesym \in \spacegroup .
\end{equation}
For the Hamiltonian $\Hspace[']$ in the primed coordinate system we get
\begin{equation}
  \spacesym \, \Hspace['] \, \spacesym^{-1} =
  \Tshift \left( \Tshift^{-1} \, \spacesym \, \Tshift \right) \Hspace
  \left( \Tshift^{-1} \, \spacesym \, \Tshift \right)^{-1} \Tshift^{-1} .
\end{equation}
Hence the primed Hamiltonian $\Hspace[']$ obeys an invariance condition analogous to Eq.\ (\ref{eq:invar:Hspace}) if $( \Tshift^{-1} \, \spacesym \, \Tshift)$ commutes with $\Hspace$
\begin{equation}
  [ \Tshift^{-1} \, \spacesym \, \Tshift, \Hspace ] = 0 .
\end{equation}
This requires in turn, given Eq.\ (\ref{eq:invar:Hspace}), that for all
$\spacesym \in \spacegroup$
\begin{subequations}
  \begin{align}
    \Tshift^{-1} \, \spacesym \, \Tshift
    & = \seitz{\openone}{ -\tshift} \, \seitz{g}{ \vek{a}}
    \, \seitz{\openone}{ \tshift} \\
    & = \seitz{g}{g\,  \tshift - \tshift + \vek{a}}
  \end{align}
\end{subequations}
is an element of the space group $\spacegroup$, so that
$g\, \tshift - \tshift$ must be equal to a lattice
vector $\vek{a}'$ of the crystal
\begin{equation}
  \label{eq:gt0:def}
  g\, \tshift - \tshift = \vek{a}'
  \quad \forall \, g \equiv \seitz{g}{0} \in \pointgroup,
\end{equation}
where $\pointgroup$ is the point group corresponding to the space group $\spacegroup$.
A nontrivial solution to this problem is a vector $\tshift$ that
is not equal to a lattice vector $\vek{a}$.  Equation (\ref{eq:gt0:def}) defines the allowed shifts $\tshift$ (up to a lattice vector) that provide alternative descriptions of a crystal structure with space group~$\spacegroup$.

We obtain nontrivial solutions to Eq.\ (\ref{eq:gt0:def}), for example, if a crystal consists of atoms at Wyckoff positions $W = A, B, \ldots$ each with multiplicity $m=1$
\footnote{In Cornwell's notation \cite{cor71}, different Wyckoff positions each with multiplicity $m=1$ are positions each forming a Bravais lattice.}.
We denote these positions in the unit cell as $\vek{t}_\mathrm{A}$, $\vek{t}_\mathrm{B}$, $\vek{t}_\mathrm{C}$, $\dots$, respectively. By definition of the space group $\spacegroup$, these positions $\vek{t}_W$ obey the condition
\begin{equation}\label{eq:t-trans}
  g \, \vek{t}_W - \vek{t}_W = \vek{a}'
  \qquad \forall \, g \in \pointgroup,
  \quad W = \mathrm{A}, \mathrm{B}, \mathrm{C}, \ldots
\end{equation}
with lattice vectors $\vek{a}'$. Hence, any linear combination of these position vectors $\{\vek{t}_W\}$ with integer prefactors (e.g., $\tshift = \vek{t}_W - \vek{t}_{W'}$ with $W \ne W'$) yields a translation $\tshift$ consistent with Eq.\ (\ref{eq:gt0:def}).

In the unprimed coordinate system the eigenfunctions $\Psi_{\vek{k}}^{I \beta} (\vek{r})$ transform according to the $\beta$th component of an IR $\Gamma_I$ of the point group $\kgroup$ of the wave vector $\vek{k}$
\begin{equation}
  \seitz{g}{0} \, \Psi_{\vek{k}}^{I \beta} (\vek{r})
  = \sum_{\beta'} \mathcal{D}_I (g)_{\beta' \beta} \, \Psi_{\vek{k}}^{I \beta'} (\vek{r})
\end{equation}
with representation matrices $\mathcal{D}_I (g)_{\beta' \beta}$.
We can evaluate the action of a symmetry operation
$g \in \kgroup$ on primed Bloch functions $\Psi_{\vek{k}}^{I' \beta} (\vek{r}')$ as follows
\begin{subequations}
  \begin{align}
    \makebox[2em][l]{$\seitz{g}{0} \, \Psi_{\vek{k}}^{I' \beta} (\vek{r}')$}
    \nonumber\\
    & = \seitz{g}{0} \, \seitz{\openone}{\tshift}
    \, \Psi_{\vek{k}}^{I \beta} (\vek{r}) \\
    & = \seitz{\openone}{\tshift}
    \, \seitz{\openone}{g \, \tshift - \tshift} \, \seitz{g}{0}
    \, \Psi_{\vek{k}}^{I \beta} (\vek{r}) \\
    & = \sum_{\beta'} \mathcal{D}_I (g)_{\beta' \beta}
    \, \seitz{\openone}{\tshift}
    \, \seitz{\openone}{g \, \tshift - \tshift}
    \, \Psi_{\vek{k}}^{I \beta'} (\vek{r}) \\
    & = \sum_{\beta'} \mathcal{D}_I (g)_{\beta' \beta}
    \, \exp [-i\vek{k} \cdot (g \, \tshift - \tshift)]
    \, \seitz{\openone}{\tshift}
    \, \Psi_{\vek{k}}^{I \beta'} (\vek{r}) \\
    & = \sum_{\beta'} \mathcal{D}_{I'} (g)_{\beta' \beta}
    \, \Psi_{\vek{k}}^{I' \beta'} (\vek{r}) ,
  \end{align}
\end{subequations}
where the primed representation matrices
$\mathcal{D}_{I'} (g)_{\beta' \beta}$ become
\begin{equation}
  \label{eq:ir:trafo}
  \mathcal{D}_{I'} (g)_{\beta' \beta} = \mathcal{D}_I (g)_{\beta' \beta}
  \, \exp [-i\vek{k} \cdot (g \, \tshift - \tshift)] .
\end{equation}
We denote the phase factors in Eq.\ (\ref{eq:ir:trafo}) by
\begin{equation}
  \label{eq:phase-d-amb}
  \mathcal{D}_\tshift^\vek{k} (g)
  = \exp [-i\vek{k} \cdot (g \, \tshift - \tshift)].
\end{equation}
Using Eq.\ (\ref{eq:vec-bg}), this becomes
\begin{subequations}
  \label{eq:phase-d-amb-inv}
  \begin{equation}
    \mathcal{D}_\tshift^\vek{k} (g) = \exp (-i \vek{b}_g \cdot \tshift )
  \end{equation}
  with a reciprocal lattice vector
  \begin{equation}
    \label{eq:phase-d-amb-inv-bg}
    \vek{b}_g = g^{-1} \vek{k} - \vek{k}.
  \end{equation}
\end{subequations}
The phase $\mathcal{D}_\tshift^\vek{k} (g)$ is therefore nontrivial when $\vek{b}_g \ne 0$, which can only happen at the border of the Brillouin zone.

We show in the next paragraph that $\Gamma_\tshift \equiv \{\mathcal{D}_\tshift^\vek{k} (g) : g \in \kgroup \}$ defines a one-dimensional IR of $\kgroup$ (for every wave vector $\vek{k}$ in the Brillouin zone).  Therefore, the IR $\Gamma_{I'}$ of a Bloch function in the primed coordinate system is given by $\Gamma_\tshift$ times the IR $\Gamma_I$ of the Bloch function in the unprimed coordinate system, so that Eq.\ (\ref{eq:ir:trafo}) becomes
\begin{equation}
  \label{eq:ir:trafo:sum}
  \Gamma_{I'} = \Gamma_\tshift \times \Gamma_I.
\end{equation}
The rearrangement lemma for IRs discussed in Appendix~\ref{sec:rearrange:lemma} applied to Eq.\ (\ref{eq:ir:trafo:sum}) shows that, unless we have the trivial case that $\Gamma_\tshift$ is the identity representation, each IR $\Gamma_I$ of $\kgroup$ in the unprimed coordinate system is mapped on an IR $\Gamma_{I'} \ne \Gamma_I$ in the primed coordinate system.  Hence we call Eq.\ (\ref{eq:ir:trafo:sum}) the rearrangement lemma for the IRs of Bloch states and $\Gamma_\tshift$ the \emph{rearrangement representation} (RAR) for the coordinate shift $\tshift$.  Examples for this rearrangement of IRs of Bloch states will be given below when we study the symmetries of the Bloch functions in monolayer MoS$_2$ (Sec.~\ref{sec:band-amb-mos2}) and trilayer graphene (Sec.~\ref{sec:crystal-tlg}).  It follows from Eq.\ (\ref{eq:phase-d-amb-inv}) that only at the border of the Brillouin zone the IR labeling of Bloch states can depend on the origin of the coordinate system \cite{cor71}.  Also, Eq.~(\ref{eq:phase-d-amb-inv}) implies $\Gamma_{-\tshift} = \Gamma_\tshift^\ast$.  Generally, the shift $\tshift$ is defined up to a lattice vector $\vek{a}$.  It follows immediately from Eq.\ (\ref{eq:phase-d-amb-inv}) that $\tshift$ and $\tilde{\tshift} \equiv \tshift + \vek{a}$ define the same RAR $\Gamma_\tshift$.

To show that $\Gamma_\tshift = \{\mathcal{D}_\tshift^\vek{k} (g) : g \in \kgroup \}$ defines a one-dimen\-sional IR of $\kgroup$ we consider two group elements $g_i \in \kgroup$ ($i = 1,2$).  According to Eq.\ (\ref{eq:gt0:def}), the transformations $g_i \, \tshift$ differ from $\tshift$ by lattice vectors $\vek{a}_i$
\begin{equation}
  g_i \, \tshift = \tshift + \vek{a}_i ,
\end{equation}
so that
\begin{equation}
  \mathcal{D}_\tshift^\vek{k} ( g_i )
    = \exp [-i\vek{k} \cdot (g_i \, \tshift - \tshift)]
    = \exp ( - i \vek{k} \cdot \vek{a}_i )
\end{equation}
and
\begin{equation}
  \mathcal{D}_\tshift^\vek{k} (g_1) \mathcal{D}_\tshift^\vek{k} (g_2) = \exp [ - i \vek{k} \cdot (\vek{a}_1 + \vek{a}_2)].
\end{equation}
We can also write $g_1 g_2 \, \tshift$ as
\begin{subequations}
  \begin{align}
    g_1 g_2 \, \tshift
    & = g_1 \, \tshift + g_1 \, \vek{a}_2\\
    & = \tshift + \vek{a}_1 + \vek{a}_2 + g_1 \, \vek{a}_2 - \vek{a}_2 .
  \end{align}
\end{subequations}
Hence
\begin{subequations}
  \begin{align}
    \mathcal{D}_\tshift^\vek{k} ( g_1 g_2 )
    & = \exp [-i\vek{k} \cdot (g_1 g_2 \, \tshift - \tshift)]\\
    & = \exp [ - i \vek{k} \cdot (\vek{a}_1 + \vek{a}_2 + g_1 \, \vek{a}_2 - \vek{a}_2)]\\
    & = \mathcal{D}_\tshift^\vek{k} (g_1) \mathcal{D}_\tshift^\vek{k} (g_2) \exp [ - i \vek{k} \cdot (g_1 \, \vek{a}_2 - \vek{a}_2)].
  \end{align}
\end{subequations}
We get similar to Eq.~(\ref{eq:k-trans}) and using Eq.\ (\ref{eq:vec-bg})
\begin{subequations}
  \begin{align}
    \exp [ - i \vek{k} \cdot (g_1 \vek{a}_2 - \vek{a}_2)]
    & = \exp [ - i (g_1^{-1} \vek{k} - \vek{k}) \cdot \vek{a}_2)] \\
    & = \exp [ - i \vek{b}_{g_1} \cdot \vek{a}_2] = 1 .
  \end{align}
\end{subequations}
This gives us finally
\begin{equation}
  \mathcal{D}_\tshift^\vek{k} ( g_1 g_2 )
  = \mathcal{D}_\tshift^\vek{k} (g_1) \mathcal{D}_\tshift^\vek{k} (g_2) ,
\end{equation}
so that, indeed, $\Gamma_\tshift = \{\mathcal{D}_\tshift^\vek{k} (g) : g \in \kgroup \}$ is a one-dimen\-sional IR of the group $\kgroup$.

The above argument \cite{cor71} is based on the full Bloch functions $\Psi_\vek{k} (\vek{r})$. We showed in Eq.\ (\ref{eq:tb-basis-factorized}) that in a TB description, these Bloch functions can be factorized as $\Phi_{\nu \vek{k}}^{W \mu} (\vek{r}) = q_\vek{k} (\vek{R}_j^{W \mu}) \, \phi_\nu^W (\delta \vek{r})$, so that the symmetry of the plane waves $q_\vek{k} (\vek{R}_j^{W \mu})$ can be discussed separately from the symmetry of the atomic orbitals $\phi_\nu^W (\delta \vek{r})$. The orbitals $\phi_\nu^W (\delta \vek{r})$ only depend on $\delta \vek{r}$ but not on the actual positions $\vek{R}_j^{W \mu}$.  Hence it follows immediately that the symmetry of the orbitals $\phi_\nu^W (\delta\vek{r})$ is independent of the coordinate system used, see also Eq.\ (\ref{eq:ao-trafo}).  Only the representation $\Gamma^q_{\vek{k} W}$ of the plane waves $q_\vek{k} (\vek{R}_j^{W \mu})$ depends, in general, on the coordinate system. A translation of the coordinate system by $\tshift$ maps the positions $\vek{R}_j^{W \mu}$ onto the atomic position $\vek{R}_j^{W' \mu} = \vek{R}_j^{W \mu} - \tshift$.  The new coordinate system is valid if and only if $g \, \tshift - \tshift$ are lattice vectors for all $g \in \kgroup$ [see Eq.\ (\ref{eq:gt0:def})], so that the plane waves transform as $\Gamma^q_{\vek{k} W'}$. We have
\begin{subequations}
  \begin{align}
    \vek{R}_{j'}^{W' \mu'}
    & \equiv g \, \vek{R}_j^{W' \mu} = g \, ( \vek{R}_j^{W \mu} - \tshift)\\
    & = g \, \vek{R}_j^{W \mu} - g \, \tshift\\
    & = \vek{R}_{j'}^{W \mu'} - g \, \tshift ,
  \end{align}
\end{subequations}
so that for the transformed Wyckoff letter $W'$, Eq.\ (\ref{eq:pw-rep}) becomes
\begin{subequations}
  \begin{align}
    \mathcal{D}_\vek{k}^{W'} (g)_{\mu' \mu}
    & = \exp [ i \vek{k} \cdot (\vek{R}_{j'}^{W' \mu'} - \vek{R}_j^{W' \mu'})]\\
    & = \exp [ i \vek{k} \cdot ( \vek{R}_{j'}^{W \mu'} - g \, \tshift
        - \vek{R}_j^{W \mu} + \tshift)]\\
    & = \exp [ i \vek{k} \cdot ( \vek{R}_{j'}^{W \mu'} - \vek{R}_j^{W \mu} )]
        \exp [ - i \vek{k} \cdot ( g \, \tshift - \tshift)]\\
    & = \mathcal{D}^q_{\vek{k} W} (g)_{\mu' \mu} \,
        \mathcal{D}_\tshift^\vek{k} (g)
  \end{align}
\end{subequations}
with $\mathcal{D}_\tshift^\vek{k} (g)$ given by Eq.\ (\ref{eq:phase-d-amb}). This gives us the rearrangement lemma for the representations of plane waves
\begin{equation}
  \label{eq:ir-shift}
  \Gamma^q_{\vek{k} W'} = \Gamma_\tshift \times \Gamma^q_{\vek{k} W}.
\end{equation}
Hence we confirm that the nontrivial case $\Gamma^q_{\vek{k} W'} \neq \Gamma^q_{\vek{k} W}$ requires that $\Gamma_\tshift = \{\mathcal{D}_\tshift^\vek{k} (g) : g \in \kgroup \}$ is not the identity representation.  As mentioned above, the symmetry of plane waves $q_\vek{k} (\vek{R}_j^{W \mu})$ is a universal problem for each space group $\spacegroup$, independent of the detailed realization of a crystal structure.  This holds, in particular, if the atoms are located at positions $\vek{R}_{W \mu j}^{I \vek{k}}$ where the $\pwbase_\vek{k}^{I\beta} (\vek{R}_{W \mu j}^{I \vek{k}})$ transforms according to only one IR $\Gamma_I$.  Hence it is possible to discuss the rearrangement lemma for the IRs of Bloch states independent of a particular crystal structure, but it depends only on the space group $\spacegroup$. Among all 230 space groups, 159 contain Wyckoff sites with origin-dependent site symmetries \cite{boy73}.  Though a necessary criterion, it is however not a sufficient criterion for a rearrangement of the IRs of Bloch states under a change of the coordinate system \cite{cor71}.

\subsection{Effect of time reversal}
\label{sec:time-reversal}

In the absence of an external magnetic field, the eigenfunctions of the TB Hamiltonian obey time-reversal symmetry $\Theta$.  This implies that if an eigenfunction $\Psi_{\vek{k}}^{I \beta} (\vek{r})$ with energy $E(\vek{k})$ transforms according to the $\beta$th component of the IR $\Gamma_I$ of $\kgroup$, the time-reversed wave function $\Theta \, \Psi_{\vek{k}}^{I \beta} (\vek{r}) = \Psi_{\vek{k}}^{I \beta} (\vek{r})^\ast = \Psi_{-\vek{k}}^{\tilde{I}, \beta'} (\vek{r})$ [which is likewise an eigenfunction of the Hamiltonian with energy  $E(-\vek{k}) = E(\vek{k})$] transforms according to the $\beta'$th component of the complex conjugate IR $\Gamma_I^\ast$ of $\kgroup[-\vek{k}] = \kgroup$.  Therefore, if the eigenfunctions of the TB Hamiltonian for some energy $E (\vek{k})$ contain atomic orbitals transforming according to an IR $\Gamma_J^\phi$ of $\kgroup$ and symmetry-adapted plane waves transforming according to $\Gamma_{J'}^q$, the eigenfunctions for wave vector $-\vek{k}$ with energy $E (-\vek{k}) = E (\vek{k})$ contain atomic orbitals transforming according to the complex conjugate IR $\Gamma_J^{\phi \ast}$ and plane waves transforming according to $\Gamma_{J'}^{q \ast}$.  Here, the symmetry-adapted atomic orbitals at $-\vek{k}$ are the complex conjugates of the corresponding atomic orbitals at $\vek{k}$.  We obtain the symmetry-adapted plane waves at $-\vek{k}$ from the corresponding plane waves at $\vek{k}$ by replacing $\vek{k} \rightarrow -\vek{k}$.

Degeneracies of Bloch states due to time-reversal symmetry are discussed in Refs.~\cite{her37, streitwolf71, bir74}.  In general, three cases must be distinguished
\footnote{The classification of representations under time reversal adopted here from Ref.~\cite{bir74} is the most convenient for physical applications, but it differs from that customary in courses on group theory, where real representations are assigned to case (a) and complex inequivalent and equivalent representations to cases (b) and (c). The two classifications agree for single-group representations, but cases (a) and (c) must be interchanged for double-group representations.}.
In case (a), eigenfunctions $\Psi$ and $\Theta \, \Psi$ of the crystal Hamiltonian $\Hspace$ are linearly dependent.  In case (b), eigenfunctions $\Psi$ and $\Theta \, \Psi$ of $\Hspace$ are linearly independent and transform according to inequivalent representations $\Gamma_I$ and $\Gamma_I^\ast$, i.e., $\chi_I (g) \ne \chi_I^\ast (g)$ for some $g \in \kgroup$.  Finally, in case (c), eigenfunctions $\Psi$ and $\Theta \, \Psi$ of $\Hspace$ are linearly independent and transform according to equivalent representations $\Gamma_I$ and $\Gamma_I^\ast$, i.e., $\chi_I (g) = \chi_I^\ast (g)$ for all $g \in \kgroup$.  In cases (b) and (c) invariance under time reversal causes additional degeneracy.
We have for symmorphic space groups \cite{her37, streitwolf71, bir74}
\begin{equation}
  \label{eq:herring}
  \frac{f}{h} \sum_{g \in \pointgroup} \chi_I (g^2)
  \; \delta_{g \vek{k} + \vek{k}, \vek{b}}
  = \left\{
    \begin{array}{rs{1em}L}
      \Theta^2 & case (a) \\[0.5ex]
      0 \hspace*{0.55em} & case (b) \\[0.5ex]
      - \Theta^2 & case (c)
    \end{array} \right. ,
\end{equation}
where $f$ is the number of points of the star of $\vek{k}$, $h$ is the order of the crystallographic point group $\pointgroup$ of the crystal, $\chi_I (g)$ are the characters of the IR $\Gamma_I$ of the point group $\kgroup$ of the wave vector $\vek{k}$, and $g \vek{k} + \vek{k}$ may be zero or a reciprocal lattice vector $\vek{b}$.  We have $\Theta^2 = +1$ for single-group representations and $\Theta^2 = -1$ for double-group representations.  The criterion (\ref{eq:herring}) applies to the IRs $\Gamma_I$ of $\kgroup$ independent of the origin of the coordinate system.  If a crystal structure permits a change of the coordinate system characterized by a vector $\tshift$ with RAR $\Gamma_\tshift$, a Bloch function transforming according to the IR $\Gamma_I$ of $\kgroup$ in the old coordinate system transforms according to $\Gamma_{I'} = \Gamma_\tshift \times \Gamma_I$ in the new coordinate, see Eq.\ (\ref{eq:ir:trafo:sum}).  Therefore, the IRs $\Gamma_I$ and $\Gamma_{I'}$ of $\kgroup$ must fall into the same category according to Eq.\ (\ref{eq:herring}).

In a more detailed analysis \cite{ras59, streitwolf71, bir74}, for each of the cases (a), (b), and (c) three possibilities must be distinguished: (1) the points $\vek{k}$ and $-\vek{k}$ are equivalent, i.e., $\vek{k} = - \vek{k} + \vek{b}$; (2) $\vek{k}$ is not equivalent to $-\vek{k}$, but the space group contains an element $R$ which maps $\vek{k}$ onto $-\vek{k}$
\begin{equation}
  \label{eq:rashba:2}
  R \vek{k} = - \vek{k};
\end{equation}
(3) the points $\vek{k}$ and $-\vek{k}$ are in different stars.  For the systems discussed below, case (1) applies to the $\vek{\Gamma}$ and $\vek{M}$ points of the BZ, whereas case (2) applies to the $\vek{K}$ points.

\section{Band Symmetries in M\lowercase{o}S$_2$}
\label{sec:bloch-sym-mos2}

Having derived a systematic theory for the symmetry of TB Bloch functions, we now apply this theory to several quasi-2D materials.  Our main focus is on monolayer MoS$_2$.  For comparison, we also discuss single-layer (SLG), bilayer (BLG), and trilayer (TLG) graphene in the next section.

\subsection{Crystal structure of M\lowercase{o}S$_2$}
\label{sec:crystal-mos2}

\begin{figure}
   \includegraphics[width=0.95\columnwidth]{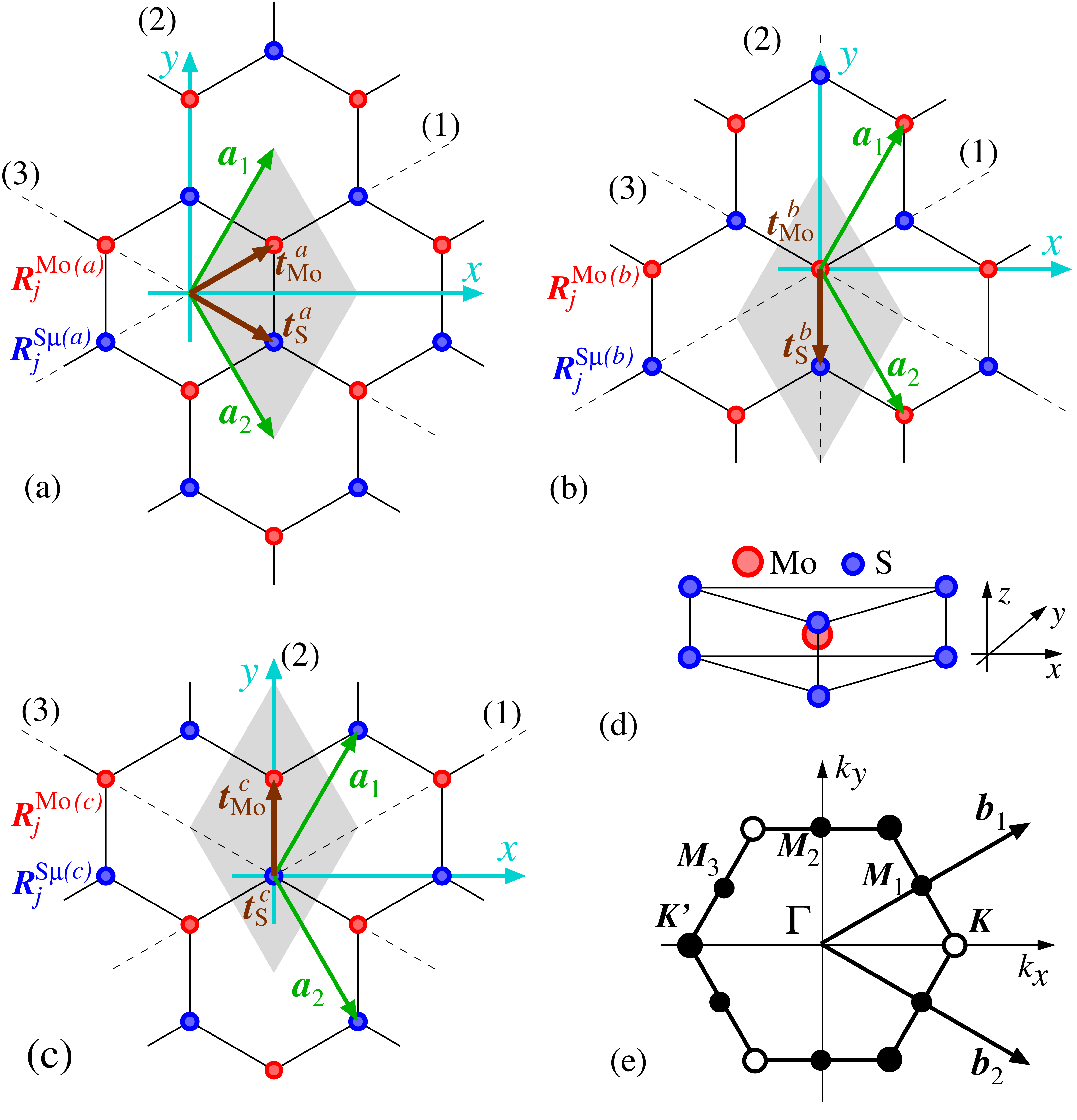}
  \caption{Crystal structure of single-layer MoS$_2$. Three coordinate systems  $\alpha =a, b, c$ are considered with (a) the origin located at the center of a hexagon, (b) origin at an Mo atom, and (c) origin at the midpoint between top and bottom S atoms. The atomic positions of Mo and S in unit cell $j$ are denoted by $\vek{R}_j^{\textrm{Mo}(\alpha)}$ and $\vek{R}_j^{\textrm{S}\mu(\alpha)}$, respectively. For the S atom, the top (bottom) atoms are labeled $\mu=1$ ($\mu = 2$). The primitive lattice vectors are denoted $\vek{a}_1$ and $\vek{a}_2$.  The shaded region shows a unit cell ($j=1$). The vectors $\vek{t}_\mathrm{Mo}^\alpha$ and $\vek{t}_\mathrm{S}^\alpha$  give the positions of the Mo and S atoms within a unit cell. The dashed axes $(1)$, $(2)$, and $(3)$ are the twofold rotation axes of the point group $D_{3h}$. (d) Three-dimensional illustration of single-layer MoS$_2$. (e) The first Brillouin zone.}
\label{fig:singlemos2}
\end{figure}

The crystal structure of single-layer MoS$_2$ is shown in Fig.~\ref{fig:singlemos2}. It is characterized by the point group $D_{3h}$ and space group $P\bar{6} m 2$ (\# 187).  Three Wyckoff positions have multiplicity $m=1$, the positions of the Mo atom, the midpoint between a pair of top and bottom S atoms, and the center of the hexagon. Hence, as discussed in Sec.~\ref{sec:band-amb}, three choices $\alpha = a,b,c$ emerge for the origin of the coordinate system: ($a$) origin at the center of the hexagon [Fig.~\ref{fig:singlemos2}(a)], ($b$) origin at a Mo atom [Fig.~\ref{fig:singlemos2}(b)], and ($c$) origin at the midpoint between a top and bottom S atom [Fig.~\ref{fig:singlemos2}(c)]. In either case, the Mo atoms have site symmetry $D_{3h}$ and Wyckoff multiplicity $m=1$.  Yet the Wyckoff letters for these positions listed in Table~\ref{tab:examples-site-sym} depend on the coordinate system \cite{boy73}.  For coordinate system ($a$), the Mo and S atoms have Wyckoff letters $e$ and $h$, respectively, whereas these letters become $a$ and $i$ in coordinate system ($b$), and $c$ and $g$ in coordinate system ($c$).  The S atoms have site symmetry $C_{3v}$ and multiplicity $m=2$.  The positions of Mo and S atoms in unit cell $j$ are denoted by $\vek{R}_j^{\textrm{Mo} (\alpha)}$ and $\vek{R}_j^{\textrm{S} \mu (\alpha)}$, respectively. For the S atom, the top (bottom) atoms are labeled $\mu=1$ ($\mu = 2$). There are yet other coordinate systems that can be used for MoS$_2$.  For example, Ref.~\cite{kor13} used a coordinate system that differs from coordinate system ($a$) by a reflection about the $xz$ plane.  Here, we do not consider these additional coordinate systems \cite{cor72}.

\begin{table*}
 \caption{Site symmetries in monolayer MoS$_2$ and SLG, BLG, and TLG. We include here the Wyckoff positions occupied by atoms as well as the unoccupied center of the hexagon denoted by $\vek{R}_j^{\mathrm{center}}$.  The coordinate systems $\alpha = a, b, c$ for MoS$_2$ and TLG are depicted in Figs.~\ref{fig:singlemos2} and \ref{fig:tlg}, respectively.}  \label{tab:examples-site-sym}
  \renewcommand{\arraystretch}{1.2} \extrarowheight0.2ex
  $\begin{array}{Ls{3em}cS{2em}cccc}
     \hline \hline
     & (\alpha) &
     \multicolumn{3}{C}{MoS$_2$\quad ($P\bar{6}m2$, \# 187, $D_{3h}$)} \\ \hline
     positions $\vek{R}_j^{W \mu (\alpha)}$ &  & \vek{R}_j^{\text{Mo} (\alpha)} & \{ \vek{R}_j^{\text{S}1(\alpha)} , \vek{R}_j^{\text{S}2(\alpha)} \} & \vek{R}_j^{\mathrm{center} (\alpha)} \\
     site symmetry & & \bar{6} m 2~(D_{3h})  &  3m~(C_{3v})  &  \bar{6}m2~(D_{3h})   \\
     multiplicity & &  1  &  2  &  1  \\
     Wyckoff letter & (a) &  e  &  h  &  a  \\
     & (b) &  a  &  i  &  c  \\
     & (c) &  c  &  g  &  e  \\
      \hline
     & & \multicolumn{3}{C}{SLG\quad ($P6/mmm$, \# 191, $D_{6h}$)} \\ \hline
     positions $\vek{R}_j^{W \mu}$ &  & \{\vek{R}_j^{c1},\vek{R}_j^{c2}\} & \vek{R}_j^{\mathrm{center}} \\
     site symmetry & &  \bar{6} m 2~(D_{3h}) &  6/mmm~(D_{6h})  \\
     multiplicity & &  2  &  1  \\
     Wyckoff letter & &  c  &  a  \\
     \hline
     & &
     \multicolumn{3}{C}{BLG\quad ($P\bar{3}m1$, \# 164, $D_{3d}$)} \\ \hline
     positions $\vek{R}_j^{W \mu}$ &  & \{  \vek{R}_j^{c1}, \vek{R}_j^{c2} \} &  \{\vek{R}_j^{d1},\vek{R}_j^{d2}\} \\
     site symmetry & &  3m~(C_{3v})  &  3m~(C_{3v})  \\
     multiplicity & &  2  &  2  \\
     Wyckoff letter & &  c  &  d  \\
      \hline
     & (\alpha) &
     \multicolumn{3}{C}{TLG\quad ($P\bar{6}m2$, \# 187, $D_{3h}$)} \\ \hline
     positions $\vek{R}_j^{W \mu (\alpha)}$ &  & \vek{R}_j^A & \vek{R}_j^ B & \{\vek{R}_j^{A'1}, \vek{R}_j^{A'2}\} &  \{ \vek{R}_j^{B'1} , \vek{R}_j^{B'2} \} \\
     site symmetry & & \bar{6} m2~(D_{3h}) & \bar{6} m2~(D_{3h}) & 3m~(C_{3v})  &  3m~(C_{3v}) \\
     multiplicity & & 1 & 1 & 2 & 2 \\
     Wyckoff letter & (a) & c & e & g & h \\
     & (b) & e & a & h & i \\
     & (c) & a & c & i & g \\ \hline \hline
   \end{array}$
\end{table*}

The Brillouin zone for single-layer MoS$_2$ is shown in Fig.~\ref{fig:singlemos2}(e).  In the following, we will focus on the $\vek{\Gamma}$ point $\vek{k} = 0$, the $\vek{K}$, and the $\vek{M}$ points.  The star of the $\vek{K}$ point includes two inequivalent wave vectors denoted $\vek{K}$ and $\vek{K}'$.  The star of the $\vek{M}$ point includes three inequivalent wave vectors denoted $\vek{M}_1$,  $\vek{M}_2$, and  $\vek{M}_3$.

\begin{figure}
   \includegraphics[width=0.95\columnwidth]{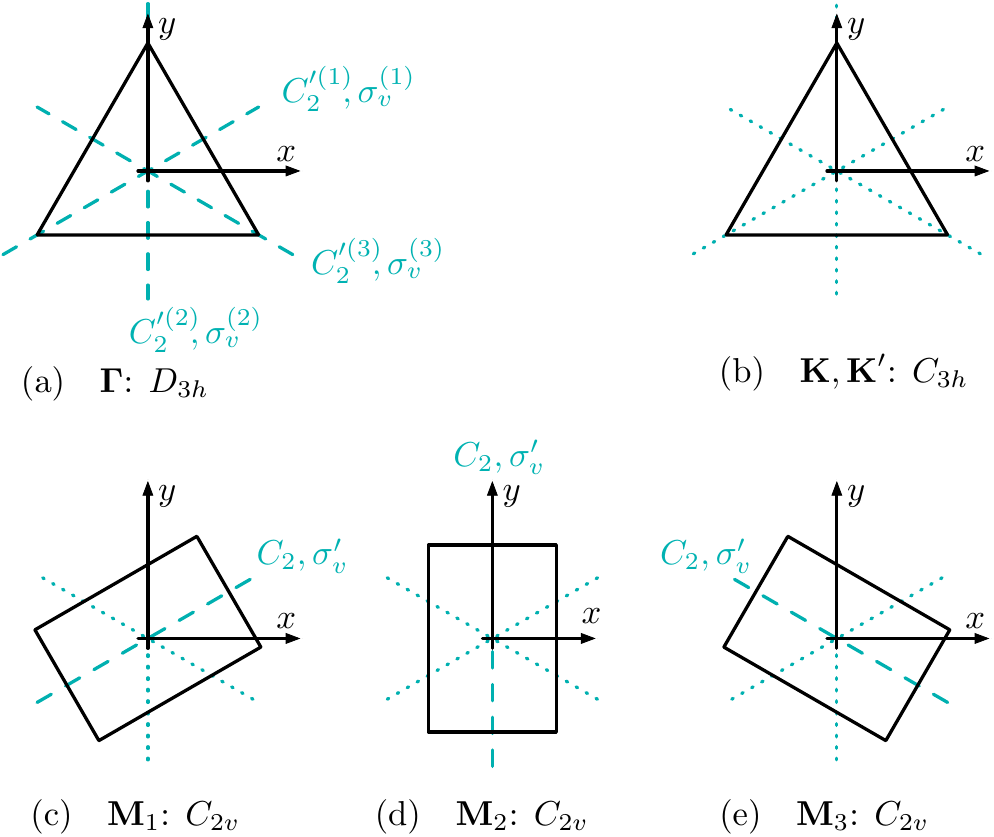}
   \caption{Groups of the wave vector in monolayer MoS$_2$ and TLG. (a) The point $\vek{\Gamma}$ has the point group $D_{3h}$ with the $z$ axis (out of plane) as the threefold rotation axis. The dashed lines ($i$) are the axes for twofold rotations $C_2'^{(i)}$ with $i=1,2,3$. The reflection $\sigma_v^{(i)}$ is about a plane that includes the dashed axis $i$ and the $z$ axis. (b) The points $\vek{K}$ and $\vek{K}'$ have the point group $C_{3h}$ with threefold rotations about the $z$ axis. The reflection plane of $\sigma_h$ is the $xy$ plane.  The dotted lines indicate the twofold rotation axes that appear in the point group $D_{3h}$ but are not symmetry elements of $C_{3h}$.
[(c)--(e)] The points $\vek{M}_1$, $\vek{M}_2$, and $\vek{M}_1$ have the point group $C_{2v}$. The dashed line is the axis of the twofold rotation $C_2$.  The reflection plane of $\sigma_v$ is the $xy$ plane.  The reflection plane of $\sigma_v'$ contains the dashed line and the $z$ axis.}
\label{fig:mos2-tlg-coord}
\end{figure}

The point group of single-layer MoS$_2$ is $D_{3h}$.  This is also the point group of the wave vector at the $\vek{\Gamma}$ point.  It contains a $120^\circ$ counterclockwise rotation $C_3$ about the $z$ axis. The reflection plane of $\sigma_h$ is the $xy$ plane and $S_3 = \sigma_h C_3$. The rotation axes of the three twofold rotations $C_2'^{(i)}$ are the axes $i=1,2,3$ shown in Fig.~\ref{fig:singlemos2}.  These axes are also shown as dashed lines in Fig.~\ref{fig:mos2-tlg-coord}(a). The reflection plane of $\sigma_v^{(i)}$ is the plane passing through the axis $i$ and the $z$ axis. The characters of $D_{3h}$ are listed in Table~\ref{tab:char-d3h} (Appendix~\ref{sec:char-tables}). We label the IRs of the crystallographic point groups following Koster \emph{et al.} \cite{kos63}.

At the $\vek{K}$ point [Fig.~\ref{fig:mos2-tlg-coord}(b)], the point group of the wave vector becomes $\kgroup[\vek{K}] = C_{3h}$ whose characters are listed in Table~\ref{tab:char-c3h} (Appendix~\ref{sec:char-tables}).  Finally, at the inequivalent points $\vek{M}_i$ ($i=1,2,3$), the group of the wave vector is $\kgroup[\vek{M}] = C_{2v}$, the character table of which is reproduced in Table~\ref{tab:char-c2v} (Appendix~\ref{sec:char-tables}). This group contains the twofold rotation $C_2$, the axis of which is indicated as dashed line in Figs.\ \ref{fig:mos2-tlg-coord}(c)--\ref{fig:mos2-tlg-coord}(e), the reflection $\sigma_v$ about the $xy$ plane, and the reflection $\sigma_v'$ for which the reflection plane includes the dashed line and the $z$ axis.

The primitive lattice vectors $\vek{a}_1$ and $\vek{a}_2$ are
\begin{equation}
  \label{eq:primitive-lattice-vectors}
  \vek{a}_1 =\frac{a}{2}\tvek{1\\\sqrt{3}},
  \quad \vek{a}_2 = \frac{a}{2} \tvek{1\\-\sqrt{3}},
\end{equation}
where $a$ is the lattice constant. Ignoring for brevity the $z$ component, the positions $\vek{t}_\mathrm{Mo}^\alpha$ of Mo and $\vek{t}_\mathrm{S}^\alpha$ of S in the unit cell are
\begin{subequations}
  \begin{align}
\vek{t}_\mathrm{Mo}^a & = \frac{a}{2} \tvek{1\\\frac{1}{\sqrt{3}}} , & \vek{t}_\mathrm{S}^a & = \frac{a}{2} \tvek{1 \\ -\frac{1}{\sqrt{3}}} , \\
\vek{t}_\mathrm{Mo}^b & = \frac{a}{2}\tvek{ 0 \\ 0 } , &
\vek{t}_\mathrm{S}^b & = \frac{a}{2}\tvek{0 \\ - \frac{2}{\sqrt{3}}} ,\\
\vek{t}_\mathrm{Mo}^c  & = \frac{a}{2} \tvek{ 0 \\ \frac{2}{\sqrt{3}}} , & \vek{t}_\mathrm{S}^c & = \frac{a}{2} \tvek{ 0 \\ 0 } ,
  \end{align}
\end{subequations}
where the superscript $\alpha = a, b, c$ denotes the coordinate system. The primitive vectors $\vek{b}_1$ and $\vek{b}_2$ of the reciprocal
lattice are
\begin{equation}
  \label{eq:rec-primitive-lattice-vectors}
  \vek{b}_1 =\frac{2 \pi}{a}\tvek{1\\1/\sqrt{3}},
  \quad \vek{b}_2 =\frac{2 \pi}{a}\tvek{1\\-1/\sqrt{3}}.
\end{equation}
The two inequivalent corner points of the Brillouin zone are
\begin{equation}
  \label{eq:K-points}
  \vek{K} =\frac{2\pi}{a}\tvek{2/3\\0},
  \quad  \vek{K}' =\frac{2\pi}{a}\tvek{-2/3\\0},
\end{equation}
and the $\vek{M}$ points are
\begin{align}
  \nonumber \vek{M}_1 & = \frac{\pi}{a}\tvek{1\\1/\sqrt{3}},
  \quad  \vek{M}_2 =\frac{2\pi}{a}\tvek{0\\1/\sqrt{3}},\\
  \vek{M}_3 & = \frac{\pi}{a}\tvek{-1\\1/\sqrt{3}},
  \label{eq:M-points}
\end{align}
see Fig.~\ref{fig:singlemos2}(e).

\subsection{Rearrangement of IRs of Bloch states in M\lowercase{o}S$_2$}
\label{sec:band-amb-mos2}

The crystal structure of monolayer MoS$_2$ can be described by three different coordinate systems $\alpha = a, b, c$ shown in Fig.~\ref{fig:singlemos2}. This provides an example for the rearrangement of the IRs of Bloch states discussed in general terms in Sec.~\ref{sec:band-amb}. The coordinate systems $\alpha$ and $\beta$ are related via a translation $\tshift_{\alpha \beta}$.  For the shifts $a \rightarrow b$, $b \rightarrow c$, and $c \rightarrow a$, the translation vectors (apart from a lattice vector) are given by
\begin{equation}\label{eq:coord-trans-mos2}
  \tshift_{a b} = \tshift_{b c} = \tshift_{c a}
  = \frac{a}{2}\tvek{1 \\ 1/\sqrt{3}}
\end{equation}
and $\tshift_{\beta \alpha} = - \tshift_{\alpha \beta}$.

For wave vectors $\vek{k}$ inside the BZ such as the $\vek{\Gamma}$ point as well as for the $\vek{M}$ points, the RARs $\Gamma_{\alpha \beta}^\vek{\Gamma}$ and $\Gamma_{\alpha \beta}^\vek{M}$ are given by Eq.\ (\ref{eq:phase-d-amb-inv}) with $\vek{b}_g =0$ for all $g \in \kgroup$.  These RARs are, therefore, given by the identity representation $\Gamma_1$, i.e., at both the $\vek{\Gamma}$ and $\vek{M}$ points the labeling of Bloch states is independent of the coordinate system. However, a shift of the coordinate system rearranges the IRs at the $\vek{K}$ point.  Using Eq.\ (\ref{eq:phase-d-amb-inv-bg}) at the $\vek{K}$ point, where the group of the wave vector is $C_{3h}$, we get
\begin{subequations}\label{eq:phase-d-amb-inv-bg-mos2-K}
  \begin{align}
    \vek{b}_E & = \vek{b}_{\sigma_h} = 0 , \\
    \vek{b}_{C_3} & = \vek{b}_{S_3} = - \vek{b}_1
    = - \frac{2 \pi}{a} \tvek{ 1 \\ 1/\sqrt{3}} , \\
    \vek{b}_{C_3^{-1}}
    & = \vek{b}_{S_3^{-1}} = - \vek{b}_2
    = - \frac{2 \pi}{a} \tvek{ 1 \\ - 1/\sqrt{3} }.
  \end{align}
\end{subequations}
For the shifts $a \rightarrow b$, $b \rightarrow c$, and $c \rightarrow a$, we thus have using Eq.\ (\ref{eq:phase-d-amb-inv})
\begin{subequations}
  \begin{align}
    \mathcal{D}_{\alpha\beta}^\vek{K} (E)
    & = \mathcal{D}_{\alpha\beta}^\vek{K} (\sigma_h)
    = \exp (-i \vek{b}_E \cdot \tshift_{\alpha\beta} ) = 1 , \\
    \mathcal{D}_{\alpha\beta}^\vek{K} (C_3)
    & = \mathcal{D}_{\alpha\beta}^\vek{K} (S_3)
    = \exp (-i \vek{b}_{C_3} \cdot \tshift_{\alpha\beta} ) = \omega^{-4} , \\
    \mathcal{D}_{\alpha\beta}^\vek{K} (C_3^{-1})
    & = \mathcal{D}_{\alpha\beta}^\vek{K} (S_3^{-1})
    = \exp (-i \vek{b}_{C_3^{-1}} \cdot \tshift_{\alpha\beta} ) = \omega^4
  \end{align}
\end{subequations}
with $\omega \equiv \exp(i\pi/6)$.  This implies $\Gamma_{ab}^\vek{K} = \Gamma_{bc}^\vek{K} = \Gamma_{ca}^\vek{K} = \Gamma_3$. Since $\tshift_{\beta\alpha} = - \tshift_{\alpha\beta}$, we have $\Gamma_{ba}^\vek{K} = \Gamma_{cb}^\vek{K} = \Gamma_{ac}^\vek{K} = \Gamma_3^\ast = \Gamma_2$. Hence, at the $\vek{K}$ point, for a Bloch state transforming in one coordinate system according to a certain IR, we can multiply this IR with either $\Gamma_2$ or $\Gamma_3$ to obtain the IR of the same Bloch state in a different coordinate system. The multiplication table for the IRs of $C_{3h}$ is reproduced in Table~\ref{tab:multiplication:C3h} (Appendix~\ref{sec:char-mult-tables}). Table~\ref{tab:ir-amb-c3h} shows how the IRs of $C_{3h}$ are rearranged when we go from one of the three coordinate systems $\alpha =a, b, c$ to a different coordinate system $\beta$. At the $\vek{K}'$ point, Eq.\ (\ref{eq:phase-d-amb}) gives
\begin{equation}
  \Gamma_{\alpha \beta}^{\vek{K}'}
  = \Gamma_{\alpha \beta}^{-\vek{K}}
  = \Gamma_{\alpha \beta}^{\vek{K} \ast}.
\end{equation}

\begin{table}
 \caption{Rearrangement of the IRs of $C_{3h}$ at the $\vek{K}$ point of monolayer MoS$_2$ when we go from one of the three coordinate systems $\alpha =a, b, c$ to a different coordinate system $\beta$ (see Fig.~\ref{fig:singlemos2}).} \label{tab:ir-amb-c3h}
  \renewcommand{\arraystretch}{1.2} \extrarowheight0.5ex
$
\begin{array}{cs{2em}*{5}{cs{1.0em}}c}
  \hline \hline
  a & \Gamma_1 & \Gamma_2 & \Gamma_3 & \Gamma_4 & \Gamma_5 & \Gamma_6 \\
  b & \Gamma_3 & \Gamma_1 & \Gamma_2 & \Gamma_6 & \Gamma_4 & \Gamma_5 \\
  c & \Gamma_2 & \Gamma_3 & \Gamma_1 & \Gamma_5 & \Gamma_6 & \Gamma_4
  \\ \hline \hline
\end{array}
$
   \end{table}

\subsection{Atomic orbitals $\phi_\nu^W(\delta \mathbf{r})$ at $\mathbf{k} = \mathbf{\Gamma}$, $\mathbf{K}$, and $\mathbf{M}$}

The conduction and valence bands in MoS$_2$ are dominated by Mo $d$ and S $p$ orbitals \cite{mat73oct}. At the points $\vek{\Gamma}$, $\vek{K}$, and $\vek{M}$, the groups of the wave vectors $\kgroup$ are $D_{3h}$, $C_{3h}$, and $C_{2v}$, respectively, with character tables reproduced in Tables~\ref{tab:char-d3h}, \ref{tab:char-c3h}, and \ref{tab:char-c2v}. We use Eq.\ (\ref{eq:projection-op}) to project these functions onto functions transforming according to the IRs of the various groups $\kgroup$. The relevant symmetry operations are defined in Fig.~\ref{fig:mos2-tlg-coord} and Table~\ref{tab:functions} considering both polar vectors $\vek{P}$ such as position $\vek{r}$ and axial vectors $\vek{A}$.  The two inequivalent points $\vek{K}$ and $\vek{K}'$ are related by a vertical reflection that transforms the component $x$ into $-x$ while keeping the $y$ and $z$ components fixed. Table~\ref{tab:mos2-orbitals} summarizes the IRs of the $p$ and $d$ orbitals.  We note that these results are fully consistent with the full rotation group compatibility tables in Ref.~\cite{kos63}.

\begin{table}
  \caption{Mapping of polar vectors $\vek{P}$ and axial vectors $\vek{A}$ under $g = C_3$, $C_3^{-1}$, $\sigma_v^{(i)}$, and $\sigma_h$, where the images are expressed in terms of the original components of these vectors.  $C_3$ is a $120^\circ$ counterclockwise rotation about the $z$ axis while $\sigma_v^{(i)}$ is a reflection about the plane containing the $z$ axis and the axis $i$ defined in Fig.~\ref{fig:mos2-tlg-coord}. We have $\omega \equiv \exp (i\pi/6)$. Note $S_3^{\pm 1} = \sigma_h \times C_3^{\pm 1}$ and $C_2^{\prime (i)} = \sigma_h \times \sigma_v^{(i)}$.}
  \label{tab:functions}
  \renewcommand{\arraystretch}{1.2} \extrarowheight0.5ex
  $\begin{array}{lS{1.5em}*{4}{c}}
     \hline\hline
     & C_3^{\pm 1} & \sigma_v^{(1)} / \sigma_v^{(3)} &
     \sigma_v^{(2)} & \sigma_h \\[0.5ex] \hline
     P_x & -\frack{1}{2}P_x \mp \frack{\sqrt{3}}{2}P_y & \frack{1}{2} P_x \pm \frack{\sqrt{3}}{2} P_y & -P_x & P_x \\
     P_y & \pm \frack{\sqrt{3}}{2}P_x - \frack{1}{2}P_y & \pm \frack{\sqrt{3}}{2}P_x - \frack{1}{2} P_y & P_y & P_y \\
     P_z & P_z & P_z & P_z & -P_z \\
     P_+ & \omega^{\pm 4} P_+ & -\omega^{\mp 4} P_- & - P_- & P_+ \\
     P_- & \omega^{\mp 4} P_- & -\omega^{\pm 4} P_+ & - P_+ & P_-\\ \hline
     A_x & -\frack{1}{2}A_x \mp \frack{\sqrt{3}}{2}A_y & -\frack{1}{2} A_x \mp \frack{\sqrt{3}}{2} A_y & A_x & -A_x \\
     A_y & \pm \frack{\sqrt{3}}{2}A_x - \frack{1}{2}A_y & \mp \frack{\sqrt{3}}{2}A_x + \frack{1}{2} A_y & -A_y & -A_y \\
     A_z & A_z & -A_z & -A_z & A_z \\
     A_+ & \omega^{\pm 4} A_+ & \omega^{\mp 4} A_- & A_- & -A_+\\
     A_- & \omega^{\mp 4} A_- & \omega^{\pm 4} A_+ & A_+ & -A_-
     \\[0.5ex] \hline\hline
  \end{array}$
\end{table}

\begin{table*}
 \caption{Symmetry-adapted $p$ and $d$ atomic orbitals for the point groups $D_{3h}$, $C_{3h}$ and $C_{2v}$ with coordinate systems defined in Fig.~\ref{fig:mos2-tlg-coord} (MoS$_2$ and TLG). For $C_{2v}$, the symmetrized atomic orbital $\phi_\nu^{[i]}$ corresponds to the coordinate systems used for the point $\vek{M}_i$ in Figs.\ \ref{fig:mos2-tlg-coord}(c)--\ref{fig:mos2-tlg-coord}(e). The orbital $[i(j)]$ takes the upper (lower) sign. The IRs of the atomic orbitals listed here are consistent with the compatibility relations in Table~\ref{tab:compatibility-ir-mos2}.} \label{tab:mos2-orbitals}
  \renewcommand{\arraystretch}{1.2} \extrarowheight0.5ex
  $\begin{array}{*{2}{cs{1.5em}cs{2.0em}}*{2}{cs{1.5em}}c}
     \hline \hline
     \multicolumn{2}{c}{D_{3h}} &
     \multicolumn{2}{c}{C_{3h}} &
     \multicolumn{3}{c}{C_{2v}} \\
     \phi_\nu & \text{IR} & \phi_\nu & \text{IR} &
     \phi_\nu^{[1(3)]} & \phi_\nu^{[2]} & \text{IR} \\ \hline
     p_z & \Gamma_4 & p_z & \Gamma_4 & p_z & p_z & \Gamma_4 \\
     \{ p_x , p_y \} & \Gamma_6 & p_x + i p_y & \Gamma_2 & p_x \mp \sqrt{3} p_y & p_x & \Gamma_2 \\
     & & p_x - i p_y& \Gamma_3 & \sqrt{3} p_x \pm p_y & p_y & \Gamma_1 \\
     \hline
     d_{z^2} & \Gamma_1 & d_{z^2} & \Gamma_1 & d_{z^2} & d_{z^2} & \Gamma_1 \\
     \{ d_{xz} , d_{yz} \} & \Gamma_5 & d_{xz} + i d_{yz} & \Gamma_5 & d_{xz} \mp \sqrt{3} d_{yz} & d_{xz} & \Gamma_3 \\
     & & d_{xz} - i d_{yz} & \Gamma_6 & \sqrt{3} d_{xz} \pm d_{yz} & d_{yz} & \Gamma_4 \\
     \{ d_{x^2 - y^2} , d_{xy} \} & \Gamma_6 & d_{x^2 - y^2} + i d_{xy} & \Gamma_3 & d_{x^2-y^2} \pm \sqrt{3} d_{xy} & d_{x^2-y^2} & \Gamma_1 \\
     & & d_{x^2 - y^2} - i d_{xy} & \Gamma_2 & \sqrt{3} d_{x^2-y^2} \mp d_{xy} & d_{xy} & \Gamma_2 \\
     \hline \hline
   \end{array}$
\end{table*}

\begin{table*}
 \caption{Compatibility relations for the IRs of $D_{3h}$ and the IRs of its subgroups $C_{3h}$ and $C_{2v}$ using the coordinate system in Fig.~\ref{fig:mos2-tlg-coord}. The three different orientations of coordinate  systems for $C_{2v}$ in Figs.\ \ref{fig:mos2-tlg-coord}(c)--\ref{fig:mos2-tlg-coord}(E) follow the same compatibility relations.} \label{tab:compatibility-ir-mos2}
  \renewcommand{\arraystretch}{1.2} \extrarowheight0.5ex
  $\begin{array}{cs{2em}*{5}{cs{1.5em}}c} \hline\hline
     D_{3h} & \Gamma_1 & \Gamma_2 & \Gamma_3 & \Gamma_4 & \Gamma_5 & \Gamma_6 \\
     \hline
     C_{3h} & \Gamma_1 & \Gamma_1 & \Gamma_4 & \Gamma_4 & \Gamma_5 + \Gamma_6 & \Gamma_2 + \Gamma_3 \\
     C_{2v} & \Gamma_1 & \Gamma_2 & \Gamma_3 & \Gamma_4 & \Gamma_3 + \Gamma_4 & \Gamma_1 + \Gamma_2 \\ \hline\hline
   \end{array}$
\end{table*}

\begin{table*}
 \caption{Symmetrized plane waves $\pwsBloch_\vek{k}^{I \beta} (\vekc{R}_j^W)$ for monolayer MoS$_2$ and SLG, BLG, and TLG at the $\vek{\Gamma}$, $\vek{K}$, and $\vek{M}$ points for the positions occupied by atoms. The symmetrized plane waves $\pwsBloch_{\vek{K}'}^{I \beta} (\vekc{R}_j^W)$ at $\vek{K}' = - \vek{K}$ are obtained from the expressions given for $\vek{K}$ by replacing $\vek{K}$ by $\vek{K}'$.  The corresponding IRs are the complex conjugates of the IRs at $\vek{K}$. The group $\kgroup$ of the wave vector and the plane wave IRs are shown. At the $\vek{K}$ point, we distinguish between the three coordinate systems $\alpha = a, b, c$ for MoS$_2$ and TLG with atomic positions $\vek{R}_j^{W (\alpha) \mu}$ depicted in Figs.~\ref{fig:singlemos2} and \ref{fig:tlg}, respectively. The IRs for the coordinate systems $\alpha$ are then denoted by $\Gamma_{i/j/k}$.  The definition of the symmetrized plane waves at the points  $\vek{M}_i$ of SLG and BLG contains prefactors $\gamma_i$, where $\gamma_1 = \gamma_3 = -1$, and $\gamma_2 = +1$.}
 \label{tab:plane-wave-examples}
  \renewcommand{\arraystretch}{1.2} \extrarowheight0.5ex
  $\begin{array}{Ls{1.0em}*{3}{s{1.2em}cs{0.8em}c}}
     \hline \hline
     & \multicolumn{2}{c}{\vek{k} = \vek{\Gamma}}
     & \multicolumn{2}{c}{\vek{k} = \vek{K}}
     & \multicolumn{2}{c}{\vek{k} = \vek{M}_1 , \vek{M}_2 , \vek{M}_3}
     \\ \hline
     MoS$_2$: & D_{3h} & & C_{3h} & & C_{2v} \\
     Mo & \Gamma_1 & \pwbase_\vek{\Gamma} (\vek{R}_j^{\textrm{Mo}})
        & \Gamma_{2/1/3} & \pwbase_\vek{K} (\vek{R}_j^{\textrm{Mo} })
        & \Gamma_1  & \pwbase_{\vek{M}_i} (\vek{R}_j^{\textrm{Mo}}) \\
     S  & \Gamma_1 , \Gamma_4 & \pwbase_\vek{\Gamma} (\vek{R}_j^{\textrm{S}1}) \pm \pwbase_\vek{\Gamma} (\vek{R}_j^{\textrm{S}2})
        & \Gamma_{3/2/1} , \Gamma_{6/5/4} & \pwbase_\vek{K} (\vek{R}_j^{\textrm{S}1 }) \pm \pwbase_\vek{K} (\vek{R}_j^{\textrm{S}2 })
        & \Gamma_1 , \Gamma_4  & \pwbase_{\vek{M}_i} (\vek{R}_j^{\textrm{S}1}) \pm \pwbase_{\vek{M}_i} (\vek{R}_j^{\textrm{S}2}) \\[0.6ex] \hline
     SLG: & D_{6h} & & D_{3h} & & D_{2h} \\
     $c$ & \Gamma_1^+ , \Gamma_3^- & \pwbase_\vek{\Gamma} (\vek{R}_j^{c1}) \pm \pwbase_\vek{\Gamma} (\vek{R}_j^{c2})
         & \Gamma_6 & \{\pwbase_\vek{K} (\vek{R}_j^{c1}),\pwbase_\vek{K} (\vek{R}_j^{c2})\}
         & \Gamma_1^+ , \Gamma_2^-  & \pwbase_{\vek{M}_i} (\vek{R}_j^{c1}) \pm \gamma_i \, \pwbase_{\vek{M}_i} (\vek{R}_j^{c2}) \\[0.6ex] \hline
     BLG: & D_{3d} & & D_3 & & C_{2h} \\
     $c$ & \Gamma_1^+ , \Gamma_2^- & \pwbase_\vek{\Gamma} (\vek{R}_j^{c1}) \pm \pwbase_\vek{\Gamma} (\vek{R}_j^{c2}) & \Gamma_1 , \Gamma_2 & \pwbase_\vek{K} (\vek{R}_j^{c1}) \pm \pwbase_\vek{K} (\vek{R}_j^{c2}) & \Gamma_1^+ , \Gamma_2^- & \pwbase_{\vek{M}_i} (\vek{R}_j^{c1}) \pm \pwbase_{\vek{M}_i} (\vek{R}_j^{c2}) \\
     $d$ & \Gamma_1^+ , \Gamma_2^- & \pwbase_\vek{\Gamma} (\vek{R}_j^{d1}) \pm \pwbase_\vek{\Gamma} (\vek{R}_j^{d2}) & \Gamma_3 & \{\pwbase_\vek{K} (\vek{R}_j^{d1}),\pwbase_\vek{K} (\vek{R}_j^{d2})\} & \Gamma_1^+ , \Gamma_2^-  & \pwbase_{\vek{M}_i} (\vek{R}_j^{d1}) \pm \gamma_i \, \pwbase_{\vek{M}_i} (\vek{R}_j^{d2}) \\[0.6ex] \hline
     TLG: & D_{3h} & & C_{3h} & & C_{2v} \\
     $A$ & \Gamma_1 & \pwbase_\vek{\Gamma} (\vek{R}_j^{\textrm{A}}) & \Gamma_{3/2/1} & \pwbase_\vek{K} (\vek{R}_j^{\textrm{A} }) & \Gamma_1  & \pwbase_{\vek{M}_i} (\vek{R}_j^{\textrm{A}}) \\
     $B$ & \Gamma_1 & \pwbase_\vek{\Gamma} (\vek{R}_j^{\textrm{B}}) & \Gamma_{2/1/3} & \pwbase_\vek{K} (\vek{R}_j^{\textrm{B} }) & \Gamma_1  & \pwbase_{\vek{M}_i} (\vek{R}_j^{\textrm{B}}) \\
     $A'$ & \Gamma_1 , \Gamma_4 & \pwbase_\vek{\Gamma} (\vek{R}_j^{\textrm{A}'1}) \pm \pwbase_\vek{\Gamma} (\vek{R}_j^{\textrm{A}'2}) & \Gamma_{1/3/2} , \Gamma_{4/6/5} & \pwbase_\vek{K} (\vek{R}_j^{\textrm{A}'1 }) \pm \pwbase_\vek{K} (\vek{R}_j^{\textrm{A}'2 }) & \Gamma_1 , \Gamma_4  & \pwbase_{\vek{M}_i} (\vek{R}_j^{\textrm{A}'1}) \pm \pwbase_{\vek{M}_i} (\vek{R}_j^{\textrm{A}'2}) \\
     $B'$ & \Gamma_1 , \Gamma_4 & \pwbase_\vek{\Gamma} (\vek{R}_j^{\textrm{B}'1}) \pm \pwbase_\vek{\Gamma} (\vek{R}_j^{\textrm{B}'2}) & \Gamma_{3/2/1} , \Gamma_{6/5/4} & \pwbase_\vek{K} (\vek{R}_j^{\textrm{B}'1 }) \pm \pwbase_\vek{K} (\vek{R}_j^{\textrm{B}'2 }) & \Gamma_1 , \Gamma_4  & \pwbase_{\vek{M}_i} (\vek{R}_j^{\textrm{B}'1}) \pm \pwbase_{\vek{M}_i} (\vek{R}_j^{\textrm{B}'2}) \\[0.6ex] \hline \hline
   \end{array}$
\end{table*}

\subsection{Transformation of plane waves $q_\mathbf{k} (\mathbf{R}_j^{\mathrm{Mo} (\alpha)})$}
\label{sec:trafo-plane-wave-mos2}

We now determine the IRs of the plane waves $q_\vek{k} (\vek{R}_j^{\textrm{Mo} (\alpha)})$ for the coordinate systems $\alpha = a, b, c$ at the $\vek{\Gamma}$, $\vek{K}$, and $\vek{M}$ points of the Brillouin zone. Since the Wyckoff letter corresponding to the positions of the Mo atoms has multiplicity $m=1$, we can use either Eq.\ (\ref{eq:phase-D-1d-r}) or Eq.\ (\ref{eq:phase-D-1d-k}) to determine the phase $\mathcal{D}_\vek{k}^{\mathrm{Mo} (\alpha)} (g)$ acquired by the plane waves $q_\vek{k} (\vek{R}_j^{\textrm{Mo} (\alpha)})$ under a transformation $g$. We can then derive the IRs of the plane waves using the projection operators (\ref{eq:projection-op}).  The results are summarized in Table~\ref{tab:plane-wave-examples}.

\subsection{Transformation of plane waves $q_\mathbf{k} (\mathbf{R}_j^{\mathrm{S} \mu (\alpha)})$}

The S atoms are located at Wyckoff positions with multiplicity $m=2$, so that we represent the plane wave at the positions $\vek{R}_j^{\mathrm{S} \mu (\alpha)}$ as a two-component spinor
\begin{equation}
  \pwBloch_\vek{k} (\vekc{R}_j^{S (\alpha)})
  = \pwbase_\vek{k} (\vek{R}_{j}^{S1(\alpha)})
  + \pwbase_\vek{k} (\vek{R}_{j}^{S2(\alpha)})
  \equiv \tvek{1 \\ 1}.
\end{equation}
We can then use Eq.\ (\ref{eq:pw-rep}) to determine the phases acquired under symmetry transformations. To obtain the plane wave IRs, it is again advantageous to consider the simplest coordinate system.  For the S atoms, this is coordinate system $(c)$ where the origin of the coordinate system is at the midpoint between the S atoms in the top and bottom layer of a unit cell. In this case, the transformation $g$ maps $\vek{R}_1^{\mathrm{S} \mu (c)}$ either onto itself or onto $\vek{R}_1^{\mathrm{S} \mu' (c)}$ with $\mu \neq \mu'$, so that Eq.\ (\ref{eq:pw-rep}) becomes
\begin{equation}
  \mathcal{D}_\vek{k}^{\mathrm{S} (c)} (g)_{\mu' \mu}
    = \exp [ i \vek{k} \cdot (\vek{R}_1^{\mathrm{S} \mu' (c) } - \vek{R}_1^{\mathrm{S} \mu' (c)})]
    = 1
  \end{equation}
for all $g \in \kgroup$. We can then determine the IRs of the plane waves for the coordinate systems $(a)$ and $(b)$ by using the respective RAR derived in Sec.~\ref{sec:band-amb-mos2}. The results are summarized in Table~\ref{tab:plane-wave-examples}.

\subsection{IRs of Bloch states in M\lowercase{o}S$_2$}
\label{sec:band-ir-mos2}

The full symmetry-adapted Bloch functions are written as products of symmetrized plane waves and symmetrized atomic orbitals. The five symmetry-adapted $d$ orbitals of the Mo atom times the plane wave $q_\vek{k} (\vek{R}_j^{\mathrm{Mo} (\alpha)})$ and the three symmetry-adapted $p$ orbitals of the S atoms times the plane waves $\pwbase_\vek{k} (\vek{R}_j^{\mathrm{S}1 (\alpha)}) \pm \pwbase_\vek{k} (\vek{R}_j^{\mathrm{S}2 (\alpha)})$ therefore comprises eleven symmetry-adapted basis functions for MoS$_2$ \cite{cap13}. The corresponding IRs are listed in Table~\ref{tab:band-ir-mos2} for the $\vek{\Gamma}$, $\vek{K}$, and $\vek{M}$ points. We list in Tables~\ref{tab:band-ir-mos2-gamma}, \ref{tab:band-ir-mos2-K}, and \ref{tab:band-ir-mos2-M} the sets of Bloch states transforming according to an IR of $\kgroup$ for the wave vectors $\vek{k} = \vek{\Gamma} , \vek{K}, \vek{M}$.  Using these symmetrized Bloch functions, the TB Hamiltonian for a wave vector $\vek{k}$ can be written in a block-diagonal form, where each block refers to the basis functions transforming according to an IR $\Gamma_I$ of $\kgroup$
\footnote{The symmetry-adapted bases derived here for MoS$_2$ define unitary transformations for block-diagonalizing a MoS$_2$ TB Hamiltonian that agree with the unitary transformations discussed in Ref.~\cite{cap13}.}.
To classify the additional degeneracy of the Bloch states due to time-reversal symmetry, we evaluate Eq.~(\ref{eq:herring}). All IRs of the space group for the stars $\{\vek{\Gamma}\}$, $\{\vek{K}\}$, and $\{\vek{M}\}$ belong to case~$(a)$.

\begin{table*}
 \caption{IRs of the plane waves ($\Gamma^q_\vek{k}$), the atomic orbitals $\phi_\nu$ ($\Gamma^\phi_\vek{k}$), and the full Bloch functions ($\Gamma^\Phi_\vek{k} = \Gamma^q_\vek{k} \times \Gamma^\phi_\vek{k}$) for the Mo and S atoms at the points $\vek{k} = \vek{\Gamma}$, $\vek{K}$, and $\vek{M}_i$. At $\vek{K}$, we distinguish between the three coordinate systems $\alpha = a, b ,c $. For the plane waves at $\vek{K}'$, the IR $\Gamma^q_{\vek{K}' (i/j/k)}$ is the complex conjugate of the IR $\Gamma^q_{\vek{K} (i/j/k)}$. At the points $\vek{M}_i$ ($i=1,2,3$), the atomic orbitals are denoted by $\phi_\nu^{[i]}$.}  \label{tab:band-ir-mos2}
  \renewcommand{\arraystretch}{1.2} \extrarowheight0.5ex
  $\begin{array}{L*{14}{|c}}
     \hline \hline
     & \multicolumn{4}{c|}{\vek{k} = \vek{\Gamma} \quad (D_{3h})}
     & \multicolumn{5}{c|}{\vek{k} = \vek{K}, \vek{K}' \quad (C_{3h})}
     & \multicolumn{5}{c}{\vek{k} = \vek{M}_1 , \vek{M}_2 , \vek{M}_3 \quad (C_{2v})} \\
     \cline{2-15}
     & \Gamma^q_{\vek{\Gamma}} & \phi_\nu & \Gamma^\phi_{\vek{\Gamma}} & \Gamma^\Phi_{\vek{\Gamma}} & \Gamma^q_{\vek{K} (a/b/c)} & \phi_\nu & \Gamma^\phi_{\vek{K}} & \Gamma^\Phi_{\vek{K} (a/b/c)} & \Gamma^\Phi_{\vek{K}' (a/b/c)} & \Gamma^q_{\vek{M}} & \phi_\nu^{[1(3)]} & \phi_\nu^{[2]} & \Gamma^\phi_{\vek{M}} & \Gamma^\Phi_{\vek{M}} \\
     \hline
     \multirow{5}{*}{Mo} & \multirow{5}{*}{$\Gamma_1$} & d_{z^2} & \Gamma_1 & \Gamma_1 & \multirow{5}{*}{$\Gamma_{2/1/3}$} & d_{z^2} & \Gamma_1 & \Gamma_{2/1/3} & \Gamma_{3/1/2} & \multirow{5}{*}{$\Gamma_1$} & d_{z^2} & d_{z^2} & \Gamma_1 & \Gamma_1\\
     \cline{3-5} \cline{7-10} \cline{12-15}
     & & \multirow{2}{*}{ $\{ d_{x^2 - y^2} , d_{xy} \}$ } & \multirow{2}{*}{$\Gamma_6$} & \multirow{2}{*}{$\Gamma_6$} & & d_{x^2 - y^2} + i d_{xy} & \Gamma_3 & \Gamma_{1/3/2} & \Gamma_{2/3/1} & & d_{x^2-y^2} \pm \sqrt{3} d_{xy} & d_{x^2-y^2} & \Gamma_1 & \Gamma_1\\
     \cline{7-10} \cline{12-15}
     & & & & & & d_{x^2 - y^2} - i d_{xy} & \Gamma_2 & \Gamma_{3/2/1} & \Gamma_{1/2/3} & & \sqrt{3} d_{x^2-y^2} \mp d_{xy} & d_{xy} & \Gamma_2 & \Gamma_2 \\
     \cline{3-5} \cline{7-10} \cline{12-15}
     & & \multirow{2}{*}{ $\{ d_{xz} , d_{yz} \}$ } & \multirow{2}{*}{$\Gamma_5$} & \multirow{2}{*}{$\Gamma_5$} & & d_{xz} + i d_{yz} & \Gamma_5 & \Gamma_{6/5/4} & \Gamma_{4/5/6} & & d_{xz} \mp \sqrt{3} d_{yz} & d_{xz} & \Gamma_3 & \Gamma_3\\
     \cline{7-10} \cline{12-15}
     & & & & & & d_{xz} - i d_{yz} & \Gamma_6 & \Gamma_{4/6/5} & \Gamma_{5/6/4} & & \sqrt{3} d_{xz} \pm d_{yz} & d_{yz} & \Gamma_4 & \Gamma_4\\
     \hline
     \multirow{6}{*}{S} & \multirow{6}{*}{\shortstack[l]{$\Gamma_1$, \\ $\Gamma_4$}} & \multirow{2}{*}{$p_z$} & \multirow{2}{*}{$\Gamma_4$} & \Gamma_4 & \multirow{6}{*}{\shortstack[l]{$\Gamma_{3/2/1}$,\\ $\Gamma_{6/5/4}$}} & \multirow{2}{*}{$p_z$} & \multirow{2}{*}{$\Gamma_4$} & \Gamma_{6/5/4} & \Gamma_{5/6/4} & \multirow{6}{*}{\shortstack[l]{$\Gamma_1$, \\ $\Gamma_4$}} & \multirow{2}{*}{$p_z$} & \multirow{2}{*}{$p_z$} & \multirow{2}{*}{$\Gamma_4$} & \Gamma_4 \\
     \cline{5-5} \cline{9-10} \cline{15-15}
     & & & & \Gamma_1 & & & & \Gamma_{3/2/1} & \Gamma_{2/3/1} & & & & & \Gamma_1 \\
     \cline{3-5} \cline{7-10} \cline{12-15}
     & & \multirow{4}{*}{ $\{ p_x , p_y \}$ } & \multirow{4}{*}{$\Gamma_6$} & \multirow{2}{*}{$\Gamma_6$} & & \multirow{2}{*}{ $p_x + ip_y$ } & \multirow{2}{*}{$\Gamma_2$} & \Gamma_{1/3/2} & \Gamma_{3/1/2} & & \multirow{2}{*}{$p_x \mp \sqrt{3} p_y$} & \multirow{2}{*}{ $p_x$ } & \multirow{2}{*}{$\Gamma_2$} & \Gamma_2 \\
     \cline{9-10} \cline{15-15}
     &&&&&&&& \Gamma_{4/6/5} & \Gamma_{6/4/5} & & & & & \Gamma_3\\
     \cline{5-5} \cline{7-10} \cline{12-15}
     &&&& \multirow{2}{*}{$\Gamma_5$} && \multirow{2}{*}{ $p_x - ip_y$ } & \multirow{2}{*}{$\Gamma_3$} & \Gamma_{2/1/3} & \Gamma_{1/2/3} & & \multirow{2}{*}{$\sqrt{3} p_x \pm p_y$} & \multirow{2}{*}{ $p_y$ } & \multirow{2}{*}{$\Gamma_1$} & \Gamma_1 \\
     \cline{9-10} \cline{15-15}
     &&&&&&&& \Gamma_{5/4/6} & \Gamma_{4/5/6} & & & & & \Gamma_4 \\
     \hline \hline
   \end{array}$
\end{table*}

\begin{table*}
 \caption{Symmetry-adapted TB Bloch functions in MoS$_2$ at $\vek{k} = \vek{\Gamma}$ with group of the wave vector $\kgroup[\vek{\Gamma}] = D_{3h}$. The Bloch functions are written as a product of the plane wave $\pwbase_\vek{\Gamma} (\vek{R}_j^{\textrm{Mo}})$ and $d$ orbitals for Mo atoms and the plane waves $\pwbase_\vek{\Gamma}^\pm (\vek{R}_j^\textrm{S}) = \pwbase_\vek{\Gamma} (\vek{R}_j^{\textrm{S}1}) \pm \pwbase_\vek{\Gamma} (\vek{R}_j^{\textrm{S}2})$ and $p$ orbitals for S atoms.  Also, $\pwbase_\vek{\Gamma} (\vek{R}_j^W) \, \{ d_\mu , d_\nu \}$ is a short-hand notation for the pair of Bloch functions $\{ \pwbase_\vek{\Gamma} (\vek{R}_j^W) \, d_\mu , \pwbase_\vek{\Gamma} (\vek{R}_j^W) \, d_\nu \}$. The last column indicates the degeneracy of Bloch states due to time-reversal symmetry discussed in Sec.~\ref{sec:time-reversal}.} \label{tab:band-ir-mos2-gamma}
  \renewcommand{\arraystretch}{1.2} \extrarowheight0.5ex
  $\begin{array}{cs{1.2em}cs{1.0em}cC}
     \hline \hline
     \text{IRs} & \text{Mo} & \text{S} & TR\\
     \hline
     \Gamma_1 &  \pwbase_\vek{\Gamma} (\vek{R}_j^{\textrm{Mo}}) \, d_{z^2}
     & \pwbase_\vek{\Gamma}^- (\vek{R}_j^\textrm{S}) \, p_z & a \\
     \Gamma_4 & & \pwbase_\vek{\Gamma}^+ (\vek{R}_j^\textrm{S}) \, p_z & a \\
     \Gamma_5 & \pwbase_\vek{\Gamma} (\vek{R}_j^{\textrm{Mo}})
     \, \{d_{xz}, d_{yz}\}
     & \pwbase_\vek{\Gamma}^- (\vek{R}_j^\textrm{S}) \, \{ p_x, p_y \} & a \\
     \Gamma_6 & \pwbase_\vek{\Gamma} (\vek{R}_j^{\textrm{Mo}})
     \, \{ d_{x^2 - y^2} , d_{xy} \}
     & \pwbase_\vek{\Gamma}^+ (\vek{R}_j^\textrm{S}) \, \{ p_x, p_y \} & a \\
     \hline \hline
   \end{array}$
\end{table*}

\begin{table*}
 \caption{Symmetry-adapted TB Bloch functions in MoS$_2$ at $\vek{k} = \vek{K}, \vek{K}'$ with group of the wave vector $\kgroup[\vek{K}] = \kgroup[\vek{K}'] = C_{3h}$. The Bloch functions are written as a product of the plane wave $\pwbase_\vek{k} (\vek{R}_j^{\textrm{Mo}})$ and $d$ orbitals for Mo atoms and the plane waves $\pwbase_\vek{k}^\pm (\vek{R}_j^\textrm{S}) = \pwbase_\vek{k} (\vek{R}_j^{\textrm{S}1}) \pm \pwbase_\vek{k} (\vek{R}_j^{\textrm{S}2})$ and $p$ orbitals for S atoms. The IRs $\Gamma_{i/j/k}$ correspond to the coordinate system $\alpha = a/b/c$ in Fig.~\ref{fig:singlemos2}.}  \label{tab:band-ir-mos2-K}
  \renewcommand{\arraystretch}{1.2} \extrarowheight0.5ex
  $\begin{array}{cs{1.2em}cs{2em}cs{0.8em}cs{2em}C}
     \hline \hline
     \vek{K} & \vek{K}' = -\vek{K} & \text{Mo} & \text{S} & TR \\ \hline
     \Gamma_{1/3/2} & \Gamma_{1/3/2}^\ast = \Gamma_{1/2/3} & \pwbase_\vek{k} (\vek{R}_j^{\textrm{Mo}}) \, (d_{x^2 - y^2} \pm i d_{xy}) & \pwbase_\vek{k}^+ (\vek{R}_j^\textrm{S}) \, (p_x \pm ip_y) & a\\
     \Gamma_{2/1/3} & \Gamma_{2/1/3}^\ast = \Gamma_{3/1/2} & \pwbase_\vek{k} (\vek{R}_j^{\textrm{Mo}}) \, d_{z^2} & \pwbase_\vek{k}^+ (\vek{R}_j^\textrm{S}) \, (p_x \mp ip_y) & a \\
     \Gamma_{3/2/1} & \Gamma_{3/2/1}^\ast = \Gamma_{2/3/1} & \pwbase_\vek{k} (\vek{R}_j^{\textrm{Mo}}) \, (d_{x^2 - y^2} \mp i d_{xy}) & \pwbase_\vek{k}^- (\vek{R}_j^\textrm{S}) \, p_z & a \\
     \Gamma_{4/6/5} & \Gamma_{4/6/5}^\ast = \Gamma_{4/5/6} & \pwbase_\vek{k} (\vek{R}_j^{\textrm{Mo}}) \, (d_{xz} \mp i d_{yz}) & \pwbase_\vek{k}^- (\vek{R}_j^\textrm{S}) \, (p_x \pm ip_y) & a \\
     \Gamma_{5/4/6} & \Gamma_{5/4/6}^\ast = \Gamma_{6/4/5} & &  \pwbase_\vek{k}^- (\vek{R}_j^\textrm{S}) \, (p_x \mp ip_y) & a \\
     \Gamma_{6/5/4} & \Gamma_{6/5/4}^\ast = \Gamma_{5/6/4} & \pwbase_\vek{k} (\vek{R}_j^{\textrm{Mo}}) \, (d_{xz} \pm i d_{yz}) & \pwbase_\vek{k}^+ (\vek{R}_j^\textrm{S}) \, p_z & a \\
     \hline \hline
   \end{array}$
\end{table*}

\begin{table*}
 \caption{Symmetry-adapted TB Bloch functions in MoS$_2$ at $\vek{k} = \vek{M}$ with group of the wave vector $\kgroup[\vek{M}] = C_{2v}$. The Bloch functions are written as a product of the plane wave $\pwbase_\vek{M} (\vek{R}_j^{\textrm{Mo}})$ and $d$ orbitals for Mo atoms and the plane waves $\pwbase_\vek{M}^\pm (\vek{R}_j^\textrm{S}) = \pwbase_\vek{M} (\vek{R}_j^{\textrm{S}1}) \pm \pwbase_\vek{M} (\vek{R}_j^{\textrm{S}2})$ and $p$ orbitals for S atoms.}  \label{tab:band-ir-mos2-M}
  \renewcommand{\arraystretch}{1.2} \extrarowheight0.5ex
  $\begin{array}{cs{2em}cs{0.8em}ccs{0.8em}cs{2em}C}
     \hline \hline
     & \multicolumn{2}{c}{\vek{M}_1 (\vek{M}_3)}
     & \multicolumn{2}{c}{\vek{M}_2} \\
     \text{IRs} & \text{Mo} & \text{S} & \text{Mo} & \text{S} & TR \\
     \hline
     \Gamma_1 &  \pwbase_\vek{M} (\vek{R}_j^{\textrm{Mo}}) \, d_{z^2} ;
     & \pwbase_\vek{M}^- (\vek{R}_j^\textrm{S}) \, p_z ;
     & \pwbase_\vek{M} (\vek{R}_j^{\textrm{Mo}}) \, d_{z^2} ;
     & \pwbase_\vek{M}^- (\vek{R}_j^\textrm{S}) \, p_z ; & a \\
     & \pwbase_\vek{M} (\vek{R}_j^{\textrm{Mo}}) \, (d_{x^2-y^2} \pm \sqrt{3} d_{xy})
     & \pwbase_\vek{M}^+ (\vek{R}_j^\textrm{S}) \, (\sqrt{3} p_x \pm p_y)
     & \pwbase_\vek{M} (\vek{R}_j^{\textrm{Mo}}) \, d_{x^2-y^2}
     & \pwbase_\vek{M}^+ (\vek{R}_j^\textrm{S}) \, p_y\\
     \Gamma_2
     & \pwbase_\vek{M} (\vek{R}_j^{\textrm{Mo}}) \, (\sqrt{3} d_{x^2-y^2} \mp d_{xy})
     & \pwbase_\vek{M}^+ (\vek{R}_j^\textrm{S}) \, (p_x \mp \sqrt{3} p_y)
     & \pwbase_\vek{M} (\vek{R}_j^{\textrm{Mo}}) \, d_{xy}
     & \pwbase_\vek{M}^+ (\vek{R}_j^\textrm{S}) \, p_x & a\\
     \Gamma_3 & \pwbase_\vek{M} (\vek{R}_j^{\textrm{Mo}}) \, (d_{xz} \mp \sqrt{3} d_{yz})
     & \pwbase_\vek{M}^- (\vek{R}_j^\textrm{S}) \, (p_x \mp \sqrt{3} p_y)
     & \pwbase_\vek{M} (\vek{R}_j^{\textrm{Mo}}) \, d_{xz}
     & \pwbase_\vek{M}^- (\vek{R}_j^\textrm{S}) \, p_x & a \\
     \Gamma_4 & \pwbase_\vek{M} (\vek{R}_j^{\textrm{Mo}}) \, (\sqrt{3} d_{xz} \pm d_{yz})
     & \pwbase_\vek{M}^+ (\vek{R}_j^\textrm{S}) \, p_z ;
     & \pwbase_\vek{M} (\vek{R}_j^{\textrm{Mo}}) \, d_{yz}
     & \pwbase_\vek{M}^+ (\vek{R}_j^\textrm{S}) \, p_z ; & a \\
     & & \pwbase_\vek{M}^- (\vek{R}_j^\textrm{S}) \, (\sqrt{3} p_x \pm p_y)
     & & \pwbase_\vek{M}^- (\vek{R}_j^\textrm{S}) \, p_y \\
     \hline \hline
   \end{array}$
\end{table*}

\cleardoublepage

\section{Band Symmetries in Few-Layer Graphene}
\label{sec:bloch-sym-graphene}

We can also apply the general formalism in Sec.~\ref{sec:bloch-sym} to identify the band symmetries in other quasi-2D materials such as SLG, BLG, and TLG.

\subsection{Crystal structure of few-layer graphene}

Like the crystal structure of monolayer MoS$_2$, the crystal structures of SLG, BLG, and TLG belong to the hexagonal crystal system. Therefore, we use the same expressions for the primitive lattice vectors [Eq.\ (\ref{eq:primitive-lattice-vectors})] and reciprocal lattice vectors [Eq.\ (\ref{eq:rec-primitive-lattice-vectors})]; and we have the same high-symmetry points in the BZ denoted $\vek{\Gamma}$, $\vek{K}$ [Eq.\ (\ref{eq:K-points})], and $\vek{M}$ [Eq.\ (\ref{eq:M-points})].  The space groups for few-layer graphene are listed in Table~\ref{tab:examples-site-sym}. This table also contains the site symmetries of the high-symmetry points for these crystal structures.

\subsubsection{Single-layer graphene}
\label{sec:crystal-slg}

\begin{figure}
   \includegraphics[width=0.95\columnwidth]{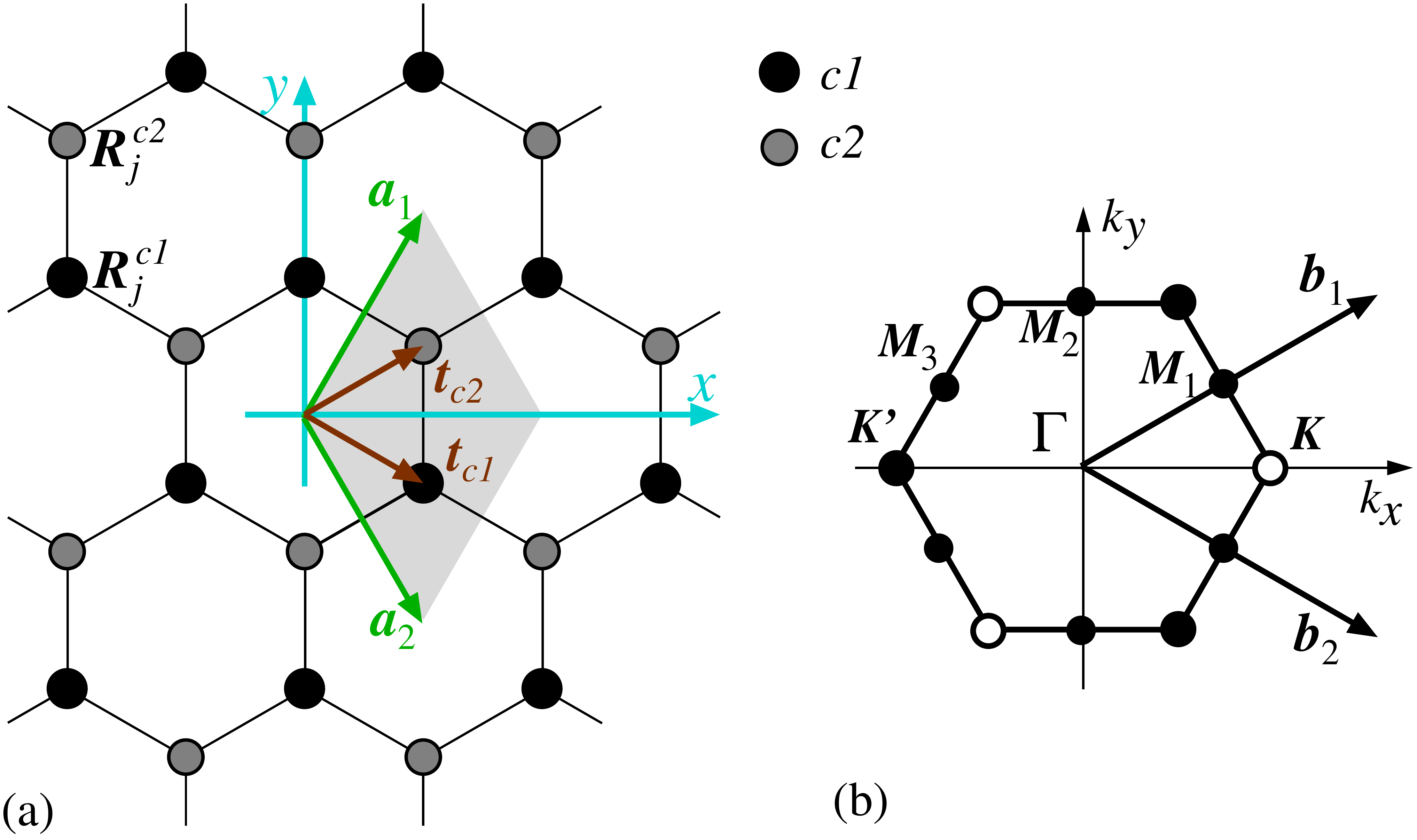}
  \caption{(a) Crystal structure of single-layer graphene characterized by the point group $D_{6h}$. The shaded region shows a unit cell ($j=1$). The C atoms are located at Wyckoff position $c$ with multiplicity $m=2$, hence the label $c1$ and $c2$ on one unit cell. The primitive lattice vectors are denoted $\vek{a}_1$ and $\vek{a}_2$. The positions of C atoms in unit cell $j$ are denoted by $\vek{R}_j^{c1}$ and $\vek{R}_j^{c2}$. The vectors $\vek{t}_{c1}$ and $\vek{t}_{c2}$ give the positions of C atoms in the unit cell. (b) The first Brillouin zone with primitive reciprocal lattice vectors $\vek{b}_1$ and $\vek{b}_2$.}
\label{fig:slg}
\end{figure}

Figure~\ref{fig:slg} shows the crystal structure of SLG. It is characterized by the point group $D_{6h}$ (space group $P6/mmm$, \# 191). The carbon atoms form two distinct Bravais lattices denoted as sublattices $A$ and $B$. The atomic positions denoted by $\{\vek{R}_j^{c1},\vek{R}_j^{c2}\}$ have site symmetries characterized by the point group $D_{3h}$ and Wyckoff letter $c$. The center of the hexagon, characterized by site symmetry $D_{6h}$ is the only Wyckoff position with multiplicity $m=1$. This point is the origin of the coordinate system for this crystal structure. The positions of the C atoms in the unit cell are given by
\begin{equation}
\vek{t}_{c_1} = \frac{a}{2}\tvek{1 \\ -\frac{1}{\sqrt{3}}},  \quad \vek{t}_{c_2} = \frac{a}{2}\tvek{1 \\ \frac{1}{\sqrt{3}}}.
\end{equation}

\begin{figure}
   \includegraphics[width=0.95\columnwidth]{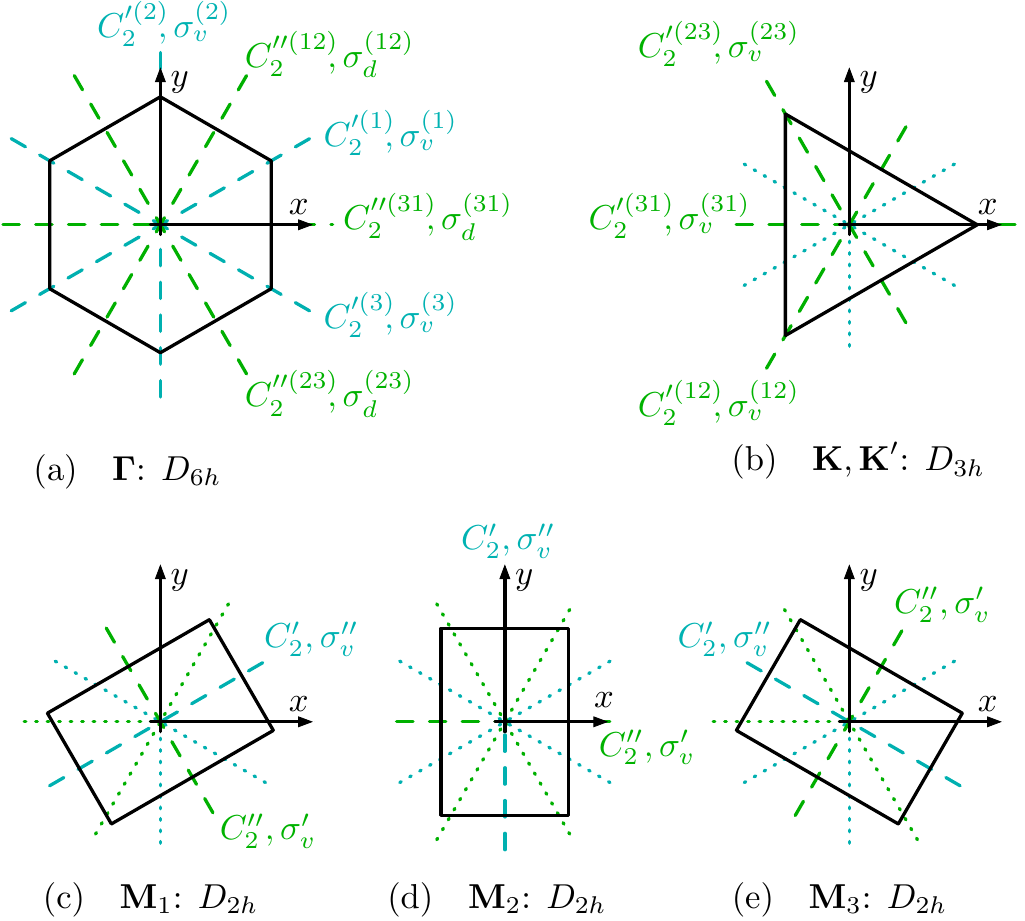}
   \caption{Groups of the wave vector in SLG. (a) The point $\vek{\Gamma}$ has the point group $D_{6h}$ with the $z$ axis (out of plane) as the axis for the $n$-fold rotations $C_n$ ($n = 2, 3, 6$) and the $xy$ plane as the reflection plane for $\sigma_h$. The dashed lines ($i$ and $ij$) are the axes for twofold rotations $C_2'^{(i)}$ and $C_2''^{(ij)}$ with $i,j = 1, 2, 3$. The reflection $\sigma_v^{(i)}$ [$\sigma_d^{(ij)}$] is about a plane that includes the corresponding dashed axis and the $z$ axis.  (b) The points $\vek{K}$ and $\vek{K}'$ have the point group $D_{3h}$ with threefold rotations about the $z$ axis. The dashed lines are the axes for the twofold rotations $C_2'^{(ij)}$. The reflection plane of $\sigma_v^{(ij)}$ contains the corresponding dashed lines and the $z$ axis. The reflection plane of $\sigma_h$ is the $xy$ plane.  The dotted lines indicate the twofold rotation axes that appear in the point group $D_{6h}$ but are not symmetry elements of $D_{3h}$.
[(c)--(e)] The points $\vek{M}_1$, $\vek{M}_2$, and $\vek{M}_1$ have the point group $D_{2h}$.  The rotation axis of $C_2$ is the $z$ axis and the reflection plane of $\sigma_v$ is the $xy$ plane.  The dashed lines are the axes of the twofold rotations $C_2'$ and $C_2''$. The reflection planes of $\sigma_v'$ and $\sigma_v''$ contain the corresponding dashed line and the $z$ axis.}
   \label{fig:slg-coord}
\end{figure}

The point group $D_{6h}$ of the crystal, which also characterizes the $\vek{\Gamma}$ point of the BZ,  contains twofold, threefold and sixfold rotations $C_2$, $C_3$ and $C_6$ where the $z$ axis is the rotation axis. The rotation axes of the three twofold rotation $C_2'^{(i)}$ and $C_2''^{(jk)}$ are the corresponding dashed lines shown in Fig.~\ref{fig:slg-coord}(a). The reflection planes for $\sigma_v^{(i)}$ and $\sigma_d^{(ij)}$ are perpendicular to the $xy$ plane passing through the corresponding dashed lines. The reflection $\sigma_h$ is along the $xy$ plane and $S_n = C_n \sigma_h$. At the $\vek{K}$ point, the group of the wave vector is $\kgroup[\vek{K}] = D_{3h}$ containing three twofold rotations $C_2^{\prime (ij)}$ about the corresponding dashed axes in Fig.~\ref{fig:slg-coord}(b). The reflection plane of $\sigma_v^{(ij)}$ is perpendicular to the $xy$ plane along the rotation axis of $C_2^{\prime (ij)}$. The threefold rotation axis is the $z$ axis, and $\sigma_h$ is a reflection about the $xy$ plane. The group of the wave vector at the $\vek{M}_i$ points is $\kgroup[\vek{M}] = D_{2h}$ containing the symmetry operations $C_2$, $C_2'$, $C_2''$, $\sigma_v$, $\sigma_v'$, and $\sigma_v''$ with rotation axes and reflection planes shown in Figs.\ \ref{fig:slg-coord}(c)--\ref{fig:slg-coord}(e). The character table for $D_{2h}$ is reproduced in Table~\ref{tab:char-d2h} (Appendix~\ref{sec:char-tables}).

\subsubsection{Bilayer graphene}
\label{sec:crystal-blg}

\begin{figure}
   \includegraphics[width=0.95\columnwidth]{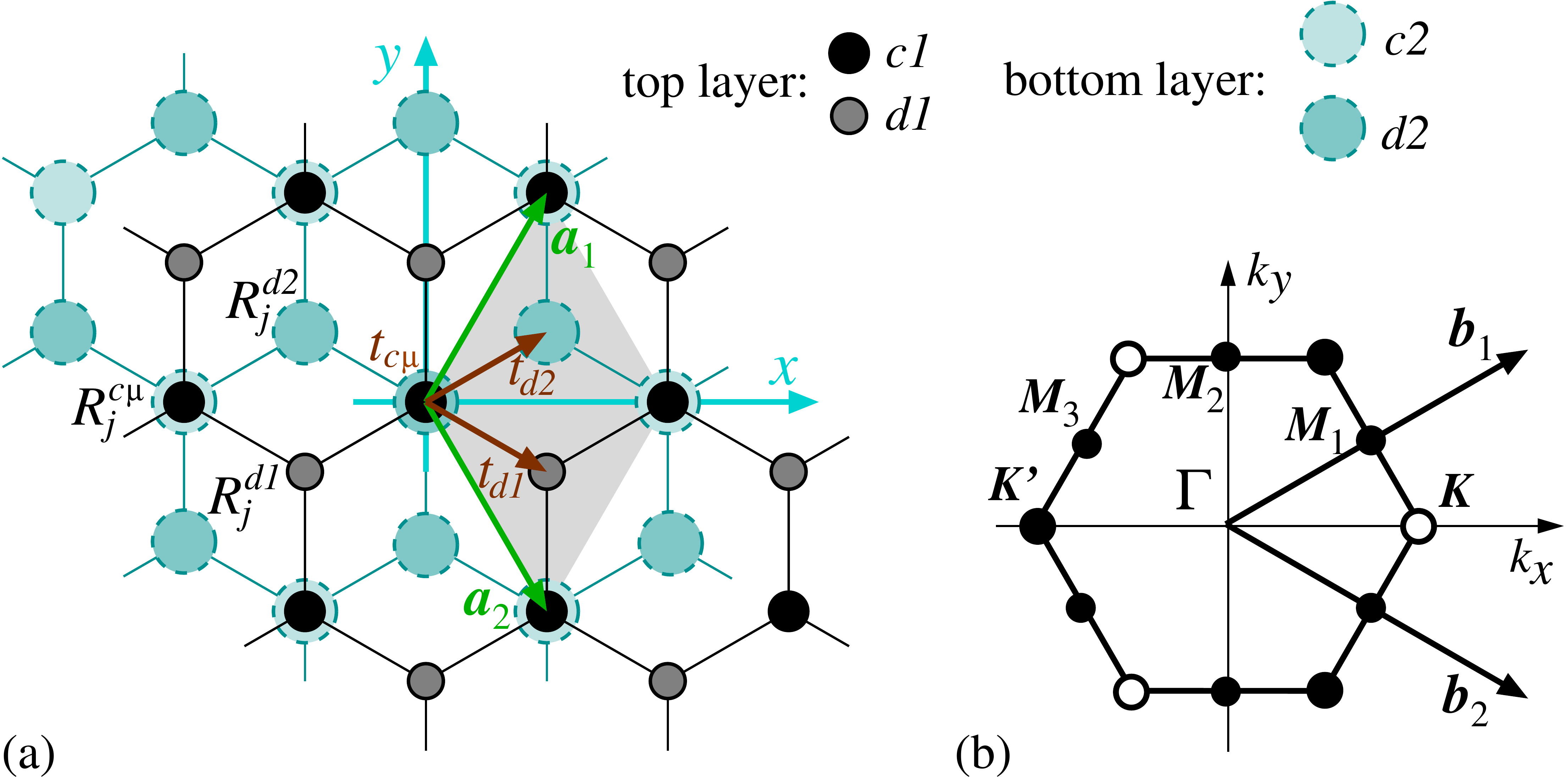}
  \caption{(a) Crystal structure of bilayer graphene characterized by the point group $D_{3d}$. The shaded region shows a unit cell ($j=1$). The C atoms are located on Wyckoff positions $c$ and $d$ each with multiplicity $m=2$. The atomic position in unit cell $j$ are denoted by $\vek{R}_j^{c \mu}$ and $\vek{R}_j^{d \mu}$ with $\mu=1$ ($\mu = 2$) corresponding to the atom at the top (bottom) layer. The vectors $\vek{t}_{c \mu}$ and $\vek{t}_{d \mu}$ give the positions of the C atoms within a unit cell. The primitive lattice vectors are denoted by $\vek{a}_1$ and $\vek{a}_2$.  (b) The first Brillouin zone with primitive reciprocal lattice vectors $\vek{b}_1$ and $\vek{b}_2$.}
\label{fig:blg}
\end{figure}

\begin{figure}
   \includegraphics[width=0.95\columnwidth]{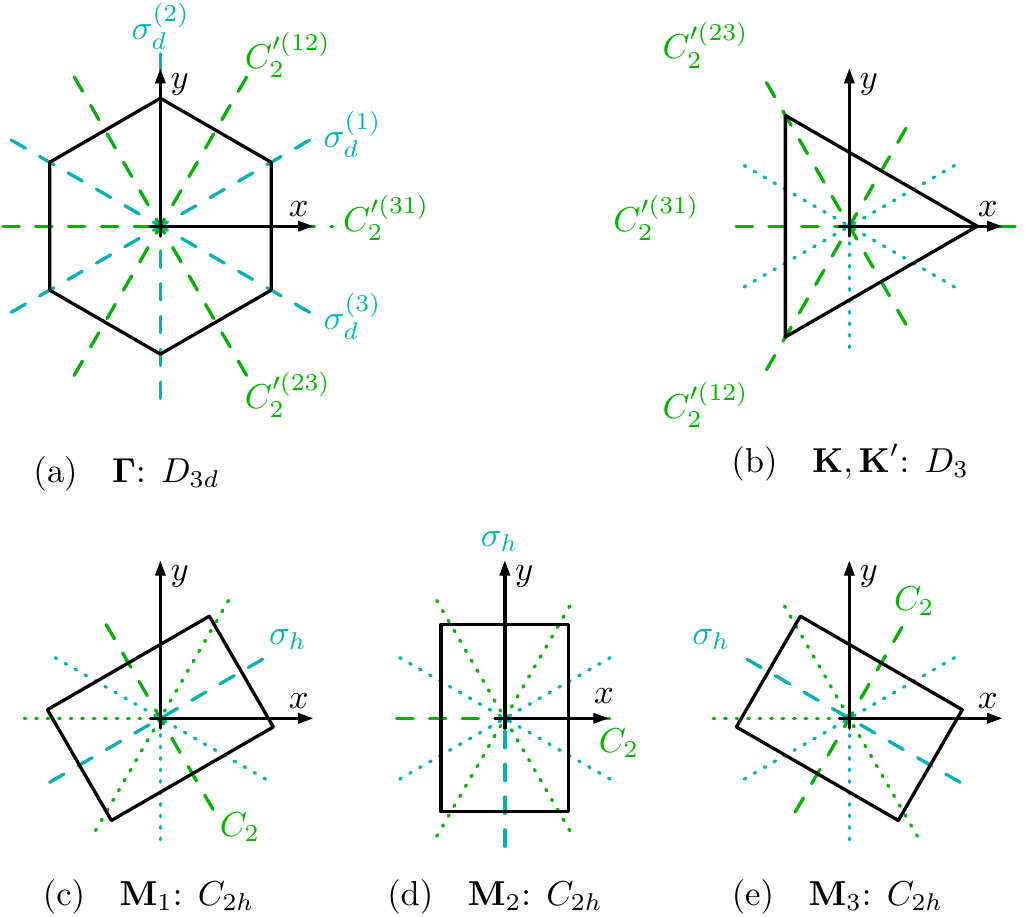}
  \caption{Groups of the wave vector in BLG. (a) The point $\vek{\Gamma}$ has the point group $D_{3d}$ with the $z$ axis (out of plane) as the axis for the threefold proper rotation $C_3$ and the sixfold improper rotation $S_6$. The green dashed lines $ij$ are the axes for twofold rotations $C_2''^{(ij)}$ with $i,j = 1, 2, 3$. The reflection $\sigma_d^{(i)}$ is about a plane that includes the corresponding blue dashed axis and the $z$ axis. (b) The points $\vek{K}$ and $\vek{K}'$ have the point group $D_3$ with threefold rotations about the $z$ axis. The dashed lines are the axes for the twofold rotations $C_2'^{(ij)}$.  The dotted lines indicate the reflection planes that appear in the point group $D_{3d}$ but are not symmetry elements of $D_3$.  [(c)--(e)] The points $\vek{M}_1$, $\vek{M}_2$, and $\vek{M}_1$ have the point group $C_{2h}$.
The green dashed line is the axis of the twofold rotation $C_2$.  The reflection plane of $\sigma_h$ contains the blue dashed line and the $z$ axis.}
\label{fig:blg-coord}
\end{figure}

The point group $D_{3d}$ (space group $P\bar{3}m1$, \# 164) characterizes BLG as shown in Fig.~\ref{fig:blg}. The only Wyckoff position with multiplicity $m=1$ is the midpoint of two C atoms on top of each other. We use this point as the origin of the coordinate system. The atomic positions in BLG are the Wyckoff positions $c$ and $d$, each with multiplicity $m=2$ and site symmetry $C_{3v}$. The two atoms in one Wyckoff letter are labeled $\mu=1$ ($\mu = 2$) for the atom in the top (bottom) layer. The layers are arranged in an $AB$ stacking, so that the atomic position $\vek{R}_j^{c1}$ is located on top of $\vek{R}_j^{c2}$. Ignoring the $z$ component, the positions of the C atoms in the unit cell are
\begin{equation}
\vek{t}_{c \mu} = \frac{a}{2}\tvek{ 0 \\ 0 }, \quad
\vek{t}_{d 1} = \frac{a}{2}\tvek{0 \\ \frac{2}{\sqrt{3}}},  \quad
\vek{t}_{d 2} = \frac{a}{2}\tvek{1 \\ \frac{1}{\sqrt{3}}} .
\end{equation}
The threefold proper and sixfold improper rotation axis is the $z$ axis. The axis of the twofold rotation $C_2'^{(ij)}$ is the corresponding dashed line in Fig.~\ref{fig:blg-coord}(a). The three reflection planes corresponding to $\sigma_d^{(i)}$ are perpendicular to the $xy$ plane passing through the corresponding dashed line. The group of the wave vector is $\kgroup[\vek{K}] = D_3$ at the $\vek{K}$ point with threefold rotations about the $z$ axis and three twofold rotations $C_2'^{(jk)}$ about the corresponding dashed line in Fig.~\ref{fig:blg-coord}(b). At the $\vek{M}_i$ points, the group of the wave vector is $\kgroup[\vek{M}] = C_{2h}$, where $\sigma_h$ is perpendicular to the $xy$ plane passing through the corresponding dashed line in Figs.\ \ref{fig:blg-coord}(c)--\ref{fig:blg-coord}(e), and the twofold rotation $C_2$ is about the corresponding dashed line. The character table for $C_{2h}$ is reproduced in Table~\ref{tab:char-c2h} (Appendix~\ref{sec:char-tables}).

\subsubsection{Trilayer graphene}
\label{sec:crystal-tlg}

\begin{figure}
   \includegraphics[width=0.95\columnwidth]{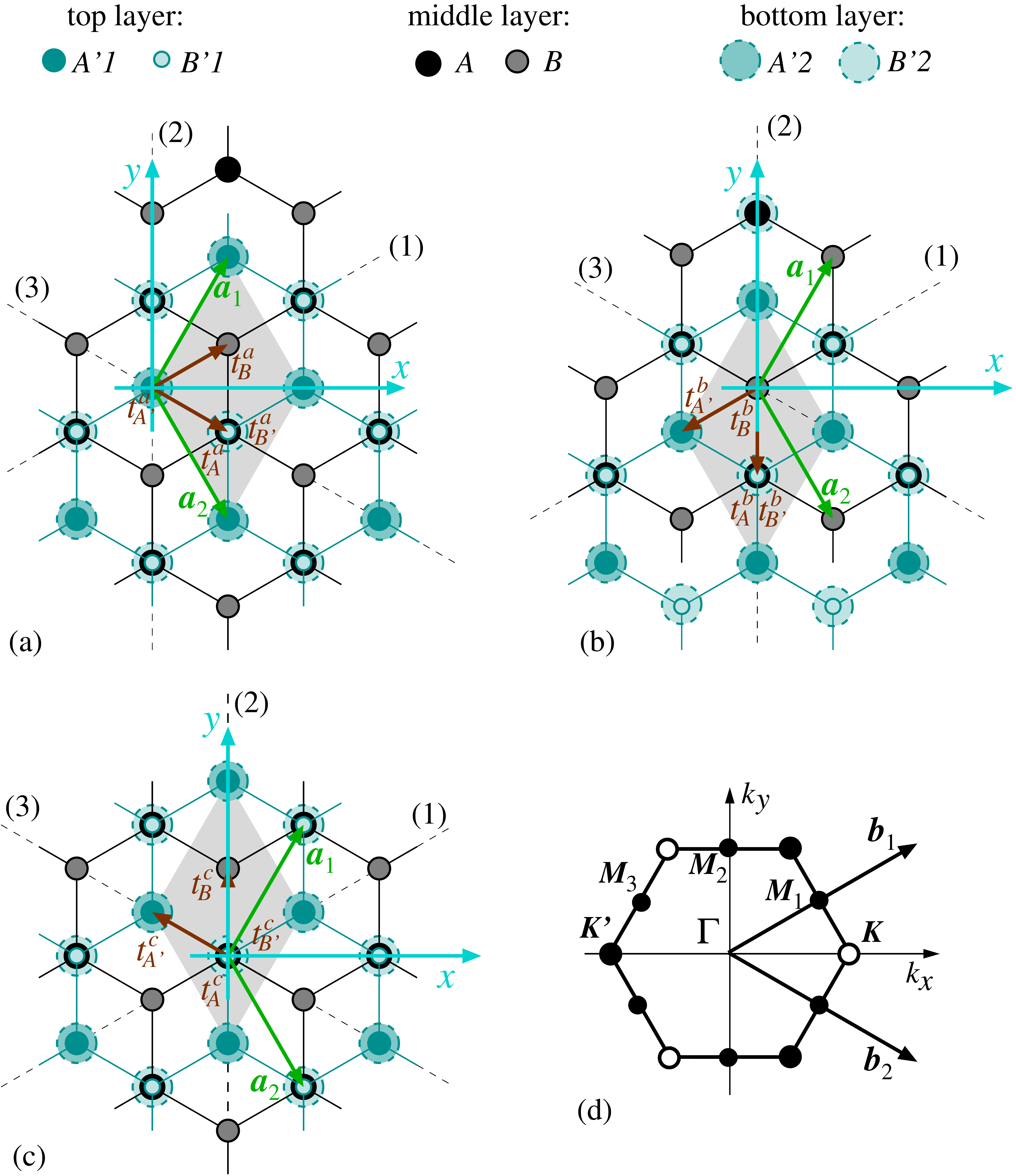}
   \caption{Crystal structure of trilayer graphene. Three coordinate systems $\alpha = a, b, c$ are considered with (a) the origin located at the atom $A$, (b) origin at atom $B$, and (c) origin at the midpoint between $B'1$ and $B'2$. The dashed axes $(1)$, $(2)$, and $(3)$ are the twofold rotation axes of the point group $D_{3h}$. The shaded region shows a unit cell ($j=1$). The vectors $\vek{t}_{A}^\alpha$, $\vek{t}_{B}^\alpha$, $\vek{t}_{A'}^\alpha$, and $\vek{t}_{B'}^\alpha$ give the positions of the C atoms labeled $A$, $B$, $A'\mu$, and $B'\mu$ within a unit cell, respectively. For the $A'$ and $B'$ atom, the top (bottom) atoms are labeled $\mu=1$ ($\mu = 2$). The positions of these atoms in unit cell $j$ are denoted by $\vek{R}_j^{A(\alpha)}$, $\vek{R}_j^{B (\alpha)}$, $\vek{R}_j^{A' \mu (\alpha)}$, and $\vek{R}_j^{B' \mu (\alpha)}$, respectively. (d) The first Brillouin zone. }
\label{fig:tlg}
\end{figure}

Last, Fig.~\ref{fig:tlg} shows the crystal structure of TLG. This system has the same space group $P \bar{6} m 2$, \# 187 as monolayer MoS$_2$ (point group $D_{3h}$). We designate the Wyckoff positions of the carbon atoms in the middle layer as $A$ and $B$ with site symmetry group $D_{3h}$, and the remaining positions as $A'\mu$ and $B'\mu$ with $\mu=1$ ($\mu=2$) for the top (bottom) layer with site symmetry group $C_{3v}$. The points $A$ and $B$ as well as the center of the hexagon in the middle layer are Wyckoff positions with multiplicity $m=1$, which we use as the origin of the three coordinate systems defined in Fig.~\ref{fig:tlg}. We therefore associate with each Wyckoff letter $W$ an index $(\alpha)$ corresponding to the coordinate systems $\alpha = a, b, c$. Ignoring the $z$ component, the positions of the C atoms in the unit cell for the three coordinate systems are
\begin{subequations}
  \begin{align}
    \vek{t}_A^a & = \frac{a}{2} \tvek{ 1 \\ -\frac{1}{\sqrt{3}} } , &
    \vek{t}_{A' \mu}^a & = \frac{a}{2} \tvek{ 0 \\ 0 }, \\
    \vek{t}_B^a & = \frac{a}{2} \tvek{ 1 \\ \frac{1}{\sqrt{3}} }, &
    \vek{t}_{B' \mu}^a & = \frac{a}{2}\tvek{ 1 \\ -\frac{1}{\sqrt{3}} } ,\\
    \vek{t}_A^b & = \frac{a}{2} \tvek{ 0 \\ -\frac{2}{\sqrt{3}} }, &
    \vek{t}_{A' \mu}^b & = \frac{a}{2} \tvek{ -1 \\ -\frac{1}{\sqrt{3}} }, \\
    \vek{t}_B^b & = \frac{a}{2} \tvek{ 0 \\ 0 }, &
    \vek{t}_{B' \mu}^b & = \frac{a}{2} \tvek{ 0 \\ -\frac{2}{\sqrt{3}} }, \\
    \vek{t}_A^c & = \frac{a}{2} \tvek{ 0 \\ 0 }, &
    \vek{t}_{A' \mu}^c & = \frac{a}{2} \tvek{ -1 \\ \frac{1}{\sqrt{3}} }, \\
    \vek{t}_B^c & = \frac{a}{2} \tvek{ 0 \\ \frac{2}{\sqrt{3}} }, &
    \vek{t}_{B' \mu}^c & = \frac{a}{2} \tvek{ 0 \\ 0 },
  \end{align}
\end{subequations}
where the superscripts $\alpha = a, b, c$ denote the coordinate system. The positions of the six atoms in unit cell $j$ are denoted by $\vek{R}_j^A$, $\vek{R}_j^B$, $\vek{R}_j^{A'\mu}$ and $\vek{R}_j^{B'\mu}$. In standard notation \cite{hahn05}, the positions $A$, $B$ , $A'\mu$ and $B'\mu$ are characterized by the Wyckoff letters $c$, $e$, $g$, and $h$ respectively in coordinate system $(a)$, by $e$, $a$, $h$, and $i$ in coordinate system $(b)$, and by $a$, $c$, $i$, and $g$ in coordinate system $(c)$, see Table~\ref{tab:examples-site-sym}.

The point group $D_{3h}$ of the crystal structure of TLG is the same as for MoS$_2$ with coordinate system shown in Fig.~\ref{fig:mos2-tlg-coord}(a). The group of the wave vector at $\vek{K}$ is $\kgroup[\vek{K}] = C_{3h}$ with coordinate system shown in Fig.~\ref{fig:mos2-tlg-coord}(b). At the three points $\vek{M}_i$ the group of the wave vector is $\kgroup[\vek{M}] = C_{2v}$ with coordinate system shown in Figs.\ \ref{fig:mos2-tlg-coord}(c)--\ref{fig:mos2-tlg-coord}(e), which contains the twofold rotation $C_2$ about the dashed line, reflection $\sigma_v'$ about the dashed line and the $z$ axis, and reflection $\sigma_v$ about the $xy$ plane.

\subsection{IRs of Bloch states in graphene}

In graphene, the bands near the Fermi level are dominated by the $p$ orbitals of the C atoms. For SLG, using the coordinate systems in Fig.~\ref{fig:slg-coord}, the IRs of the symmetry-adapted $p$ orbitals are listed in Table~\ref{tab:p-orbitals}  for the points $\vek{\Gamma}$, $\vek{K}$, and $\vek{M}$ with groups of the wave vector $D_{6h}$, $D_{3h}$, and $D_{2h}$, respectively. For BLG, Fig.~\ref{fig:blg-coord} shows the coordinate systems used for the points $\vek{\Gamma}$, $\vek{K}$, and $\vek{M}$ with group of the wave vector $D_{3d}$, $D_3$, and $C_{2h}$, respectively. The IRs of the $p$ orbitals at these points are listed in Table~\ref{tab:p-orbitals}. The coordinate systems used for TLG are the same as the ones for MoS$_2$ (Fig.~\ref{fig:mos2-tlg-coord}), so that the IRs of the $p$ orbitals can be taken from Table~\ref{tab:mos2-orbitals}.  The symmetry-adapted plane waves with the corresponding IRs are summarized in Table~\ref{tab:plane-wave-examples}.

\begin{table*}
\caption{Symmetry-adapted $p$ orbitals for the point groups $D_{6h}$, $D_{3h}$, and $D_{2h}$ for the coordinate systems shown in Fig.~\ref{fig:slg-coord} (SLG) and $D_{3d}$, $D_3$, and $C_{2h}$ for the coordinate system shown in Fig.~\ref{fig:blg-coord} (BLG). For $D_{2h}$ and $C_{2h}$, the symmetrized atomic orbital $\phi_\nu^{[i]}$ corresponds to the coordinate systems used for the point $\vek{M}_i$ in Figs.\ \ref{fig:slg-coord}(c)--\ref{fig:slg-coord}(e) and \ref{fig:blg-coord}(c)--\ref{fig:blg-coord}(e), respectively.  The orbital $[i(j)]$ takes the upper (lower) sign.} \label{tab:p-orbitals}
  \renewcommand{\arraystretch}{1.2} \extrarowheight0.5ex
  $\begin{array}{*{2}{cs{1.2em}cs{2.0em}}*{2}{cs{1.2em}}cs{2.0em}
                 *{2}{cs{1.2em}cs{2.0em}}*{2}{cs{1.2em}}c}
     \hline \hline
     \multicolumn{2}{c}{D_{6h}} & \multicolumn{2}{c}{D_{3h}} & \multicolumn{3}{c}{D_{2h}} & \multicolumn{2}{c}{D_{3d}} & \multicolumn{2}{c}{D_3} & \multicolumn{3}{c}{C_{2h}} \\
     \phi_\nu & \text{IR} & \phi_\nu & \text{IR} & \phi_\nu^{[1(3)]} & \phi_\nu^{[2]} & \text{IR} & \phi_\nu & \text{IR} & \phi_\nu & \text{IR} & \phi_\nu^{[1(3)]} & \phi_\nu^{[2]} & \text{IR} \\
     \hline
     p_z & \Gamma_2^- & p_z & \Gamma_4 & p_z & p_z & \Gamma_3^- & p_z & \Gamma_2^- & p_z & \Gamma_2 & p_z & p_z & \Gamma_2^- \\
     \{ p_x , p_y \} & \Gamma_5^- & \{ p_x , p_y \} & \Gamma_6 &
     p_x \mp \sqrt{3} p_y & p_x & \Gamma_4^- &
     \{ p_x , p_y \} & \Gamma_3^- & \{ p_x , p_y \} & \Gamma_3 &
     p_x \mp \sqrt{3} p_y & p_x & \Gamma_1^- \\
     &&&& \sqrt{3} p_x \pm p_y & p_y & \Gamma_2^- &
     &&&& \sqrt{3} p_x \pm p_y & p_y & \Gamma_2^- \\
     \hline \hline
   \end{array}$
\end{table*}

The IRs of the full Bloch functions for SLG, BLG, and TLG are listed in Table~\ref{tab:band-ir-graphene}. The sets of Bloch states transforming as an IR in $\kgroup$ for the wave vectors $\vek{k} = \vek{\Gamma} , \vek{K}, \vek{M}$ are listed in Tables~\ref{tab:band-ir-slg-gamma}, \ref{tab:band-ir-slg-K}, and \ref{tab:band-ir-slg-M} for SLG, Tables~\ref{tab:band-ir-blg-gamma}, \ref{tab:band-ir-blg-K}, and \ref{tab:band-ir-blg-M} for BLG, and Tables~\ref{tab:band-ir-tlg-gamma}, \ref{tab:band-ir-tlg-K}, and \ref{tab:band-ir-tlg-M} for TLG.

\begin{table*}
 \caption{IRs of the plane waves ($\Gamma^q_\vek{k}$), the atomic orbitals $\phi_\nu$ ($\Gamma^\phi_\vek{k}$), and the full Bloch functions ($\Gamma^\Phi_\vek{k} = \Gamma^q_\vek{k} \times \Gamma^\phi_\vek{k}$) for SLG, BLG, and TLG at the points $\vek{k} = \vek{\Gamma}$, $\vek{K}$, and $\vek{M}_i$. The positions of C atoms in SLG are characterized by the Wyckoff letter $c$ with multiplicity $m=2$. For BLG, the atomic positions have Wyckoff letters $c$ and $d$ each with multiplicity $m=2$. For TLG, the Wyckoff letters of the atomic positions are denoted $A$ and $B$ with multiplicity $m=1$, and $A'$ and $B'$ with multiplicity $m=2$. For TLG at $\vek{K}$, we need to distinguish between the three coordinate systems $\alpha = a, b ,c $.  The IRs of $\kgroup[\vek{K}]$ and $\kgroup[\vek{K}']$ in SLG and BLG are real, so that $\Gamma^q_\vek{K} = \Gamma^q_{\vek{K}'}$ and $\Gamma^\phi_\vek{K} = \Gamma^\phi_{\vek{K}'}$, whereas in TLG the IR $\Gamma^q_{\vek{K}' (i/j/k)}$ is the complex conjugate of $\Gamma^q_{\vek{K} (i/j/k)}$. At the points $\vek{M}_i$ ($i=1,2,3$), the atomic orbitals are denoted by $\phi_\nu^{[i]}$.} \label{tab:band-ir-graphene}
  \renewcommand{\arraystretch}{1.2} \extrarowheight0.3ex
  $\begin{array}{L*{13}{|c}}
     \hline \hline
     & \multicolumn{4}{c|}{\vek{k} = \vek{\Gamma}} & \multicolumn{4}{c|}{\vek{k} = \vek{K}} & \multicolumn{5}{c}{\vek{k} = \vek{M}_1 , \vek{M}_2 , \vek{M}_3} \\
     \cline{2-14}
     & \Gamma^q_{\vek{\Gamma}} & \phi_\nu & \Gamma^\phi_{\vek{\Gamma}} & \Gamma^\Phi_{\vek{\Gamma}} & \Gamma^q_{\vek{K}} & \phi_\nu & \Gamma^\phi_{\vek{K}} & \Gamma^\Phi_{\vek{K}} & \Gamma^q_{\vek{M}} & \phi_\nu^{[1(3)]} & \phi_\nu^{[2]} & \Gamma^\phi_{\vek{M}} & \Gamma^\Phi_{\vek{M}} \\
     \hline \hline
     SLG &
     \multicolumn{4}{c|}{D_{6h}} &
     \multicolumn{4}{c|}{D_{3h}} &
     \multicolumn{5}{c}{D_{2h}} \\
     \hline
     \multirow{6}{*}{$c$} & \multirow{6}{*}{ \shortstack[l]{$\Gamma_1^+$, \\ $\Gamma_3^-$ }} & \multirow{2}{*}{$p_z$} & \multirow{2}{*}{$\Gamma_2^-$} & \Gamma_2^- & \multirow{6}{*}{$\Gamma_6$} & \multirow{2}{*}{$p_z$} & \multirow{2}{*}{$\Gamma_4$} & \multirow{2}{*}{$\Gamma_5$} & \multirow{6}{*}{\shortstack[l]{$\Gamma_1^+$, \\ $\Gamma_2^-$}} & \multirow{2}{*}{$p_z$} & \multirow{2}{*}{$p_z$} & \multirow{2}{*}{$\Gamma_3^-$} & \Gamma_3^- \\
     \cline{5-5} \cline{14-14}
     & & & & \Gamma_3^+ & & & & & & & & & \Gamma_4^+ \\
     \cline{3-5} \cline{7-9} \cline{11-14}
     & & \multirow{4}{*}{ $\{ p_x , p_y \}$ } & \multirow{4}{*}{$\Gamma_5^-$} & \multirow{2}{*}{$\Gamma_5^-$} & & \multirow{4}{*}{ $\{ p_x , p_y \}$ } & \multirow{4}{*}{$\Gamma_6$} & \Gamma_1 & & \multirow{2}{*}{$p_x \mp \sqrt{3} p_y$} & \multirow{2}{*}{$p_x$} & \multirow{2}{*}{$\Gamma_4^-$} & \Gamma_4^- \\
     \cline{9-9} \cline{14-14}
     &&&&&&&& \Gamma_2 && &&& \Gamma_3^+ \\
     \cline{5-5} \cline{9-9} \cline{11-14}
     &&&& \multirow{2}{*}{$\Gamma_6^+$} &&& & \multirow{2}{*}{$\Gamma_6$} & & \multirow{2}{*}{$\sqrt{3} p_x \pm p_y$} & \multirow{2}{*}{$p_y$} & \multirow{2}{*}{$\Gamma_2^-$} & \Gamma_2^- \\
     \cline{14-14}
     &&&&&&& && & &&& \Gamma_1^+ \\
     \hline \hline
     BLG &
     \multicolumn{4}{c|}{D_{3d}} &
     \multicolumn{4}{c|}{D_3} &
     \multicolumn{5}{c}{C_{2h}} \\
     \hline
     \multirow{6}{*}{$c$} & \multirow{6}{*}{ \shortstack[l]{$\Gamma_1^+$, \\ $\Gamma_2^-$ }} & \multirow{2}{*}{$p_z$} & \multirow{2}{*}{$\Gamma_2^-$} & \Gamma_2^- & \multirow{6}{*}{\shortstack[l]{$\Gamma_1$, \\ $\Gamma_2$}} & \multirow{2}{*}{$p_z$} & \multirow{2}{*}{$\Gamma_2$} & \Gamma_2 & \multirow{6}{*}{\shortstack[l]{$\Gamma_1^+$, \\ $\Gamma_2^-$}} & \multirow{2}{*}{$p_z$} & \multirow{2}{*}{$p_z$} & \multirow{2}{*}{$\Gamma_2^-$} & \Gamma_2^- \\
     \cline{5-5} \cline{14-14}
     & && & \Gamma_1^+ & && & \Gamma_1 &&& & & \Gamma_1^+ \\
     \cline{3-5} \cline{7-9} \cline{11-14}
     & & \multirow{4}{*}{ $\{ p_x , p_y \}$ } & \multirow{4}{*}{$\Gamma_3^-$} & \multirow{2}{*}{$\Gamma_3^-$} & & \multirow{4}{*}{ $\{ p_x , p_y \}$ } & \multirow{4}{*}{$\Gamma_3$} & \multirow{2}{*}{$\Gamma_3$} & & \multirow{2}{*}{$p_x \mp \sqrt{3} p_y$} & \multirow{2}{*}{$p_x$} & \multirow{2}{*}{$\Gamma_1^-$} & \Gamma_1^- \\
     \cline{14-14}
     &&&&&&& & &&&&& \Gamma_2^+ \\
     \cline{5-5} \cline{9-9} \cline{11-14}
     &&&& \multirow{2}{*}{$\Gamma_3^+$} &&&& \multirow{2}{*}{$\Gamma_3$} & & \multirow{2}{*}{$\sqrt{3} p_x \pm p_y$} & \multirow{2}{*}{$p_y$} & \multirow{2}{*}{$\Gamma_2^-$} & \Gamma_2^- \\
     \cline{14-14}
     &&&&&&&&&& & & & \Gamma_1^+ \\
     \hline
     \multirow{6}{*}{$d$} & \multirow{6}{*}{ \shortstack[l]{$\Gamma_1^+$, \\ $\Gamma_2^-$ }} & \multirow{2}{*}{$p_z$} & \multirow{2}{*}{$\Gamma_2^-$} & \Gamma_2^- & \multirow{6}{*}{$\Gamma_3$} & \multirow{2}{*}{$p_z$} & \multirow{2}{*}{$\Gamma_2$} & \multirow{2}{*}{$\Gamma_3$} & \multirow{6}{*}{\shortstack[l]{$\Gamma_1^+$, \\ $\Gamma_2^-$}} & \multirow{2}{*}{$p_z$} & \multirow{2}{*}{$p_z$} & \multirow{2}{*}{$\Gamma_2^-$} & \Gamma_2^- \\
     \cline{5-5} \cline{14-14}
     && & & \Gamma_1^+ & &&&& & & & & \Gamma_1^+ \\
     \cline{3-5} \cline{7-9} \cline{11-14}
     & & \multirow{4}{*}{ $\{ p_x , p_y \}$ } & \multirow{4}{*}{$\Gamma_3^-$} & \multirow{2}{*}{$\Gamma_3^-$} & & \multirow{4}{*}{ $\{ p_x , p_y \}$ } & \multirow{4}{*}{$\Gamma_3$} & \Gamma_1 & & \multirow{2}{*}{$p_x \mp \sqrt{3} p_y$} & \multirow{2}{*}{$p_x$} & \multirow{2}{*}{$\Gamma_1^-$} & \Gamma_1^- \\
     \cline{9-9} \cline{14-14}
     &&&&&&&& \Gamma_2 & &&& & \Gamma_2^+ \\
     \cline{5-5} \cline{9-9} \cline{11-14}
     &&&& \multirow{2}{*}{$\Gamma_3^+$} &&&& \multirow{2}{*}{$\Gamma_3$} & & \multirow{2}{*}{$\sqrt{3} p_x \pm p_y$} & \multirow{2}{*}{$p_y$} & \multirow{2}{*}{$\Gamma_2^-$} & \Gamma_2^- \\
     \cline{14-14}
     &&&&&& & &&&&& & \Gamma_1^+ \\
     \hline \hline
     \multirow{2}{*}{TLG}
     & \multicolumn{4}{c|}{D_{3h}}
     & \multicolumn{4}{c|}{C_{3h}}
     & \multicolumn{5}{c}{C_{2v}} \\
     \cline{2-14}
     & \Gamma^q_{\vek{\Gamma}} & \phi_\nu & \Gamma^\phi_{\vek{\Gamma}} & \Gamma^\Phi_{\vek{\Gamma}} & \Gamma^q_{\vek{K} (a/b/c)} & \phi_\nu & \Gamma^\phi_{\vek{K}} & \Gamma^\Phi_{\vek{K} (a/b/c)} \quad \Gamma^\Phi_{\vek{K}' (a/b/c)} & \Gamma^q_{\vek{M}} & \phi_\nu^{[1(3)]} & \phi_\nu^{[2]} & \Gamma^\phi_{\vek{M}} & \Gamma^\Phi_{\vek{M}} \\
     \hline
     \multirow{3}{*}{$A$} & \multirow{3}{*}{$\Gamma_1$} & p_z & \Gamma_4 & \Gamma_4 & \multirow{3}{*}{$\Gamma_{3/2/1}$} & p_z & \Gamma_4 & \Gamma_{6/5/4} \quad \quad \Gamma_{5/6/4} & \multirow{3}{*}{$\Gamma_1$} & p_z & p_z & \Gamma_4 & \Gamma_4\\
     \cline{3-5} \cline{7-9} \cline{11-14}
     & & \multirow{2}{*}{ $\{ p_x , p_y \}$ } & \multirow{2}{*}{$\Gamma_6$} & \multirow{2}{*}{$\Gamma_6$} & & p_x + ip_y & \Gamma_2 & \Gamma_{1/3/2} \quad \quad \Gamma_{3/1/2} & & p_x \mp \sqrt{3} p_y & p_x & \Gamma_2 & \Gamma_2\\
     \cline{7-9} \cline{11-14}
     & &&& & & p_x - ip_y & \Gamma_3 & \Gamma_{2/1/3}\quad \quad \Gamma_{1/2/3} & & \sqrt{3} p_x \pm p_y & p_y& \Gamma_1 & \Gamma_1 \\
     \hline
     \multirow{3}{*}{$B$} & \multirow{3}{*}{$\Gamma_1$} & p_z & \Gamma_4 & \Gamma_4 & \multirow{3}{*}{$\Gamma_{2/1/3}$} & p_z & \Gamma_4 & \Gamma_{5/4/6} \quad \quad \Gamma_{6/4/5} & \multirow{3}{*}{$\Gamma_1$} & p_z & p_z & \Gamma_4 & \Gamma_4\\
     \cline{3-5} \cline{7-9} \cline{11-14}
     & & \multirow{2}{*}{ $\{ p_x , p_y \}$ } & \multirow{2}{*}{$\Gamma_6$} & \multirow{2}{*}{$\Gamma_6$} & & p_x + ip_y & \Gamma_2 & \Gamma_{3/2/1} \quad \quad \Gamma_{1/2/3} & & p_x \mp \sqrt{3} p_y & p_x & \Gamma_2 & \Gamma_2\\
     \cline{7-9} \cline{11-14}
     & & & &&& p_x - ip_y & \Gamma_3 & \Gamma_{1/3/2} \quad \quad \Gamma_{2/3/1} & & \sqrt{3} p_x \pm p_y & p_y& \Gamma_1 & \Gamma_1 \\
     \hline
     \multirow{6}{*}{$A'$} & \multirow{6}{*}{\shortstack[l]{ $\Gamma_1$, \\ $\Gamma_4$}} & \multirow{2}{*}{$p_z$} & \multirow{2}{*}{$\Gamma_4$} & \Gamma_4 & \multirow{6}{*}{\shortstack[l]{$\Gamma_{1/3/2}$, \\ $\Gamma_{4/6/5}$}} & \multirow{2}{*}{$p_z$} & \multirow{2}{*}{$\Gamma_4$} & \Gamma_{4/6/5} \quad \quad \Gamma_{4/5/6} & \multirow{6}{*}{\shortstack[l]{$\Gamma_1$, \\ $\Gamma_4$}} & \multirow{2}{*}{$p_z$} & \multirow{2}{*}{$p_z$} & \multirow{2}{*}{$\Gamma_4$} & \Gamma_4 \\
     \cline{5-5} \cline{9-9} \cline{14-14}
     & & && \Gamma_1 & & && \Gamma_{1/3/2} \quad \quad \Gamma_{1/2/3} &&& & & \Gamma_1 \\
     \cline{3-5} \cline{7-9} \cline{11-14}
     & & \multirow{4}{*}{ $\{ p_x , p_y \}$ } & \multirow{4}{*}{$\Gamma_6$} & \multirow{2}{*}{$\Gamma_6$} & & \multirow{2}{*}{ $ p_x + ip_y $ } & \multirow{2}{*}{$\Gamma_2$} & \Gamma_{2/1/3} \quad \quad \Gamma_{2/3/1} & & \multirow{2}{*}{$p_x \mp \sqrt{3} p_y$} & \multirow{2}{*}{$p_x$} & \multirow{2}{*}{$\Gamma_2$} & \Gamma_2 \\
     \cline{9-9} \cline{14-14}
     &&&&&&&& \Gamma_{5/4/6} \quad \quad \Gamma_{5/6/4} & &&& & \Gamma_3\\
     \cline{5-5} \cline{7-9} \cline{11-14}
     &&&& \multirow{2}{*}{$\Gamma_5$} && \multirow{2}{*}{ $ p_x - ip_y $ } & \multirow{2}{*}{$\Gamma_3$} & \Gamma_{3/2/1} \quad \quad \Gamma_{3/1/2} & & \multirow{2}{*}{$\sqrt{3} p_x \pm p_y$} & \multirow{2}{*}{$p_y$} & \multirow{2}{*}{$\Gamma_1$} & \Gamma_1 \\
     \cline{9-9} \cline{14-14}
     &&&&&&&& \Gamma_{6/5/4} \quad \quad \Gamma_{6/4/5} &&& & & \Gamma_4 \\
     \hline
     \multirow{6}{*}{$B'$} & \multirow{6}{*}{\shortstack[l]{$\Gamma_1$, \\ $\Gamma_4$}} & \multirow{2}{*}{$p_z$} & \multirow{2}{*}{$\Gamma_4$} & \Gamma_4 & \multirow{6}{*}{\shortstack[l]{$\Gamma_{3/2/1}$,\\ $\Gamma_{6/5/4}$}} & \multirow{2}{*}{$p_z$} & \multirow{2}{*}{$\Gamma_4$} & \Gamma_{6/5/4} \quad \quad \Gamma_{5/6/4} & \multirow{6}{*}{\shortstack[l]{$\Gamma_1$, \\ $\Gamma_4$}} & \multirow{2}{*}{$p_z$} & \multirow{2}{*}{$p_z$} & \multirow{2}{*}{$\Gamma_4$} & \Gamma_4 \\
     \cline{5-5} \cline{9-9} \cline{14-14}
     & && & \Gamma_1 & && & \Gamma_{3/2/1} \quad \quad \Gamma_{2/3/1} & &&& & \Gamma_1 \\
     \cline{3-5} \cline{7-9} \cline{11-14}
     & & \multirow{4}{*}{ $\{ p_x , p_y \}$ } & \multirow{4}{*}{$\Gamma_6$} & \multirow{2}{*}{$\Gamma_6$} & & \multirow{2}{*}{ $ p_x + ip_y $ } & \multirow{2}{*}{$\Gamma_2$} & \Gamma_{1/3/2} \quad \quad \Gamma_{3/1/2} & & \multirow{2}{*}{$p_x \mp \sqrt{3} p_y$} & \multirow{2}{*}{$p_x$} & \multirow{2}{*}{$\Gamma_2$} & \Gamma_2 \\
     \cline{9-9} \cline{14-14}
     &&&&&&&& \Gamma_{4/6/5} \quad \quad \Gamma_{6/4/5} & &&& & \Gamma_3\\
     \cline{5-5} \cline{7-9} \cline{11-14}
     &&&& \multirow{2}{*}{$\Gamma_5$} && \multirow{2}{*}{ $ p_x - ip_y $ } & \multirow{2}{*}{$\Gamma_3$} & \Gamma_{2/1/3} \quad \quad \Gamma_{1/2/3} & & \multirow{2}{*}{$\sqrt{3} p_x \pm p_y$} & \multirow{2}{*}{$p_y$} & \multirow{2}{*}{$\Gamma_1$} & \Gamma_1 \\
     \cline{9-9} \cline{14-14}
     &&&&&&&& \Gamma_{5/4/6} \quad \quad \Gamma_{4/5/6} &&& & & \Gamma_4 \\
     \hline \hline
   \end{array}$
\end{table*}

\begin{table}
 \caption{Symmetry-adapted TB Bloch functions in SLG at $\vek{k} = \vek{\Gamma}$ with group of the wave vector $\kgroup[\vek{\Gamma}] = D_{6h}$. The C atoms are located at Wyckoff position $c$ of multiplicity 2. The Bloch functions are written as a product of the plane waves $\pwbase_\vek{\Gamma}^\pm (\vek{R}_j^c) = \pwbase_\vek{\Gamma} (\vek{R}_j^{c1}) \pm \pwbase_\vek{\Gamma} (\vek{R}_j^{c2})$ and the $p$ orbitals of the C atoms. Also, $\pwbase_\vek{\Gamma} (\vek{R}_j^W) \, \{ p_x , p_y \}$ is a short-hand notation for the pair of Bloch functions $\{ \pwbase_\vek{\Gamma} (\vek{R}_j^W) \, p_x , \pwbase_\vek{\Gamma} (\vek{R}_j^W) \, p_y \}$.  The last column indicates the degeneracy of Bloch states due to time-reversal symmetry discussed in Sec.~\ref{sec:time-reversal}.}  \label{tab:band-ir-slg-gamma}
  \renewcommand{\arraystretch}{1.2} \extrarowheight0.5ex
  $\begin{array}{cS{2em}cC}
     \hline \hline
     \text{IRs} & \text{Bloch function} & TR\\
     \hline
     \Gamma_2^- & \pwbase_\vek{\Gamma}^- (\vek{R}_j^c) \, p_z & a \\
     \Gamma_3^+ & \pwbase_\vek{\Gamma}^+ (\vek{R}_j^c) \, p_z & a \\
     \Gamma_5^- & \pwbase_\vek{\Gamma}^+ (\vek{R}_j^c) \, \{ p_x, p_y \} & a \\
     \Gamma_6^+ & \pwbase_\vek{\Gamma}^- (\vek{R}_j^c) \, \{ p_x, p_y \} & a \\
     \hline \hline
   \end{array}$
\end{table}

\begin{table}
 \caption{Symmetry-adapted TB Bloch functions in SLG at $\vek{k} = \vek{K}$ with group of the wave vector $\kgroup[\vek{K}] = D_{3h}$. The same form of the symmetry-adapted TB Bloch functions and corresponding IRs work for $\vek{k} = \vek{K}'$, that is we replace $\vek{K}$ with $\vek{K}'$.}  \label{tab:band-ir-slg-K}
  \renewcommand{\arraystretch}{1.2} \extrarowheight0.5ex
  $\begin{array}{cS{2em}cC}
     \hline \hline
     \text{IRs} & \text{Bloch function} & TR \\
     \hline
     \Gamma_1 & \pwbase_\vek{K} (\vek{R}_j^{c1}) \, p_y + \pwbase_\vek{K} (\vek{R}_j^{c2}) \, p_x & a \\
     \Gamma_2 & \pwbase_\vek{K} (\vek{R}_j^{c1}) \, p_y - \pwbase_\vek{K} (\vek{R}_j^{c2}) \, p_x & a \\
     \Gamma_5 & \{\pwbase_\vek{K} (\vek{R}_j^{c1}) \, p_z, \pwbase_\vek{K} (\vek{R}_j^{c2}) \, p_z\} & a \\
     \Gamma_6 & \{\pwbase_\vek{K} (\vek{R}_j^{c1}) \, p_x, \pwbase_\vek{K} (\vek{R}_j^{c2}) \, p_y\} & a \\
     \hline \hline
   \end{array}$
\end{table}

\begin{table}
 \caption{Symmetry-adapted TB Bloch functions in SLG at $\vek{k} = \vek{M}$ with group of the wave vector $\kgroup[\vek{M}] = D_{2h}$.}  \label{tab:band-ir-slg-M}
  \renewcommand{\arraystretch}{1.2} \extrarowheight0.5ex
  $\begin{array}{cS{2em}ccC}
     \hline \hline
     \text{IRs} & \vek{M}_1 \, (\vek{M}_3) & \vek{M}_2 & TR \\
     \hline
     \Gamma_1^+ & \pwbase_\vek{M}^+ (\vek{R}_j^c) (\sqrt{3} p_x \pm p_y) & \pwbase_\vek{M}^- (\vek{R}_j^c) \, p_y & a \\
     \Gamma_2^- & \pwbase_\vek{M}^- (\vek{R}_j^c) (\sqrt{3} p_x \pm p_y) & \pwbase_\vek{M}^+ (\vek{R}_j^c) \, p_y & a \\
     \Gamma_3^+ & \pwbase_\vek{M}^+ (\vek{R}_j^c) (p_x \mp \sqrt{3} p_y) & \pwbase_\vek{M}^- (\vek{R}_j^c) \, p_x & a \\
     \Gamma_3^- & \pwbase_\vek{M}^- (\vek{R}_j^c) \, p_z & \pwbase_\vek{M}^+ (\vek{R}_j^c) \, p_z & a \\
     \Gamma_4^+ & \pwbase_\vek{M}^+ (\vek{R}_j^c) \, p_z & \pwbase_\vek{M}^- (\vek{R}_j^c) \, p_z & a \\
     \Gamma_4^- & \pwbase_\vek{M}^- (\vek{R}_j^c) (p_x \mp \sqrt{3} p_y) & \pwbase_\vek{M}^+ (\vek{R}_j^c) \, p_x & a \\
     \hline \hline
   \end{array}$
\end{table}

\begin{table}
  \caption{Symmetry-adapted TB Bloch functions in BLG at $\vek{k} = \vek{\Gamma}$ with group of the wave vector $\kgroup[\vek{\Gamma}] = D_{3d}$. The C atoms are located at Wyckoff positions $W = c, d$ each of multiplicity 2.  $\pwbase_\vek{\Gamma}^\pm (\vek{R}_j^W) \, \{ p_x , p_y \}$ is a short-hand notation for the pair of Bloch functions $\{ \pwbase_\vek{\Gamma}^\pm (\vek{R}_j^W) \, p_x , \pwbase_\vek{\Gamma}^\pm (\vek{R}_j^W) \, p_y \}$.}  \label{tab:band-ir-blg-gamma}
  \renewcommand{\arraystretch}{1.2} \extrarowheight0.5ex
  $\begin{array}{cS{2em}cC}
     \hline \hline
     \text{IRs} & W = c, d & TR\\
     \hline
     \Gamma_1^+ & \pwbase_\vek{\Gamma}^- (\vek{R}_j^W) \, p_z & a \\
     \Gamma_2^- & \pwbase_\vek{\Gamma}^+ (\vek{R}_j^W) \, p_z & a \\
     \Gamma_3^+ & \pwbase_\vek{\Gamma}^- (\vek{R}_j^W) \, \{p_x, p_y\} & a \\
     \Gamma_3^- & \pwbase_\vek{\Gamma}^+ (\vek{R}_j^W) \, \{p_x, p_y\} & a \\
     \hline \hline
   \end{array}$
\end{table}

\begin{table}
  \caption{Symmetry-adapted TB Bloch functions in BLG at $\vek{k} = \vek{K}$ with group of the wave vector $\kgroup[\vek{K}] = D_3$. $\pwbase_\vek{K}^\pm (\vek{R}_j^W) \, \{ p_x , p_y \}$ is a short-hand notation for the pair of Bloch functions $\{ \pwbase_\vek{K}^\pm (\vek{R}_j^W) \, p_x , \pwbase_\vek{K}^\pm (\vek{R}_j^W) \, p_y \}$. The same form of the symmetry-adapted TB Bloch functions and corresponding IRs work for $\vek{k} = \vek{K}'$, that is we replace $\vek{K}$ with $\vek{K}'$.}  \label{tab:band-ir-blg-K}
  \renewcommand{\arraystretch}{1.2} \extrarowheight0.5ex
  $\begin{array}{cs{1.2em}ccs{1.2em}C}
     \hline \hline
     \text{IRs} & c & d & TR \\
     \hline
     \Gamma_1 & \pwbase_\vek{K}^- (\vek{R}_j^c) \, p_z & \pwbase_\vek{K} (\vek{R}_j^{d1}) \, p_y + \pwbase_\vek{K} (\vek{R}_j^{d2}) \, p_x & a \\
     \Gamma_2 & \pwbase_\vek{K}^+ (\vek{R}_j^c) \, p_z & \pwbase_\vek{K} (\vek{R}_j^{d1}) \, p_y - \pwbase_\vek{K} (\vek{R}_j^{d2}) \, p_x & a \\
     \Gamma_3 & \pwbase_\vek{K}^- (\vek{R}_j^c) \, \{p_x, p_y\} ; & \{ \pwbase_\vek{K} (\vek{R}_j^{d1}) \, p_z , \pwbase_\vek{K} (\vek{R}_j^{d2}) \, p_z \} ; & a \\
     & \pwbase_\vek{K}^+ (\vek{R}_j^c) \, \{p_x, p_y\} & \{ \pwbase_\vek{K} (\vek{R}_j^{d1}) \, p_x, \pwbase_\vek{K} (\vek{R}_j^{d2}) \, p_y \} \\
     \hline \hline
   \end{array}$
\end{table}

\begin{table*}
 \caption{Symmetry-adapted TB Bloch functions in BLG at $\vek{k} = \vek{M}$ with group of the wave vector $\kgroup[\vek{M}] = C_{2h}$.}  \label{tab:band-ir-blg-M}
  \renewcommand{\arraystretch}{1.2} \extrarowheight0.5ex
  $\begin{array}{cs{2em}ccs{1.5em}ccs{2em}C}
     \hline \hline
     & \multicolumn{2}{c}{\vek{k} = \vek{M}_1 \, (\vek{M}_3)} & \multicolumn{2}{c}{\vek{k} = \vek{M}_2} \\
     \text{IRs} & c & d & c & d & TR \\
     \hline
     \Gamma_1^+
     & \pwbase_\vek{M}^- (\vek{R}_j^c) \, p_z ;
     & \pwbase_\vek{M}^+ (\vek{R}_j^d) \, p_z ;
     & \pwbase_\vek{M}^- (\vek{R}_j^c) \, p_z ;
     & \pwbase_\vek{M}^- (\vek{R}_j^d) \, p_z ; & a \\
     & \pwbase_\vek{M}^- (\vek{R}_j^c) \, (\sqrt{3} p_x \pm p_y)
     & \pwbase_\vek{M}^+ (\vek{R}_j^d) \, (\sqrt{3} p_x \pm p_y)
     & \pwbase_\vek{M}^- (\vek{R}_j^c) \, p_y
     & \pwbase_\vek{M}^- (\vek{R}_j^d) \, p_y \\
     \Gamma_1^-
     & \pwbase_\vek{M}^+ (\vek{R}_j^c) \, (p_x \mp \sqrt{3} p_y)
     & \pwbase_\vek{M}^- (\vek{R}_j^d) \, (p_x \mp \sqrt{3} p_y)
     & \pwbase_\vek{M}^+ (\vek{R}_j^c) \, p_x
     & \pwbase_\vek{M}^+ (\vek{R}_j^d) \, p_x & a \\
     \Gamma_2^+
     & \pwbase_\vek{M}^- (\vek{R}_j^c) \, (p_x \mp \sqrt{3} p_y)
     & \pwbase_\vek{M}^+ (\vek{R}_j^d) \, (p_x \mp \sqrt{3} p_y)
     & \pwbase_\vek{M}^- (\vek{R}_j^c) \, p_x
     & \pwbase_\vek{M}^- (\vek{R}_j^d) \, p_x & a \\
     \Gamma_2^-
     & \pwbase_\vek{M}^+ (\vek{R}_j^c) \, p_z ;
     & \pwbase_\vek{M}^- (\vek{R}_j^d) \, p_z ;
     & \pwbase_\vek{M}^+ (\vek{R}_j^c) \, p_z ;
     & \pwbase_\vek{M}^+ (\vek{R}_j^d) \, p_z ; & a \\
     & \pwbase_\vek{M}^+ (\vek{R}_j^c) \, (\sqrt{3} p_x \pm p_y)
     & \pwbase_\vek{M}^- (\vek{R}_j^d) \, (\sqrt{3} p_x \pm p_y)
     & \pwbase_\vek{M}^+ (\vek{R}_j^c) \, p_y
     & \pwbase_\vek{M}^+ (\vek{R}_j^d) \, p_y \\
     \hline \hline
   \end{array}$
\end{table*}

\begin{table*}
 \caption{Symmetry-adapted TB Bloch functions in TLG at $\vek{k} = \vek{\Gamma}$ with group of the wave vector $\kgroup[\vek{\Gamma}] = D_{3h}$. The C atoms are located at Wyckoff positions $W = A, B$ $(W = A', B')$ of multiplicity 1 (2). $\pwbase_\vek{\Gamma} (\vek{R}_j^W) \, \{ p_x , p_y \}$ is a short-hand notation for the pair of Bloch functions $\{ \pwbase_\vek{\Gamma} (\vek{R}_j^W) \, p_x , \pwbase_\vek{\Gamma} (\vek{R}_j^W) \, p_y \}$.}  \label{tab:band-ir-tlg-gamma}
  \renewcommand{\arraystretch}{1.2} \extrarowheight0.5ex
  $\begin{array}{cs{2em}cs{1.2em}cs{2em}C}
     \hline \hline
     \text{IRs} & W = A, B & W = A', B' & TR \\
     \hline
     \Gamma_1 & & \pwbase_\vek{\Gamma}^- (\vek{R}_j^W) \, p_z & a \\
     \Gamma_4 & \pwbase_\vek{\Gamma} (\vek{R}_j^W) \, p_z & \pwbase_\vek{\Gamma}^+ (\vek{R}_j^W) \, p_z & a \\
     \Gamma_5 & & \pwbase_\vek{\Gamma}^- (\vek{R}_j^W) \, \{ p_x, p_y \} & a \\
     \Gamma_6 & \pwbase_\vek{\Gamma} (\vek{R}_j^W) \, \{ p_x, p_y \} & \pwbase_\vek{\Gamma}^+ (\vek{R}_j^W) \, \{ p_x, p_y \} & a \\
     \hline \hline
   \end{array}$
\end{table*}

\begin{table*}
 \caption{Symmetry-adapted TB Bloch functions in TLG at $\vek{k} = \vek{K}, \vek{K}'$ with group of the wave vector $\kgroup[\vek{K}] = \kgroup[\vek{K}'] = C_{3h}$. The IRs $\Gamma_{i/j/k}$ correspond to the coordinate system $\alpha = a/b/c$ in Fig.~\ref{fig:tlg}. The IRs at $\vek{K}'$ are the complex conjugates of the IRs at $\vek{K}$.}  \label{tab:band-ir-tlg-K}
  \renewcommand{\arraystretch}{1.2} \extrarowheight0.5ex
  $\begin{array}{cs{1em}c*{2}{s{1.6em}cs{1.2em}c}s{2em}C}
     \hline \hline
     \vek{K} & \vek{K}' = -\vek{K} & A & B & A' & B' & TR\\
     \hline
     \Gamma_{1/3/2} & \Gamma_{1/3/2}^\ast = \Gamma_{1/2/3} & \pwbase_\vek{K} (\vek{R}_j^A) \, (p_x \pm ip_y) & \pwbase_\vek{K} (\vek{R}_j^B) \, (p_x \mp ip_y) & \pwbase_\vek{K}^- (\vek{R}_j^{A'}) \, p_z & \pwbase_\vek{K}^+ (\vek{R}_j^{B'}) \, (p_x \pm ip_y) & a \\
     \Gamma_{2/1/3} & \Gamma_{2/1/3}^\ast = \Gamma_{3/1/2} & \pwbase_\vek{K} (\vek{R}_j^A) \, (p_x \mp ip_y) & & \pwbase_\vek{K}^+ (\vek{R}_j^{A'}) \, (p_x \pm ip_y) & \pwbase_\vek{K}^+ (\vek{R}_j^{B'}) \, (p_x \mp ip_y) & a \\
     \Gamma_{3/2/1} & \Gamma_{3/2/1}^\ast = \Gamma_{2/3/1} & & \pwbase_\vek{K} (\vek{R}_j^B) \, (p_x \pm ip_y) & \pwbase_\vek{K}^+ (\vek{R}_j^{A'}) \, (p_x \mp ip_y) & \pwbase_\vek{K}^- (\vek{R}_j^{A'}) \, p_z & a \\
     \Gamma_{4/6/5} & \Gamma_{4/6/5}^\ast = \Gamma_{4/5/6} & & & \pwbase_\vek{K}^+ (\vek{R}_j^{A'}) \, p_z & \pwbase_\vek{K}^- (\vek{R}_j^{B'}) \, (p_x \pm ip_y) & a \\
     \Gamma_{5/4/6} & \Gamma_{5/4/6}^\ast = \Gamma_{6/4/5} & & \pwbase_\vek{K} (\vek{R}_j^B) \, p_z & \pwbase_\vek{K}^- (\vek{R}_j^{A'}) \, (p_x \pm ip_y) & \pwbase_\vek{K}^- (\vek{R}_j^{B'}) \, (p_x \mp ip_y) & a \\
     \Gamma_{6/5/4} & \Gamma_{6/5/4}^\ast = \Gamma_{5/6/4} & \pwbase_\vek{K} (\vek{R}_j^A) \, p_z & & \pwbase_\vek{K}^- (\vek{R}_j^{A'}) \, (p_x \mp ip_y) & \pwbase_\vek{K}^+ (\vek{R}_j^{B'}) \, p_z & a \\
     \hline \hline
   \end{array}$
\end{table*}

\begin{table*}
 \caption{Symmetry-adapted TB Bloch functions in TLG at $\vek{k} = \vek{M}$ with group of the wave vector $\kgroup[\vek{M}] = C_{2v}$.}  \label{tab:band-ir-tlg-M}
  \renewcommand{\arraystretch}{1.2} \extrarowheight0.5ex
  $\begin{array}{c*{2}{s{1.6em}cs{1.2em}c}s{2em}C}
     \hline \hline
     & \multicolumn{2}{c}{\vek{k} = \vek{M}_1 \, (\vek{M}_3)} & \multicolumn{2}{c}{\vek{k} = \vek{M}_2} \\
     \text{IRs} & W = A, B & W = A', B' & W = A, B & W = A', B' & TR \\
     \hline
     \Gamma_1
     & \pwbase_\vek{M} (\vek{R}_j^W) \, (\sqrt{3} p_x \pm p_y)
     & \pwbase_\vek{M}^+ (\vek{R}_j^W) \, p_z ;
     & \pwbase_\vek{M} (\vek{R}_j^W) \, p_y
     & \pwbase_\vek{M}^- (\vek{R}_j^W) \, p_z ; & a \\
     & & \pwbase_\vek{M}^+ (\vek{R}_j^W) \, (\sqrt{3} p_x \pm p_y)
     & & \pwbase_\vek{M}^+ (\vek{R}_j^W) \, p_y \\
     \Gamma_2
     & \pwbase_\vek{M} (\vek{R}_j^W) \, (p_x \mp \sqrt{3} p_y)
     & \pwbase_\vek{M}^+ (\vek{R}_j^W) \, (p_x \mp \sqrt{3} p_y)
     & \pwbase_\vek{M} (\vek{R}_j^W) \, p_x
     & \pwbase_\vek{M}^+ (\vek{R}_j^W) \, p_x & a \\
     \Gamma_3
     & & \pwbase_\vek{M}^- (\vek{R}_j^W) \, (p_x \mp \sqrt{3} p_y)
     & & \pwbase_\vek{M}^- (\vek{R}_j^W) \, p_x & a \\
     \Gamma_4
     & \pwbase_\vek{M} (\vek{R}_j^W) \, p_z
     & \pwbase_\vek{M}^- (\vek{R}_j^W) \, p_z ;
     & \pwbase_\vek{M} (\vek{R}_j^W) \, p_z
     & \pwbase_\vek{M}^+ (\vek{R}_j^W) \, p_z ; & a \\
     & & \pwbase_\vek{M}^- (\vek{R}_j^W) \, (\sqrt{3} p_x \pm p_y)
     & & \pwbase_\vek{M}^- (\vek{R}_j^W) \, p_y \\
     \hline \hline
   \end{array}$
\end{table*}

\cleardoublepage
\section{Selection Rules: Effect of Band IR Rearrangement}
\label{sec:sel-rules}

We show in the following that the selection rules for the observable matrix elements of a Hermitian operator $\mathcal{O}$ taken between Bloch states are not affected by the rearrangement of band IRs discussed in Sec.~\ref{sec:band-amb} provided the perturbation of a crystal represented by the operator $\mathcal{O}$ preserves translational invariance.  This condition for the operator $\mathcal{O}$ is certainly obeyed by the dipole operator $\mathcal{O} = \vek{r}$ representing optical transitions, but it does not apply to localized perturbations such as point defects for which anyway the specific location of the defect plays a crucial role \cite{bir74}.  We consider a translation $\tshift$ of the coordinate system as introduced in Sec.~\ref{sec:band-amb}, so that  $\Tshift = \seitz{\openone}{\tshift}$ is a unitary operator with $T^\dagger = T^{-1}$.  This transforms both the states and operators. A state $\ket{\psi_{J \beta}}$ in the old coordinate system transforming according to the IR $\Gamma_J$ becomes $\ket{\psi_{J' \beta'}} = \Tshift \ket{\psi_{J \beta}}$ transforming according to $\Gamma_{J'} = \Gamma_\tshift \times \Gamma_J$, with $\Gamma_\tshift = \{ \mathcal{D}_\tshift^\vek{k} (g) = \exp (-i \vek{b}_g \cdot \tshift ) : g \in \kgroup \}$ [Eq.~(\ref{eq:phase-d-amb-inv})].  Similarly, $\bra{\psi_{I \alpha}}$ transforming according to the IR $\Gamma_I^\ast$ becomes $\bra{\psi_{I' \alpha'}} = \bra{\psi_{I \alpha}} \Tshift^\dagger$ transforming according to $\Gamma_{I'}^\ast = \Gamma_\tshift^\ast \times \Gamma_I^\ast$.  Invariance of the operator $\mathcal{O}$ under translations $\Tshift$ implies $[\mathcal{O}, \Tshift] = 0$, or
\begin{equation}
  \label{eq:op-trans}
  \mathcal{O}' \equiv \Tshift \, \mathcal{O} \, \Tshift^{-1} = \mathcal{O} ,
\end{equation}
so that both $\mathcal{O}$ and $\mathcal{O}'$ transform according to the same representation (which need not be irreducible)
\begin{equation}
  \Gamma_\mathcal{O} = \Gamma_{\mathcal{O}'} .
\end{equation}
Hence, we get (note $\Gamma_\tshift^\ast \times \Gamma_\tshift = \Gamma_1$)
\begin{equation}
 \Gamma_{I'}^\ast \times \Gamma_{\mathcal{O}'} \times \Gamma_{J'}
 = \Gamma_I^\ast \times \Gamma_{\mathcal{O}} \times \Gamma_J ,
\end{equation}
i.e., in the unprimed and primed coordinate system we get the same selection rules.  In the primed coordinate system, the matrix elements become
\begin{subequations}\label{eq:me-trans}
  \begin{align}
    \mathcal{O}_{\alpha' \beta'}'
    & = \braket{\psi_{I' \alpha'} | \mathcal{O}' | \psi_{J' \beta'}}\\
    & = \braket{\psi_{I \alpha} | \Tshift^{-1} \Tshift \, \mathcal{O}
        \, \Tshift^{-1} \Tshift \,| \psi_{J \beta}}\\
    & = \braket{\psi_{I \alpha} | \mathcal{O} | \psi_{J \beta}}\\
    & = \mathcal{O}_{\alpha \beta} .
\end{align}
\end{subequations}
Note that for multidimensional IRs $\Gamma_J$ and $\Gamma_{J'} = \Gamma_\tshift \times \Gamma_J$, we can always choose the representation matrices such that  $\Tshift \ket{\psi_{J \beta}} = \ket{\psi_{J' \beta}}$, and similarly $\bra{\psi_{I \alpha}} \Tshift^\dagger = \bra{\psi_{I' \alpha}}$, so that Eq.\ (\ref{eq:me-trans}) becomes $\mathcal{O}_{\alpha \beta}' = \mathcal{O}_{\alpha \beta}$.

\section{Theory of Invariants}
\label{sec:theory-inv}

\subsection{General theory}

The Bloch eigenstates $\Psi_{n \vek{k}}^{I \beta} (\vek{r})$ at a wave vector $\vek{k}$ transform according to an IR $\Gamma_I$ of the group of the wave vector $\kgroup$.  The knowledge of these IRs suffices to construct the general form of the effective Hamiltonian characterizing the Bloch eigenstates near the expansion point $\vek{k}$ consistent with the symmetry operations in $\kgroup$. This method is known as the theory of invariants \cite{bir74}. The Hamiltonian can be expressed in terms of a general tensor operator denoted as $\vekc{K}$ that may depend on, e.g., the kinetic wave vector $\vek{\kappa}$ measured from $\vek{k}$, external electric and magnetic fields $\vekc{E}$ and $\vekc{B}$, strain $\vek{\epsilon}$ and spin $\vek{S}$. For all group elements $g \in \kgroup$, the Hamiltonian $\mathcal{H} (\vekc{K})$ obeys the invariance condition \cite{bir74}
\begin{equation}\label{eq:invariance-condition}
\mathcal{D} (g) \mathcal{H} (g^{-1} \vekc{K}) \mathcal{D}^{-1} (g) = \mathcal{H} (\vekc{K}).
\end{equation}
According to the theory of invariants, each block $\mathcal{H}_{I I'}$ of the matrix $\mathcal{H}$ corresponding to a pair of bands transforming according to the IRs $\Gamma_I$ and $\Gamma_{I'}$ has the form
\begin{equation}\label{eq:hamiltonian-blocks}
  \mathcal{H}_{I I'} (\vekc{K})
  = \sum_{i,J}a_{iJ}^{I I'} \sum_{l=1}^{L_J} X_l^J \mathcal{K}^{i,J*}_l ,
\end{equation}
where $J$ labels the $L_J$-dimensional IRs $\Gamma_J$ contained in the product representation $\Gamma_I^\ast \times \Gamma_{I'}$, the basis matrices $X_l^J$ and the irreducible tensor operators $\mathcal{K}_J^i$ constructed from the perturbations $\vekc{K}$ transform according to the IR $\Gamma_J$, and $a_{iJ}^{I I'}$ are constant prefactors. The index $i$ labels the irreducible tensor operators transforming as $\Gamma_J$.  In general, we have multiple blocks $\mathcal{H}_{I I'} (\vekc{K})$ corresponding to different bands $n$ and $n'$ transforming according to $\Gamma_I$ and $\Gamma_{I'}$.  To simplify the notation, we drop these additional indices.

\subsection{Effect of band IR rearrangements}

A translation of the coordinate system by $\tshift$ changes the IR of an eigenfunction $\Psi_{n\vek{k}}^I (\vek{r})$ from $\Gamma_I$ to $\Gamma_J = \Gamma_\tshift \times \Gamma_I$, see Eq.\ (\ref{eq:ir:trafo:sum}). We have
\begin{subequations}
  \begin{align}\label{eq:product-ir-tensor}
    \Gamma_J^\ast \times \Gamma_{J'}
    & =  (\Gamma_\tshift \times \Gamma_I)^\ast \times (\Gamma_\tshift \times \Gamma_{I'})\\
    & =  \Gamma_I^\ast \times \Gamma_{I'} ,
  \end{align}
\end{subequations}
where we used $\Gamma_\tshift^\ast \times \Gamma_\tshift = \Gamma_1$ for one-dimensional IRs $\Gamma_\tshift$.  Therefore, the IRs contained in $\Gamma_J^\ast \times \Gamma_{J'}$ are equal to the IRs contained in $\Gamma_I^\ast \times \Gamma_{I'}$.  This implies that translations $\tshift$ of the coordinate system resulting in a rearrangement of the IRs assigned to the eigenfunctions do not affect the invariant expansion of the Hamiltonian $\mathcal{H} (\vekc{K})$.

\subsection{Invariant Hamiltonian for M\lowercase{o}S$_2$}

As an application of the theory of invariants, we consider monolayer MoS$_2$.  In this material, the lowest conduction and highest valence band are at the points $\vek{K}$ and $\vek{K}'$ \cite{xia12}, hence we focus on these points (where $\kgroup[\vek{K}] = C_{3h}$).  Table \ref{tab:functions} lists the mapping of axial ($\vek{A}$) and polar vectors ($\vek{P}$) under the relevant symmetry operations. This allows one to confirm the examples for basis functions listed for $C_{3h}$ in Table~\ref{tab:char-c3h}.  Crystal momentum $\vek{\kappa}$ and an electric field $\vekc{E}$ transform like polar vectors, whereas spin $\vek{S}$ and a magnetic field $\vekc{B}$ transform like axial vectors.  Hence we immediately obtain from Table~\ref{tab:char-c3h} the lowest-order tensor operators listed in the second column of Table~\ref{tab:invariants}.  In general, we obtain higher-order tensor operators using the Clebsch-Gordan coefficients that are tabulated in, e.g., Ref.~\cite{kos63}.  However, this procedure is greatly simplified if all IRs of the relevant group are one-dimensional, which holds for $C_{3h}$.  In such a case, the higher-order tensor operators can be constructed using the multiplication table for the IRs that is reproduced for $C_{3h}$ in Table~\ref{tab:multiplication:C3h}.  Irreducible tensor operators are generally not unique.  Given two irreducible tensor operators $\mathcal{K}$ and  $\mathcal{K}'$ transforming according to the same IR $\Gamma_I$, any linear combinations of these tensors transforms likewise irreducibly according to $\Gamma_I$ \cite{win03}.  We exploit this freedom to choose linear combinations of irreducible tensors such as $\kappa_-^3$ and $\kappa_+^3$ that have also a well-defined behavior under time reversal symmetry (see Sec.~\ref{sec:invar-time}).

We can also consider the effects of strain. When stress deforms a crystalline solid, the symmetry of the system is altered which changes the energy spectrum of the material. Suppose under a deformation a point $\vek{r}$ in a solid undergoes a displacement $\vek{u} (\vek{r})$. For small homogeneous strain, the symmetric strain tensor is defined as ($i,j = x,y,z$) \cite{lan70}
\begin{equation}
\epsilon_{ij}=\frac{1}{2}\left( \frac{\partial u_i}{\partial r_j}+\frac{\partial u_j}{\partial r_i}+\frac{\partial u_k}{\partial r_i}\frac{\partial u_k}{\partial r_j}\right).
\end{equation}
However, considering MoS$_2$ as a quasi-2D material, strain due to a perpendicular stress component is not relevant.  The components $\epsilon_{ij}$ transform like the symmetrized products $\lbrace \kappa_i, \kappa_j \rbrace$ \cite{bir74} so that we get the lowest-order operators listed in the second column of Table~\ref{tab:invariants} while mixed higher-order tensor operators are listed in the third column of Table~\ref{tab:invariants}.

Since the IRs of $C_{3h}$ are all one-dimensional, the Hamiltonian blocks (\ref{eq:hamiltonian-blocks}) at the points $\vek{K}$ and $\vek{K}'$ of MoS$_2$ are one-dimensional. The $1\times 1$ basis matrices $X_1^J$ can be absorbed into the prefactors $a_{iJ}^{I I'}$. Hence, Eq.~(\ref{eq:hamiltonian-blocks}) can be simplified to
\begin{equation}\label{eq:1d-ham-blocks}
  \mathcal{H}_{I I'} (\vekc{K}) = \sum_ia_i^{I I'} \mathcal{K}^{i,J*}_l .
\end{equation}
The IRs corresponding to the eleven bands at $\vek{K}$ and $\vek{K}'$ that are dominated by the Mo~$d$ and S~$p$ orbitals are listed in Table \ref{tab:mos2-K-band-IR}. Here we focus on constructing a generic $6 \times 6$ Hamiltonian consisting of a sequence of bands transforming as $\Gamma_{1/3/2}, \Gamma_{2/1/3}, \Gamma_{3/2/1}, \Gamma_{4/6/5}, \Gamma_{5/4/6}$, and $\Gamma_{6/5/4}$, respectively (realized, e.g., by the bands $v_1, c_1, c_3, c_2, v_3$, and $v_2$; additional bands will replicate the behavior obtained for these bands). For these bands, the Hamiltonian matrix elements $\mathcal{H}^{\vek{K}}(\vekc{K})_{I I'}$ contain tensors $\mathcal{K}$ transforming as
\begin{equation}
  \label{eq:multiplication:C3h}
  (\Gamma_I^\ast \times \Gamma_{I'}) = \tvek[*{11}{c}]{%
    \Gamma_1 & \Gamma_2 & \Gamma_3 & \Gamma_4 & \Gamma_5 & \Gamma_6 \\
    \Gamma_3 & \Gamma_1 & \Gamma_2 & \Gamma_6 & \Gamma_4 & \Gamma_5 \\
    \Gamma_2 & \Gamma_3 & \Gamma_1 & \Gamma_5 & \Gamma_6 & \Gamma_4 \\
    \Gamma_4 & \Gamma_5 & \Gamma_6 & \Gamma_1 & \Gamma_2 & \Gamma_3 \\
    \Gamma_6 & \Gamma_4 & \Gamma_5 & \Gamma_3 & \Gamma_1 & \Gamma_2 \\
    \Gamma_5 & \Gamma_6 & \Gamma_4 & \Gamma_2 & \Gamma_3 & \Gamma_1 },
\end{equation}
see Table~\ref{tab:multiplication:C3h} (Appendix~\ref{sec:char-mult-tables}).

\begin{table*}
\caption{Irreducible tensor operators for the point group $C_{3h}$. For off-diagonal terms in the Hamiltonian $\mathcal{H} (\vekc{K})$, using the phase conventions in Table \ref{tab:tr-mos2-K}, the bold-face tensor operators give rise to invariants with purely real prefactors, while the remaining tensor operators give rise to invariants with purely imaginary prefactors.  At the same time, these phase conventions and time-reversal symmetry imply that invariants appearing on the diagonal of $\mathcal{H} (\vekc{K})$ (which transform as $\Gamma_1$) can only be formed from tensor operators listed in bold, the remaining tensor operators correspond to invariants that are forbidden by time reversal symmetry. Tensors transforming according to $\Gamma_3$ and $\Gamma_6$ are the Hermitean adjoint of the tensors transforming according to $\Gamma_2$ and $\Gamma_5$, respectively. Notation: $V_\pm = V_x \pm iV_y$ with $\vek{V} = \vek{\kappa}, \vekc{B}, \vekc{E}$, and $\epsilon_\pm = \epsilon_{xx}-\epsilon_{yy}\pm 2i\epsilon_{xy}$, $\epsilon_\| =\epsilon_{xx}+\epsilon_{yy}$.}
\label{tab:invariants}

\extrarowheight 0.3ex
$\begin{array}{ccl} \hline \hline
  & \text{lowest order} & \text{higher order} \\ \hline
  \Gamma_1 & \bm{\openone}; \bm{\mathcal{B}_z}; \bm{S_z}; \bm{\epsilon_\|} &
  \bm{\kappa^2}; \bm{\kappa_-^3 + \kappa_+^3};
  i\kappa_-^3 - i\kappa_+^3; \bm{\kappa^4};
  \kappa_- \mathcal{E}_+ + \kappa_+ \mathcal{E}_- ;
  \bm{i\kappa_- \mathcal{E}_+ - i\kappa_+ \mathcal{E}_-};
  \bm{\kappa^2 \mathcal{B}_z};
  \bm{\kappa^2 S_z};
  \\ &&
  \bm{\kappa_- \epsilon_- + \kappa_+ \epsilon_+};
  i\kappa_- \epsilon_- - i\kappa_+ \epsilon_+;
  \bm{\kappa^2 \epsilon_\|};
  \bm{\mathcal{B}_zS_z};
  \bm{\mathcal{B}_-S_+ + \mathcal{B}_+S_-};
  i\mathcal{B}_-S_+ - i\mathcal{B}_+S_-;
  \bm{\mathcal{B}_z\epsilon_\|};
  \\ &&
  \mathcal{E}_- \epsilon_- + \mathcal{E}_+ \epsilon_+;
  \bm{i\mathcal{E}_- \epsilon_- - i\mathcal{E}_+ \epsilon_+};
  \bm{S_z \epsilon_\|};
  \bm{\kappa_-\mathcal{B}_-S_- + \kappa_+\mathcal{B}_+S_+};
  i\kappa_-\mathcal{B}_-S_- - i\kappa_+\mathcal{B}_+S_+;
  \\ &&
  \mathcal{E}_z (\kappa_+S_- + \kappa_-S_+);
  \bm{i\mathcal{E}_z (\kappa_+S_- - \kappa_-S_+)};
  \bm{\mathcal{B}_z (\kappa_- \epsilon_- + \kappa_+ \epsilon_+)};
  i\mathcal{B}_z (\kappa_- \epsilon_- - \kappa_+ \epsilon_+);
  \\ &&
  \kappa_+ \mathcal{E}_+ \epsilon_- + \kappa_- \mathcal{E}_- \epsilon_+;
  \bm{i\kappa_+ \mathcal{E}_+ \epsilon_- -i\kappa_- \mathcal{E}_- \epsilon_+}
  \\
  \hline \Gamma_2 & \bm{\kappa_+}; \mathcal{E}_+; \bm{\epsilon_-} &
  \bm{\kappa_-^2}; \bm{\kappa^2\kappa_+}; \bm{\kappa^2\kappa_-^2};
  \bm{\kappa_+^4};
  \kappa_- \mathcal{E}_-;
  \kappa^2 \mathcal{E}_+;
  \bm{\kappa_+^2 \mathcal{E}_-};
  \bm{\kappa_+\mathcal{B}_z};
  \bm{\kappa_-^2 \mathcal{B}_z};
  \bm{\kappa_+S_z};
  \bm{\kappa_-^2 S_z};
  \bm{\kappa_- \epsilon_+};
  \bm{\kappa^2 \epsilon_-};
  \\ &&
  \mathcal{E}_+ S_z;
  \mathcal{E}_zS_+;
  \bm{\mathcal{E}_-\epsilon_+};
  \mathcal{B}_z\mathcal{E}_+;
  \mathcal{B}_+ \mathcal{E}_z;
  \bm{\mathcal{B}_-S_-};
  \bm{\kappa_+\mathcal{B}_+S_-};
  \kappa_+\mathcal{E}_+\epsilon_+
  \\
  \hline \Gamma_4  & \mathcal{E}_z &
  \bm{\kappa_+\mathcal{B}_- + \kappa_-\mathcal{B}_+};
  i\kappa_+\mathcal{B}_- - i\kappa_-\mathcal{B}_+;
  \bm{\kappa_+S_- + \kappa_-S_+};
  i\kappa_+S_- - i\kappa_-S_+;
  \\ &&
  \mathcal{B}_- \mathcal{E}_+ + \mathcal{B}_+ \mathcal{E}_-;
  \bm{i\mathcal{B}_- \mathcal{E}_+ -i\mathcal{B}_+ \mathcal{E}_-};
  \mathcal{E}_+S_- + \mathcal{E}_-S_+;
  \bm{i\mathcal{E}_+S_- - i\mathcal{E}_-S_+};
  \\ &&
  \mathcal{B}_-\epsilon_- + \mathcal{B}_+ \epsilon_+;
  \bm{i\mathcal{B}_-\epsilon_- - i\mathcal{B}_+\epsilon_+};
  S_-\epsilon_- + S_+ \epsilon_+;
  \bm{iS_-\epsilon_- - iS_+ \epsilon_+}
  \\
  \hline \Gamma_5 & \bm{\mathcal{B}_+}; \bm{S_+} &
  \kappa_+ \mathcal{E}_z;
  \kappa_-^2 \mathcal{E}_z;
  \bm{\kappa_-\mathcal{B}_-};
  \bm{\kappa_-S_-};
  \bm{\kappa^2\mathcal{B}_+};
  \bm{\kappa^2S_+};
  \bm{\kappa_+^2\mathcal{B}_-};
  \bm{\kappa_+^2S_-};
  \bm{\mathcal{E}_z\mathcal{E}_+};
  \mathcal{E}_-\mathcal{B}_-;
  \mathcal{E}_-S_-;
  \\ &&
  \bm{\mathcal{B}_+\mathcal{B}_z};
  \bm{\mathcal{B}_+ S_z};
  \bm{\mathcal{B}_zS_+};
  \bm{S_zS_+};
  \mathcal{E}_z\epsilon_-;
  \bm{\mathcal{B}_-\epsilon_+};
  \bm{S_-\epsilon_+};
  \kappa_+\mathcal{E}_-S_+;
  \bm{\kappa_-\mathcal{E}_z\epsilon_+};
  \bm{\kappa_+\mathcal{B}_+\epsilon_+};
  \bm{\kappa_+S_+\epsilon_+}
  \\
\hline \hline
\end{array}$
\end{table*}

\begin{table}
  \caption{IRs $\Gamma_{i/j/k}$ for the eleven TB bands \cite{cap13} in MoS$_2$ obtained from Mo $d$ and S $p$ orbitals at the $\vek{K}$ point for coordinate systems $\alpha = a/b/c$ in Fig.~\ref{fig:singlemos2}. We denote the bands by $c_4, \ldots, c_1 , v_1, \ldots, v_7$ arranged in order of decreasing energy. The seven bands $c_3, \ldots, v_4$ are used in Ref.~\cite{kor13}, while the three bands $c_3$, $c_1$, and $v_1$ are used in Ref.~\cite{liu13a}. The main (second) atomic orbital for each band is given in the third (fourth) column (compare Table~\ref{tab:band-ir-mos2-K}).}
  \label{tab:mos2-K-band-IR}
  $\extrarowheight0.3ex
  \begin{array}{cs{2em}cs{2em}cc}
     \hline \hline
     \text{band} & \text{IR} & \mbox{main orbital} & \mbox{second orbital} \\
     \hline
     c_4 & \Gamma_{6/5/4} & d_{xz} + i d_{yz} & p_z \\
     c_3 & \Gamma_{3/2/1} & d_{x^2 - y^2} - i d_{xy} & p_z \\
     c_2 & \Gamma_{4/6/5} & d_{xz} - i d_{yz} & p_x + ip_y \\
     c_1 & \Gamma_{2/1/3} & d_{z^2} & p_x - ip_y \\
     \hline
     v_1 & \Gamma_{1/3/2} & d_{x^2 - y^2} + i d_{xy} & p_x + ip_y \\
     v_2 & \Gamma_{6/5/4} & p_z & d_{xz} + i d_{yz} \\
     v_3 & \Gamma_{5/4/6} & p_x - ip_y \\
     v_4 & \Gamma_{3/2/1} & p_z & d_{x^2 - y^2} - i d_{xy} \\
     v_5 & \Gamma_{1/3/2} & p_x + ip_y & d_{x^2 - y^2} + i d_{xy} \\
     v_6 & \Gamma_{2/1/3} & p_x - ip_y & d_{z^2} \\
     v_7 & \Gamma_{4/6/5} & p_x + ip_y & d_{xz} - i d_{yz} \\ \hline \hline
   \end{array}$
\end{table}

\subsection{Time reversal}
\label{sec:invar-time}

At the points $\vek{K}$ and $\vek{K}' = -\vek{K}$ with $\kgroup[\vek{K}] = C_{3h}$, the Bloch functions $\Psi_{\vek{K}}^I$ and the corresponding time reversed functions $\Theta \, \Psi_{\vek{K}}^I$ are linearly dependent on each other [case $(a)$ according to Eq.\ (\ref{eq:herring}), see also Table~\ref{tab:band-ir-mos2-K}].  Also, the eigenfunctions at $\vek{K}$ can be mapped onto the eigenfunctions at $- \vek{K}$ by a vertical reflection $R = \sigma_v^{(2)}$ or a $180^{\circ} $ rotation $C_2'^{(2)}$, which is case (2) as defined in Eq.\ (\ref{eq:rashba:2}). [These two operations $R$ are elements of the group $D_{3h}$, the point group of MoS$_2$, see Fig.~\ref{fig:mos2-tlg-coord}(a).] Hence, for a Bloch state $\Psi_{\vek{K}}^I$ of band $I$, the time reversed state $\Theta \, \Psi_{\vek{K}}^I$ and the spatially transformed state $R \, \Psi_{\vek{K}}^I$, with $R = \sigma_v^{(2)}, C_2'^{(2)}$, obey the linear relation
\begin{equation}\label{eq:time-space-rel}
  \Theta \, \Psi_{\vek{K}}^I
  =  (\Psi_{\vek{K}}^I) ^\ast
  = t_R^I R \, \Psi_{\vek{K}}^I ,
\end{equation}
where $t_R^I$ is a phase factor (a unitary matrix if the dimensions of the IRs $\Gamma_I$ and $\Gamma_I^\ast$ was larger than one) that depends on the choice for the operation $R$. For the phase conventions used in Table \ref{tab:tr-mos2-K}, we have
\begin{subequations}
  \begin{align}
    t_{\sigma_v^{(2)}}^I & = \left\{
      \begin{array}{rs{1em}l}
        1, & I = 1, 2, 3 \\
        -1, & I = 4, 5, 6 ,
      \end{array}
      \right.  \\
    t_{C_2'^{(2)}}^I & = 1, \quad I = 1, \ldots, 6 .
  \end{align}
\end{subequations}
The matrix $\mathcal{H} (\vekc{K})$ must then satisfy the additional condition
\begin{equation}\label{eq:tr-cond}
(\mathcal{t}_R)^{-1} \, \mathcal{H} (R^{-1} \, \vekc{K}) \, \mathcal{t}_R
= \mathcal{H}^\ast (\zeta \, \vekc{K})
= \mathcal{H}^t (\zeta \, \vekc{K}) .
\end{equation}
where $\mathcal{t}_R$ is a diagonal matrix with elements $(\mathcal{t}_R)_{II} = t_R^I$, $\zeta=+1$ ($-1$) for quantities that are even (odd) under time reversal such as $\vekc{E}$ and $\vek{\epsilon}$ ($\vek{\kappa}$, $\vekc{B}$, and $\vek{S}$), $\ast$ denotes complex conjugation and $t$ transposition. The components of polar vectors $\vek{P}$ and axial vectors $\vek{A}$ transform under $\sigma_v^{(2)}$ and $C_2'^{(2)}$ as follows (see Table~\ref{tab:functions})
\begin{subequations}\label{eq:polaraxial}
\begin{align}
  \sigma_v^{(2)} \, P_\pm & = -P_\mp, &
  C_2'^{(2)} \, P_\pm & = -P_\mp, \\
  \sigma_v^{(2)} \, P_z   & = P_z, &
  C_2'^{(2)} \, P_z   & = -P_z, \\
  \sigma_v^{(2)} \, A_\pm & = A_\mp, &
  C_2'^{(2)} \, A_\pm & = -A_\mp, \\
  \sigma_v^{(2)} \, A_z   & = -A_z,  &
  C_2'^{(2)} \, A_z   & = -A_z .
\end{align}
\end{subequations}

\begin{table*}
\caption{Linear relation between the Bloch functions at $\vek{K}' = - \vek{K}$ obtained via time reversal $\Theta$ and via a spatial transformation $R$ applied to a Bloch function $\Psi_{\vek{K}}^I$ at wave vector $\vek{k} = \vek{K}$. The $\vek{K}$ point in the Brillouin zone is mapped onto $- \vek{K}$ by a vertical reflection $R = \sigma_v^{(2)}$ and by a rotation $R = C_2'^{(2)}$, see Fig.~\ref{fig:mos2-tlg-coord}(a).}
\label{tab:tr-mos2-K}

  \renewcommand{\arraystretch}{1.5} \extrarowheight0.2ex
  \arraycolsep 0.5em
$\begin{array}{c*{2}{s{1.6em}cs{1.2em}c}} \hline \hline
   \mathrm{IR} & \Psi_{\vek{K}}^I & \Theta \,\Psi_{\vek{K}}^I
   & \sigma_v^{(2)} \, \Psi_{\vek{K}}^I & C_2'^{(2)} \, \Psi_{\vek{K}}^I \\
   \hline
   \Gamma_1
   & \pwbase_\vek{K} (\vek{R}_j^{\textrm{Mo}}) \, (d_{x^2 - y^2} + i d_{xy});
   & \pwbase_{\vek{K}'} (\vek{R}_j^{\textrm{Mo}}) \, (d_{x^2 - y^2} - i d_{xy});
   & \Theta \,\Psi_{\vek{K}}^1 & \Theta \,\Psi_{\vek{K}}^1 \\
   & i \pwbase_\vek{K}^+ (\vek{R}_j^\textrm{S}) \, (p_x + ip_y)
   & -i \pwbase_{\vek{K}'}^+ (\vek{R}_j^\textrm{S}) \, (p_x - ip_y) \\
   \Gamma_2
   & \pwbase_\vek{K} (\vek{R}_j^{\textrm{Mo}}) \, d_{z^2};
   & \pwbase_{\vek{K}'} (\vek{R}_j^{\textrm{Mo}}) \, d_{z^2};
   & \Theta \,\Psi_{\vek{K}}^2 & \Theta \,\Psi_{\vek{K}}^2 \\
   & i \pwbase_\vek{K}^+ (\vek{R}_j^\textrm{S}) \, (p_x - ip_y)
   & - i \pwbase_{\vek{K}'}^+ (\vek{R}_j^\textrm{S}) \, (p_x + ip_y) \\
   \Gamma_3
   & \pwbase_\vek{K} (\vek{R}_j^{\textrm{Mo}}) \, (d_{x^2 - y^2} - i d_{xy});
   & \pwbase_{\vek{K}'} (\vek{R}_j^{\textrm{Mo}}) \, (d_{x^2 - y^2} + i d_{xy});
   & \Theta \,\Psi_{\vek{K}}^3 & \Theta \,\Psi_{\vek{K}}^3 \\
   & \pwbase_\vek{K}^- (\vek{R}_j^\textrm{S}) \, p_z
   & \pwbase_{\vek{K}'}^- (\vek{R}_j^\textrm{S}) \, p_z \\
   \Gamma_4
   & \pwbase_\vek{K} (\vek{R}_j^{\textrm{Mo}}) \, (d_{xz} - i d_{yz});
   & \pwbase_{\vek{K}'} (\vek{R}_j^{\textrm{Mo}}) \, (d_{xz} + i d_{yz});
   & - \Theta \,\Psi_{\vek{K}}^4 & \Theta \,\Psi_{\vek{K}}^4 \\
   & \pwbase_\vek{K}^- (\vek{R}_j^\textrm{S}) \, (p_x + ip_y)
   &  \pwbase_{\vek{K}'}^- (\vek{R}_j^\textrm{S}) \, (p_x - ip_y) \\
   \Gamma_5
   & \pwbase_\vek{K}^- (\vek{R}_j^\textrm{S}) \, (p_x - ip_y)
   & \pwbase_{\vek{K}'}^- (\vek{R}_j^\textrm{S}) \, (p_x + ip_y)
   & - \Theta \,\Psi_{\vek{K}}^5 & \Theta \,\Psi_{\vek{K}}^5 \\
   \Gamma_6
   & \pwbase_\vek{K} (\vek{R}_j^{\textrm{Mo}}) \, (d_{xz} + i d_{yz});
   & \pwbase_{\vek{K}'} (\vek{R}_j^{\textrm{Mo}}) \, (d_{xz} - i d_{yz});
   & - \Theta \,\Psi_{\vek{K}}^6 & \Theta \,\Psi_{\vek{K}}^6 \\
   & i \pwbase_\vek{K}^+ (\vek{R}_j^\textrm{S}) \, p_z
   & -i \pwbase_{\vek{K}'}^+ (\vek{R}_j^\textrm{S}) \, p_z \\ \hline \hline
\end{array}$
\end{table*}

It follows from Eq.\ (\ref{eq:tr-cond}) whether off-diagonal terms in $\mathcal{H} (\vekc{K})$ have real or imaginary prefactors.  Tensor operators that give rise to invariants with real prefactors are marked in bold in Table~\ref{tab:invariants}.  Furthermore, condition (\ref{eq:tr-cond}) provides a general criterion which terms are allowed by time-reversal symmetry on the diagonal of the Hamiltonian [when the tensor operators must transform according to the identity IR $\Gamma_1$, see Eq.\ (\ref{eq:multiplication:C3h})].  Here, our phase conventions imply that invariants (with real prefactors) can only be formed from tensor operators marked in bold in Table~\ref{tab:invariants}.  Thus, for example, on the diagonal of the Hamiltonian the third-order trigonal term $\kappa_+^3 + \kappa_-^3 = 2 \kappa_x(\kappa_x^2 - 3\kappa_y^2)$, as well as the field-dependent terms $i\kappa_- \mathcal{E}_+ - i\kappa_+ \mathcal{E}_- = 2 (\kappa_y \mathcal{E}_x - \kappa_x \mathcal{E}_y)$ and $\mathcal{B}_+S_- + \mathcal{B}_-S_+ = 2 (\mathcal{B}_x S_x + \mathcal{B}_y S_y)$ are allowed by symmetry and thus present in the Hamiltonian.  However, the terms $i\kappa_-^3 -i\kappa_+^3 = 2 \kappa_y (3\kappa_x^2 - \kappa_y^2)$,  $\kappa_+ \mathcal{E}_- + \kappa_- \mathcal{E}_+ = 2 (\kappa_x \mathcal{E}_x + \kappa_y \mathcal{E}_y)$ and $i\mathcal{B}_-S_+ -i\mathcal{B}_+S_- = 2 (\mathcal{B}_y S_x - \mathcal{B}_x S_y)$ are allowed by spatial symmetry; but these terms are forbidden by time-reversal symmetry and hence do not appear in the Hamiltonian. (They are allowed, though, as off-diagonal terms coupling different bands transforming according to the same IR, when these terms will have imaginary prefactors.)

Using the transformation $R$, one can derive the effective Hamiltonian for the valley $\vek{K}'$ using the transformation
\begin{equation}\label{eq:valleytransform}
\mathcal{H}^{\vek{K}'} (\vekc{K}) = \mathcal{H}^\vek{K} (R^{-1} \, \vekc{K}).
\end{equation}
Alternatively, time reversal can be used, see Eq.\ (\ref{eq:tr-cond}). Note that the definition of $\mathcal{H}^{\vek{K}'} (\vekc{K})$ relative to $\mathcal{H}^{\vek{K}} (\vekc{K})$ depends on the phase conventions used for the basis functions $\Psi_{\vek{K}'}^I$ relative to the phase conventions used for $\Psi_{\vek{K}}^I$.

\subsection{Analysis of the invariant expansion}

The invariant Hamiltonian at the point $\vek{K}$ of the BZ becomes in lowest order of the wave vector $\vek{\kappa}$ and spin-orbit coupling (spin $\vek{S}$)
\begin{equation}
  \label{eq:hamiltonian}
  \mathcal{H} (\vek{\kappa}, \vek{S})
  = \mathcal{H}_0 + \mathcal{H}_\kappa + \mathcal{H}_\mathrm{so}
\end{equation}
with
\begin{subequations}
  \begin{align}
    \mathcal{H}_0 & =
    \tvek[c*{5}{s{0.2em}c}]
    {E_1 & 0 & 0 & 0 & 0 & 0\\
    0 & E_2 & 0 & 0 & 0 & 0 \\
    0 & 0 & E_3 & 0 & 0 & 0 \\
    0 & 0 & 0 & E_4 & 0 & 0 \\
    0 & 0 & 0 & 0 & E_5 & 0 \\
    0 & 0 & 0 & 0 & 0 & E_6} ,
    \label{eq:hamiltonian-en} \\
    \mathcal{H}_{\vek{\kappa}} & =
    \tvek[c*{5}{s{0.2em}c}]
    {0 & \gamma_{12} \kappa_+ & \gamma_{13} \kappa_- & 0 & 0 & 0\\
    \gamma_{12} \kappa_- & 0 & \gamma_{23} \kappa_+ & 0 & 0 & 0 \\
    \gamma_{13} \kappa_+ & \gamma_{23} \kappa_- & 0 & 0 & 0 & 0 \\
    0 & 0 & 0 & 0 & \gamma_{45} \kappa_+ & \gamma_{46} \kappa_- \\
    0 & 0 & 0 & \gamma_{45} \kappa_- & 0 & \gamma_{56} \kappa_+ \\
    0 & 0 & 0 & \gamma_{46} \kappa_+ & \gamma_{56} \kappa_- & 0} ,
    \label{eq:hamiltonian-k} \\
    \mathcal{H}_\mathrm{so}
    & = \tvek[c*{5}{s{0.2em}c}]
        {\lambda_{11} S_z & 0 & 0 & 0 & \lambda_{15} S_+ & \lambda_{16} S_-\\
    0 & \lambda_{22} S_z & 0 & \lambda_{24} S_- & 0 & \lambda_{26} S_+ \\
    0 & 0 & \lambda_{33} S_z & \lambda_{34} S_+ & \lambda_{35} S_- & 0 \\
    0 & \lambda_{24} S_+ & \lambda_{34} S_- & \lambda_{44} S_z & 0 & 0 \\
    \lambda_{15} S_- & 0 & \lambda_{35} S_+ & 0 & \lambda_{55} S_z & 0 \\
    \lambda_{16} S_+ & \lambda_{26} S_- & 0 & 0 & 0 & \lambda_{66} S_z} ,
    \label{eq:hamiltonian-s}
  \end{align}
\end{subequations}
where $E_I$, $\gamma_{ij}$, and $\lambda_{ij}$ are material-dependent real parameters.

The lowest-order strain-dependent terms become
\begin{equation}\label{eq:hamiltonian-st}
  \mathcal{H}_{\vek{\epsilon}} =
  \tvek[c*{5}{s{0.2em}c}]
  {\xi_{11} \epsilon_\| & \xi_{12} \epsilon_- & \xi_{13} \epsilon_+ & 0 & 0 & 0\\
    \xi_{12} \epsilon_+ & \xi_{22} \epsilon_\| & \xi_{23} \epsilon_- & 0 & 0 & 0 \\
    \xi_{13} \epsilon_- & \xi_{23} \epsilon_+ & \xi_{33} \epsilon_\| & 0 & 0 & 0 \\
    0 & 0 & 0 & \xi_{44} \epsilon_\| & \xi_{45} \epsilon_- & \xi_{46} \epsilon_+ \\
    0 & 0 & 0 & \xi_{45} \epsilon_+ & \xi_{55} \epsilon_\| & \xi_{56} \epsilon_- \\
    0 & 0 & 0 & \xi_{46} \epsilon_- & \xi_{56} \epsilon_+ & \xi_{66} \epsilon_\|}
\end{equation}
with real parameters $\xi_{ij}$.  The effect of electric and magnetic fields $\vekc{E}$ and $\vekc{B}$ can also be included in the Hamiltonian (\ref{eq:hamiltonian}).  For the electric field, we add on the diagonal a scalar potential $e\vekc{E}\cdot \vek{r}$ and we replace crystal momentum by kinetic momentum $\hbar \vek{\kappa}=-i\hbar\vek{\nabla} +e\vek{A}$, where $\vek{A}$ is the vector potential due to the magnetic field.  In addition, we may have terms that depend explicitly on the fields $\vekc{E}$ and $\vekc{B}$. The in-plane components $\mathcal{E}_x$ and $\mathcal{E}_y$ transform spatially like the wave vector components $\kappa_x$ and $\kappa_y$. However, $\vekc{E}$ is even under time reversal symmetry, whereas $\vek{\kappa}$ is odd. In lowest order, the $\vekc{E}$-dependent terms become
\begin{equation}\label{eq:hamiltonian-El}
      \mathcal{H}_{\vek{\mathcal{E}}} =
    \tvek[c*{5}{s{0.2em}c}]
    {0 & \eta_{12} \mathcal{E}_+ & \eta_{13} \mathcal{E}_- & \eta_{14} \mathcal{E}_z & 0 & 0\\
    \eta^\ast_{12} \mathcal{E}_- & 0 & \eta_{23} \mathcal{E}_+ & 0 & \eta_{25} \mathcal{E}_z & 0 \\
    \eta^\ast_{13} \mathcal{E}_+ & \eta^\ast_{23} \mathcal{E}_- & 0 & 0 & 0 & \eta_{36} \mathcal{E}_z \\
    \eta^\ast_{14} \mathcal{E}_z & 0 & 0 & 0 & \eta_{45} \mathcal{E}_+ & \eta_{46} \mathcal{E}_- \\
    0 & \eta^\ast_{25} \mathcal{E}_z & 0 & \eta^\ast_{45} \mathcal{E}_- & 0 & \eta_{56} \mathcal{E}_+ \\
    0 & 0 & \eta^\ast_{36} \mathcal{E}_z & \eta^\ast_{46} \mathcal{E}_+ & \eta^\ast_{56} \mathcal{E}_- & 0}
\end{equation}
with imaginary prefactors $\eta_{ij}$. Since $\vekc{B}$
transforms in the same way as the spin $\vek{S}$ both under spatial transformations and time reversal, the $\vekc{B}$-dependent terms are similar to Eq.~(\ref{eq:hamiltonian-s})
\begin{equation}\label{eq:hamiltonian-B}
      \mathcal{H}_{\vekc{B}}
    = \tvek[c*{5}{s{0.2em}c}]
        {\beta_{11} \mathcal{B}_z & 0 & 0 & 0 & \beta_{15} \mathcal{B}_+ & \beta_{16} \mathcal{B}_-\\
    0 & \beta_{22} \mathcal{B}_z & 0 & \beta_{24} \mathcal{B}_- & 0 & \beta_{26} \mathcal{B}_+ \\
    0 & 0 & \beta_{33} \mathcal{B}_z & \beta_{34} \mathcal{B}_+ & \beta_{35} \mathcal{B}_- & 0 \\
    0 & \beta_{24} \mathcal{B}_+ & \beta_{34} \mathcal{B}_- & \beta_{44} \mathcal{B}_z & 0 & 0 \\
    \beta_{15} \mathcal{B}_- & 0 & \beta_{35} \mathcal{B}_+ & 0 & \beta_{55} \mathcal{B}_z & 0 \\
    \beta_{16} \mathcal{B}_+ & \beta_{26} \mathcal{B}_- & 0 & 0 & 0 & \beta_{66} \mathcal{B}_z}
\end{equation}
with real parameters $\beta_{ij}$.

We can project the multiband Hamiltonian (\ref{eq:hamiltonian}) on a subspace containing the bands of interest using quasidegenerate perturbation theory or L\"owdin partitioning \cite{win03}.  This gives rise to higher-order terms in the effective Hamiltonian that may include also mixed terms proportional to products of $\vek{\kappa}$, $\vek{S}$, $\vek{\epsilon}$, $\vekc{E}$, and $\vekc{B}$.  In particular, in the presence of a magnetic field $\vekc{B}$, the components of crystal momentum $\hbar \vek{\kappa} = - i\hbar\vek{\nabla} + e\vek{A}$ do not commute,  $\left[ \kappa_x, \kappa_y \right]=(e/i\hbar)\mathcal{B}_z$, so that antisymmetrized products of $\kappa_x$ and $\kappa_y$ appearing in higher-order perturbation theory give rise to terms proportional to $\mathcal{B}_z$.  Similarly, terms proportional to in-plane electric fields $\mathcal{E}_x, \mathcal{E}_y$ appear because of the presence of the scalar potential $e\vekc{E} \cdot \vek{r}$ and the relation $\left[ r_i, \kappa_j \right]=i\delta_{ij}$.

\section{Conclusions}
\label{sec:conclusion}

Starting from a TB approach, we have developed a comprehensive theory to derive the IRs characterizing the Bloch eigenstates in a crystal by decomposing the TB basis functions into localized symmetry-adapted atomic orbitals and crystal-periodic symmetry-adapted plane waves.  Both the symmetry-adapted atomic orbitals and the symmetry-adapted plane waves can easily be tabulated, thus accelerating the design and exploration of new materials.  The symmetry-adapted basis functions block-diagonalize the TB Hamiltonian, which naturally facilitates further analysis of the band structure.  While our analysis focused for clarity on symmorphic space groups, our theory can readily be generalized to nonsymmorphic groups.

The present work was motivated by the goal to develop a systematic theory of effective multiband Hamiltonians for the dynamics of Bloch electrons in external fields that break the symmetry of the crystal structure.  Yet our general symmetry analysis of Bloch states will likely be useful for other applications, too.

\begin{acknowledgments}
  We appreciate stimulating discussions with G.~Burkard,
  A.~Korm\'anyos, and U.~Z\"ulicke This work was supported by the
  NSF under Grant No.\ DMR-1310199.  Work at Argonne was supported
  by DOE BES under Contract No.\ DE-AC02-06CH11357.
\end{acknowledgments}

\appendix

\section{Projection operators}

A general method to identify the IRs of the eigenstates of a Hamiltonian uses projection operators \cite{wig59, streitwolf71, bir74, dre08}. Given a symmetry group $\group$ with IRs $\Gamma_I$, we can project a general function $f(\vek{r})$ onto its components $f_{I \beta} (\vek{r})$ transforming according to the $\beta$th component of the IR $\Gamma_I$ of $\group$.  Here the projection operators $\Pi_{I \beta}$ are
\begin{equation}
  \label{eq:projection-op}
  \Pi_{I \beta} \equiv
  \frac{n_I}{h} \sum_g \: \mathcal{D}_I(g)^\ast_{\beta \beta} \, P(g) ,
\end{equation}
where $h$ is the order of the group $\group$, $n_I$ is the dimensionality of the IR $\Gamma_I$, $\mathcal{D}_I(g)$ are the representation matrices for $\Gamma_I$, and $P(g)$ are the symmetry operators corresponding to $g \in \group$. Often, we denote $P(g)$ simply as $g$.  If we are not interested in a particular component $\beta$ of $\Gamma_I$, we can use the ``coarse-grained'' projection operators
\begin{equation}
  \label{eq:projection-op-char}
  \Pi_I \equiv \sum_\beta \: \Pi_{I \beta}
  = \frac{n_I}{h} \sum_g \: \chi_I(g)^\ast \, P(g),
\end{equation}
where $\chi_I (g) \equiv \trace \mathcal{D}_I(g)$ are the characters for $\Gamma_I$.  For one-dimensional IRs, the operators $\Pi_I$ become equivalent to $\Pi_{I \beta}$.  The projection operators obey the completeness relation
\begin{equation}
  \sum_{I, \beta} \: \Pi_{I \beta} = \sum_I \: \Pi_I = \openone .
\end{equation}

\section{Rearrangement lemma for IRs}
\label{sec:rearrange:lemma}

Generally, if $\Gamma_0$ is a one-dimensional IR of a group $\group$, then for any IR $\Gamma_I$ of $\group$, the product representation $\Gamma_{I'} = \Gamma_0 \times \Gamma_I$ is irreducible.  If two IRs $\Gamma_I$ and $\Gamma_J$ of $\group$ are (in)equivalent, then the IRs $\Gamma_{I'} = \Gamma_0 \times \Gamma_I$ and $\Gamma_{J'} = \Gamma_0 \times \Gamma_J$ are also (in)equivalent. Hence, a multiplication of the IRs of $\group$ by $\Gamma_0$ simply rearranges the sequence of IRs of $\group$. These statements can be proven as follows: For the element $g \in \group$, we denote the characters for the IRs $\Gamma_0$ and $\Gamma_I$ by $\chi_0(g)$ and $\chi_I(g)$, respectively. Hence the character of the representation $\Gamma_{I'}$ becomes $\chi_{I'} (g) = \chi_0(g) \, \chi_I(g)$. Since $\Gamma_0$ is one-dimensional, its characters are also its unitary representation matrices obeying $|\chi_0(g)|^2 = 1$. Using the orthogonality relations for characters we get
\begin{subequations}
  \begin{align}
    \sum_g |\chi_{I'} (g) |^2
    & = \sum_g |\chi_0(g) \, \chi_I(g)|^2\\
    & = \sum_g |\chi_0(g)|^2 \, |\chi_I(g)|^2\\
    & = \sum_g |\chi_I(g)|^2 = h ,
  \end{align}
\end{subequations}
where $h$ is the order of the group. Hence $\Gamma_{I'} = \Gamma_0 \times \Gamma_I$ is irreducible if $\Gamma_I$ is irreducible. Now consider two IRs $\Gamma_{I'} = \Gamma_0 \times \Gamma_I$ and $\Gamma_{J'} = \Gamma_0 \times \Gamma_J$. Using again the orthogonality relations for characters we get
\begin{subequations}
  \begin{align}
    \sum_g \chi_{I'}^\ast(g) \chi_{J'} (g)
    & = \sum_g \chi_I^\ast(g) \, \chi_0^\ast(g) \, \chi_0(g) \, \chi_J(g)\\
    & = \sum_g \chi_I^\ast(g) \, \chi_J(g) = h\, \delta_{IJ},
  \end{align}
\end{subequations}
so that $\Gamma_{I'}$ and $\Gamma_{J'}$ are indeed (in)equivalent if $\Gamma_I$ and $\Gamma_J$ are (in)equivalent.

\section{Group tables}
\label{sec:char-tables}

Character tables and basis functions for the groups $D_{3h}$,
$D_{2h}$, $C_{3h}$, $C_{2v}$, and $C_{2h}$ are reproduced in
Tables~\ref{tab:char-d3h} -- \ref{tab:char-c2h} following the
designations by Koster \emph{et al.} \cite{dre08}. The second column
in each table gives the designations for the IRs used by Dresselhaus
\emph{et al.} \cite{dre08}.  Examples of basis functions that
transform irreducibly according to the different IRs are listed in
the last column.  These functions are expressed in terms of the
cartesian components of polar (\vek{P}) and axial (\vek{A}) vectors
using the coordinate systems defined for the respective groups in
Figs.~\ref{fig:mos2-tlg-coord}, \ref{fig:slg-coord}, and \ref{fig:blg-coord}.  Finally, we reproduce in Table~\ref{tab:multiplication:C3h} the multiplication table for the IRs of the group $C_{3h}$ \cite{kos63}.

\begin{table}[H]
  \caption{Character table and basis functions for the group $D_{3h}$ \cite{kos63}. The coordinate system for the basis functions is defined in Fig.~\ref{fig:mos2-tlg-coord}(a).}
  \label{tab:char-d3h}
  \centerline{\extrarowheight0.4ex
  $\begin{array}{*{9}c}
     \hline \hline
      & & E & 2C_3 & 3C_2'^{(i)} & \sigma_h & 2S_3 & 3\sigma_v^{(jk)} & \mathrm{bases} \\
     \hline
     \Gamma_1 & A_1'  & 1 & 1 & 1 & 1 & 1 & 1 & \openone \\
     \Gamma_2 & A_2'  & 1 & 1 & -1 & 1 & 1 & -1 & A_z \\
     \Gamma_3 & A_1'' & 1 & 1 & 1 & -1 & -1 & -1 & P_z A_z \\
     \Gamma_4 & A_2'' & 1 & 1 & -1 & -1 & -1 & 1 & P_z \\
     \Gamma_5 & E''   & 2 & -1 & 0 & -2 & 1 & 0 & A_x, A_y \\
     \Gamma_6 & E'    & 2 & -1 & 0 & 2 & -1 & 0 & P_x, P_y \\
     \hline \hline
  \end{array}$}
\end{table}

\begin{table}[H]
  \caption{Character table and basis functions for the group $D_{2h}$ \cite{kos63}. The coordinate system for the basis functions is defined in Fig.~\ref{fig:slg-coord}(c-e).}
  \label{tab:char-d2h}
  \centerline{\extrarowheight0.4ex
  $\begin{array}{*{12}c}
     \hline \hline
      & & E & C_2 & C_2' & C_2'' & I & \sigma_v & \sigma_v' & \sigma_v'' & \mathrm{bases}~[c(e)] & [d] \\
     \hline
     \Gamma_1^+ & A_{1g} & 1 & 1 & 1 & 1 & 1 & 1 & 1 & 1 & \openone & \openone \\
     \Gamma_2^+ & B_{2g} & 1 & -1 & 1 & -1 & 1 & -1 & 1 & -1 & \sqrt{3} A_x \pm A_y & A_y \\
     \Gamma_3^+ & B_{1g} & 1 & 1 & -1 & -1 & 1 & 1 & -1 & -1 & A_z & A_z \\
     \Gamma_4^+ & B_{3g} & 1 & -1 & -1 & 1 & 1 & -1 & -1 & 1 & A_x \mp \sqrt{3} A_y & A_x \\
     \Gamma_1^- & A_{1u} & 1 & 1 & 1 & 1 & -1 & -1 & -1 & -1 & & P_x P_y P_z \\
     \Gamma_2^- & B_{2u} & 1 & -1 & 1 & -1 & -1 & 1 & -1 & 1 & \sqrt{3} P_x \pm P_y & P_y \\
     \Gamma_3^- & B_{1u} & 1 & 1 & -1 & -1 & -1 & -1 & 1 & 1 &P_z & P_z \\
     \Gamma_4^- & B_{3u} & 1 & -1 & -1 & 1 & -1 & 1 & 1 & -1 & P_x \mp \sqrt{3} P_y & P_x \\
     \hline \hline
  \end{array}$}
\end{table}

\begin{table}[H]
  \caption{Character table and basis functions for the group $C_{3h}$ with $\omega \equiv \exp (i\pi/6)$ \cite{kos63}. The coordinate system for the basis functions is defined in Fig.~\ref{fig:mos2-tlg-coord}(b).}
  \label{tab:char-c3h}
  \centerline{\extrarowheight0.4ex
  $\begin{array}{*{9}c}
     \hline \hline
             & & E & C_3       & C_3^{-1}     & \sigma_h & S_3       & S_3^{-1} & \mathrm{bases}\\
     \hline
     \Gamma_1 & A' & 1 & 1 & 1 & 1 & 1 & 1 & \openone; A_z \\
     \Gamma_2 & E_1' & 1 & \omega^4 & \omega^{-4} & 1 & \omega^4 & \omega^{-4} & P_+\\
     \Gamma_3 & E_2' & 1 & \omega^{-4} & \omega^4  & 1 & \omega^{-4} & \omega^4 & P_- \\
     \Gamma_4 & A'' & 1 & 1 & 1 & -1 & -1 & -1 & P_z \\
     \Gamma_5 & E_1'' & 1 & \omega^4 & \omega^{-4} & -1 & -\omega^4 & -\omega^{-4} & A_+ \\
     \Gamma_6 & E_2'' & 1 & \omega^{-4} & \omega^4 & -1 & -\omega^{-4} & -\omega^4 & A_- \\
     \hline \hline
  \end{array}$}
\end{table}

\begin{table}[H]
  \caption{Character table and basis functions for the group $C_{2v}$ \cite{kos63}. The coordinate system for the basis functions is defined in Fig.~\ref{fig:mos2-tlg-coord}(c-e).}
  \label{tab:char-c2v}
  \centerline{\extrarowheight0.4ex
  $\begin{array}{*{8}c}
     \hline \hline
     & & E & C_2 & \sigma_v & \sigma_v' & \mathrm{bases}~[c(e)] & [d] \\
     \hline
     \Gamma_1 & A_1 & 1 & 1 & 1 & 1 & \openone; \sqrt{3} P_x \pm P_y & \openone; P_y \\
     \Gamma_2 & B_1 & 1 & -1 & 1 & -1 & P_x \mp \sqrt{3} P_y; A_z & P_x; A_z \\
     \Gamma_3 & A_2 & 1 & 1 & -1 & -1 & \sqrt{3} A_x \pm A_y & A_y \\
     \Gamma_4 & B_2 & 1 & -1 & -1 & 1 & P_z; A_x \mp \sqrt{3} A_y  & P_z; A_x \\
     \hline \hline
   \end{array}$}
\end{table}

\begin{table}[H]
  \caption{Character table and basis functions for the group $C_{2h}$ \cite{kos63}. The coordinate system for the basis functions is defined in Fig.~\ref{fig:blg-coord}(c-e).}
  \label{tab:char-c2h}
  \centerline{\extrarowheight0.4ex
  $\begin{array}{*{8}c}
     \hline \hline
      & & E & C_2 & I & \sigma_h & \mathrm{bases}~[c(e)] & [d] \\
     \hline
     \Gamma_1^+ & A_g & 1 & 1 & 1 & 1 & \openone; A_x \mp \sqrt{3} A_y & \openone; A_x \\
     \Gamma_2^+ & B_g & 1 & -1 & 1 & -1 & \sqrt{3} A_x \pm A_y; A_z & A_y; A_z \\
     \Gamma_1^- & A_u & 1 & 1 & -1 & -1 & P_x \mp \sqrt{3} P_y & P_x \\
     \Gamma_2^- & B_u & 1 & -1 & -1 & 1 & \sqrt{3} P_x \pm P_y ; P_z & P_y ; P_z \\
     \hline \hline
  \end{array}$}
\end{table}

\label{sec:char-mult-tables}
\begin{table}[H]
    \caption{Multiplication table for the IRs of the group $C_{3h}$ \cite{kos63}.}
  \label{tab:multiplication:C3h}
  \centerline{\extrarowheight0.4ex
  $\begin{array}{*{5}{cs{1em}}cs{2em}c} \hline \hline
     \Gamma_1 & \Gamma_2 & \Gamma_3 & \Gamma_4 & \Gamma_5 & \Gamma_6
     & \Gamma_i \times \Gamma_j \\
     \hline
     \Gamma_1 & \Gamma_2 & \Gamma_3 & \Gamma_4 & \Gamma_5 & \Gamma_6 & \Gamma_1\\
              & \Gamma_3 & \Gamma_1 & \Gamma_5 & \Gamma_6 & \Gamma_4 & \Gamma_2\\
              &          & \Gamma_2 & \Gamma_6 & \Gamma_4 & \Gamma_5 & \Gamma_3\\
              &          &          & \Gamma_1 & \Gamma_2 & \Gamma_3 & \Gamma_4\\
              &          &          &          & \Gamma_3 & \Gamma_1 & \Gamma_5\\
              &          &          &          &          & \Gamma_2 & \Gamma_6\\ \hline \hline
   \end{array}$}
\end{table}


%

\end{document}